\documentclass[annual]{acmsiggraph}

\TOGonlineid{}

\TOGvolume{}
\TOGnumber{}

\TOGarticleDOI{}

\TOGprojectURL{}
\TOGvideoURL{}
\TOGdataURL{}
\TOGcodeURL{}

\title{A reciprocal formulation of non-exponential radiative transfer.   1: Sketch and motivation}

\author{ Eugene d'Eon \\
8i }

\pdfauthor{ }

\keywords{Linear transport theory,
Kinetic theory,
Generalized linear Boltzmann equation,
Reciprocity,
Stochastic media}

\usepackage{smrdefaults}

\usepackage{color}

\usepackage{multirow}

\usepackage{rotating}
\newcommand{\new}[1]{#1}
\newcommand{\newtwo}[1]{#1}
\newcommand{\remove}[1]{}

\newcommand{\mfp}{\langle s \rangle}  
\newcommand{\mfpsqr}{\langle s^2 \rangle} 
\newcommand{\smin}{\hat{s}}  

\newcommand{\vect}[1]{\vec{#1}}

\newcommand{\e}{\text{e}}

\newcommand{\pos}{\mathbf{x}}

\newcommand{\n}{\vect{\mathbf{n}}}

\newcommand{\dir}{\Omega}

\def\rm{R^{-}}

\usepackage{paralist}
\usepackage{subfigure}
\usepackage{rotating}
\usepackage{amsgen}

\usepackage{alltt}
\usepackage{ulem}

\begin{document}

\normalem




\maketitle


\begin{abstract}
  Previous proposals to permit non-exponential free-path statistics in radiative transfer have not included support for volume and boundary sources that are spatially uncorrelated from the scattering events in the medium.  Birth-collision free paths are treated identically to collision-collision free paths and application of this to general, bounded scenes with inclusions leads to non-reciprocal transport.  Beginning with reciprocity as a desired property, we propose a new way to integrate non-exponential transport theory into general scenes.  We distinguish between the free-path-length statistics between correlated medium particles and the free-path-length statistics beginning at locations not correlated to medium particles, such as boundary surfaces, inclusions and uncorrelated sources.  Reciprocity requires that the uncorrelated free-path distributions are simply the normalized transmittance of the correlated free-path distributions.  
  The combination leads to an equilibrium imbedding of a previously derived generalized transport equation into bounded domains.
  We compare predictions of this approach to Monte Carlo simulation of multiple scattering from negatively-correlated suspensions of monodispersive hard spheres in bounded two-dimensional domains and demonstrate improved performance relative to previous work.  We also derive new, exact, reciprocal, single-scattering solutions for plane-parallel half-spaces over a variety of non-exponential media types.
\end{abstract}




\keywordlist






\section{Introduction}

  After more than 125 years of continued utility in many fields~\cite{mishchenko13}, radiative transfer is undergoing a nascent generalization, termed \emph{Generalized Radiative Transfer (GRT)}, to consider light and neutral particle transport in random volumes with scattering centers that are correlated in their relative positions.  Practical motivations for this extension include observations of non-exponential attenuation laws for light in atmospheric scattering and other settings~\cite{kostinski01,davis04,davis2011two,davis2011radiation}, and similarly for neutrons in pebble-bed reactors~\cite{larsen11,vasques13} and in using linear transport methodology as a condensed-history accelerator for light transport in discrete random media~\cite{moon07}.  Non-exponential transport should be preferred in the general case, if only marginally, given that the approach preserves a physical property of random systems not preserved by the atomic mix approximation and therefore should generally provide better predictions for random configurations of finite-sized scatterers.

  Larsen and Vasques~\shortcite{larsen11} presented a \emph{generalized linear Boltzmann equation} (GLBE) for infinite medium problems where the relative positions of particle birth and medium collisions are correlated.  \new{Similar approaches have also appeared~\cite{golse12,rukolaine2016generalized} and have been shown to be basically equivalent~\cite{larsen17}}.  By extending the concept of total macroscopic cross section $\Sigma_{tc}(s)$ to include a dependence on the path length parameter $s$---the distance since the previous collision or birth---new non-exponential distributions
  \begin{equation}\label{eq:pc}
  	p_c(s) = \Sigma_{tc}(s) \e^{-\int_0^s \Sigma_{tc}(s') ds'}
  \end{equation}
  give the free-path-length statistics between pairs of \emph{correlated} medium events.  The phase function and single-scattering albedo were not similarly generalized, remaining independent of $s$.  The GLBE was extended shortly after~\cite{vasques13} to support asymmetric scattering in anisotropic random media with a macroscopic cross section $\Sigma_{tc}(s,\dir)$ depending on both $s$ and the direction of flight $\dir$.

  With the macroscopic cross section $\Sigma_{tc}(s)$ no longer a constant, the trivial conversion from collision rate density to radiance is lost and so an integro-differential form of the GLBE can only be written provided the spectrum of radiances $L(\pos,\dir)$ over the memory variable $s$ is known
  \begin{equation}
  	L(\pos,\dir) = \int_0^\infty L(\pos,\dir,s) ds.
  \end{equation}
  Here $L(\pos,\dir)$ is the classical mono-energetic time-independent quantity of radiance (vector-flux) describing the density of energy in flight at a specific position and direction.  This spectral decomposition, much like one over energy or wavelength, is required because the incremental probability of experiencing an interaction while traveling an incremental distance $ds$ is $\Sigma_{tc}(s) ds$.
  Given this decomposition, the GLBE becomes
  \begin{multline}\label{eq:GLBE}
  	\frac{\partial L}{\partial s}(\pos,\dir,s) + \dir \cdot \nabla L(\pos,\dir,s) + \Sigma_{tc}(s) L(\pos,\dir,s)  \\ = \delta(s)  c  \int_{4\pi} \int_0^\infty P(\dir' \cdot \dir) \Sigma_{tc}(s') L(\pos,\dir',s') ds' d\dir' + \delta(s) \frac{Q(\pos)}{4 \pi}
  \end{multline}
  where the $\delta(s)$ terms impart the initialization $s=0$ to all particles born by $Q$ or in-scattered by the integral.
  Here, $c$ is the single-scattering albedo, $P$ the phase function, and $Q$ an isotropically-emitting correlated volume source. 
  A traditional integral form of the transport equation over collision rate densities without the appearance of $s$ was also presented~\cite{larsen11} and prescribes Monte Carlo estimators for transport problems using the generalized free-path sampling $p_c(s)$ after birth and between medium interactions. The \emph{correlated-origin transmittance} function for the medium
  \begin{equation}
    X_c(s) = 1 - \int_0^s p_c(s') ds'
  \end{equation}
  gives the probability to fly uncollided from a medium collision to a detector or medium boundary a distance $s$ away.  Expected-value estimators for radiance, fluence or escape probability can therefore be generalized by replacing the classical exponential transmittance $e^{-\Sigma_t s}$ with $X_c(s)$.  While this provides a complete and consistent formulation for infinite medium transport problems with correlated emission, there are a number of subtle limitations that are important to clarify when extending this approach more generally to bounded scenes.

  \paragraph{Equivalence of birth and collision}
  The summation in Eq.~\ref{eq:GLBE} of the source term $Q$ with the integral of incident collisions demands that both newborn and scattered photons continue to their next event with the same correlation-driven statistics, dictated by $\Sigma_{tc}(s)$.  As defined by Larsen and Vasques, $\Sigma_{tc}(s)$ is estimated or chosen such that the correlated free-path distribution $p_c(s)$ preserves the ensemble average of distances between medium collisions.  This distribution is not equivalent to the distribution of free paths where the origin of the path is an independent random starting location in the medium.  Thus, as written, the GLBE only describes volume emission from locations that are correlated to the scattering particles in the same way that they are correlated to themselves.  While highly appropriate for neutron transport \new{and thermal emission in cloud interiors}, it is of broader interest to support uncorrelated emission and to further include incident energy at medium boundaries, and support for reflective boundaries and other general inclusions in the medium in a way that is uncorrelated to the scattering centers in the volume.  

  The appropriate way to achieve uncorrelated sources is not universally agreed.  While Frank et al.~\shortcite{frank15} caution that application of the correlated free path distribution $p_c(s)$ with initialization $s = 0$ at boundary interactions is not necessarily appropriate, Davis et al.~\shortcite{davis14,davis18} and others~\cite{wrenninge17,larsen17} have taken exactly this approach in atmospheric scattering and computer graphics, at the cost of abandoning Helmholtz reciprocity.  
  We propose an alternative approach that uses two distinct free path distributions for the medium: one for medium-correlated origins and one for medium-uncorrelated origins.  For medium-correlated origins, we use the same free-path distributions as proposed by Larsen and Vasques to preserve the ensemble-average distances between collisions and accurate diffusion asymptotics for bulk scattering.  For the uncorrelated case, we assume that a new form of weak reciprocity holds in bounded media and then determine the uncorrelated sampling procedure at boundary sources that achieves it.  We find this leads to a simple, consistent formulation of bounded non-exponential transport.

  The rest of the paper is as follows.  In the next section we derive our reciprocal imbedding of the GLBE into bounded media and discuss the application of the uncorrelated-free-path distribution to scenes with imbedded inclusions and uncorrelated emitters.  In Section~\ref{sec:MC} we describe the Monte Carlo simulation methods used to compute transport statistics in blue noise random media where scattering centers are required to be separated by a minimum distance, relating to hard sphere packings in two-phase random media and more generally to repelling particles.  These methods are used in \new{S}ection~\ref{sec:blue} to study the correlated and uncorrelated free paths in infinite configurations of blue noise media and simple approximate analytic forms of these statistics are derived.  These distributions are then used to make predictions using our transport formalism, which are compared to Monte Carlo simulations of transmission through purely-absorbing slabs and low-order scattering from half-spaces.  In \new{S}ection~\ref{sec:determ} we include additional deterministic predictions of our formalism for other families of non-exponential free path distributions.



  \section{Bounded Generalized Boltzmann \new{T}ransport}

    We would like to extend the GLBE of Larsen and Vasques to include medium boundaries in a way that maintains reciprocity.  We define a weak form of reciprocity for correlated media transport and a sampling procedure and associated transport formalism that exhibits it.  The relationship between medium-correlated and medium-uncorrelated emission and detection are discussed.

    \subsection{Terminology}
    Throughout the paper we will distinguish between events that begin with correlated (``c'') and uncorrelated (``u'') origins and use these labels for two distinct varieties of distributions, means and radiances.  To be clear, in both cases the statistics of the random walks are influenced by the correlated nature of the underlying medium.  These labels refer to the path-step \emph{origins} and their statistical relationship to the particles in the medium.

    \subsection{Reciprocity Thought Experiment}

\begin{figure}
		  \centering
		  \includegraphics[width=.6\linewidth]{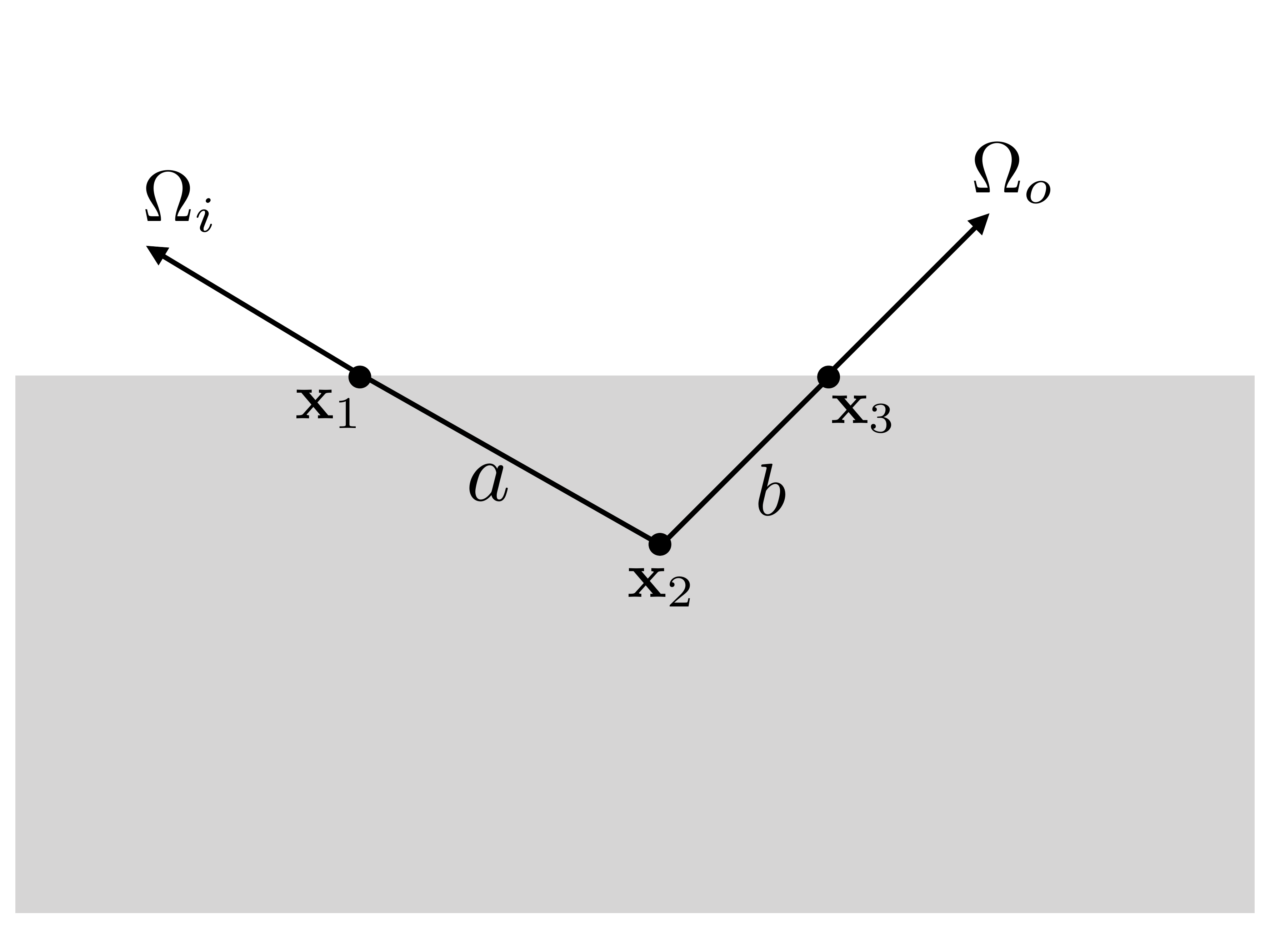}
		  \caption{Thought experiment for reciprocal single-scattering from a non-classical half-space.}
		  \label{fig-ss} 
	\end{figure}

    Consider the set of single-scattering paths within a homogeneous half-space with vacuum boundary conditions and light arriving from a direction $\dir_i$ and leaving along $\dir_o$ (Figure~\ref{fig-ss}).
	We would like the extension of the GLBE to this scenario to predict the ensemble average of single-scattering over all random realizations inside the medium under the assumptions of geometrical optics and neglecting coherent backscattering.  Under these assumptions, single-scattering from any specific realization is reciprocal, so we expect the ensemble-averaged transport to be as well, with one key difference: we require only that the \emph{total} path throughput between the two medium-uncorrelated boundary locations is the same regardless of transport direction (ie. whether $a$ or $b$ was sampled first).  We do not require identical collision density at $\pos_2$ nor identical transmittance along the segments of lengths $a$ and $b$.

	\subsection{A weak reciprocity condition for correlated medium transport}
	We briefly give a general definition of this non-traditional form of reciprocity before returning to our thought experiment.  The form of reciprocity we seek does not require identical throughput along every path segment nor at every path vertex.  Instead, we require a weaker condition: identical path throughput for both forward and adjoint directions along any transport path or subpath with two medium-uncorrelated end points.  If either endpoint of a path or subpath is a birth/collision/detection event that is statistically correlated to other vertices in the path space (regardless of whether those vertices are in the current path or not), then we do not require reciprocity over that path or subpath.  
	This is very similar to the non-traditional reciprocity exhibited when expressing Smith microsurface scattering from random height fields as a generalized form of linear transport with asymmetric cross sections ($\Sigma_t(\dir) \neq \Sigma_t(-\dir)$)~\cite{heitz15,dupuy16}.

	Returning to our thought experiment, we begin with what is already determined in the forward direction by the formalism of Larsen and Vasques: the single scattering albedo at the collision $c$, the \emph{phase function} $P(\dir_i \cdot \dir_o)$, and the probability of escaping the medium along the exitant path of length $b$.  The escape probability must be $X_c(b)$ because we must select subsequent scattering events in the medium with $p_c(s)$ to preserve the ensemble-average path lengths between the correlated scattering centers.
    The unknown to be determined is then the path-length distribution for sampling path length $a$ when entering the medium.  Because this sampling distribution is used only at boundary locations that are \emph{uncorrelated} to the scattering centers in the volume we denote it $p_u(s)$, the \emph{uncorrelated free-path-length distribution} for the medium.  We will later use this free-path distribution for sampling all paths inside of random media that have an uncorrelated starting position.

    To determine what $p_u(s)$ should be in order to satisfy our reciprocity condition, we compare the analog probability densities for the single-scattering path to occur in both the forward and adjoint directions.  In the forward direction we find that the analog probability $A_f \, da \, d\Omega_o$ of a photon taking this path is
    \begin{equation}
		A_f \, da \, d\Omega_o =  c \, P(\dir_i \cdot \dir_o) d\Omega_o \, X_c(b) \, p_u(a) da,
	\end{equation}
	which involves of the density of first colliding at a distance $a$ from the boundary, $p_u(a) da$, the probability that the collision is a scattering event, $c$, the density of selecting angle $\dir_o$, $P(\dir_i \cdot \dir_o) d\Omega_o$, and the probability of escaping along a path of length $b$, $X_c(b)$. In the adjoint direction we find \new{for the adjoint probability $A_a \, db\, d\Omega_i$},
	\begin{equation}
		A_a \, db \, d\Omega_i =  c \, P(\dir_o \cdot \dir_i) d\Omega_i \, X_c(a) \, p_u(b) db.
	\end{equation}
	We propose determining $p_u(s)$ by requiring that the two analog densities are equal for all directions $\dir_i, \dir_o$ that create a connected path (thus, for all pairs of path lengths $a,b > 0$).  \new{Assuming that the phase function $P$ is reciprocal and given that $c$ is also}, reciprocity is only attained if uncorrelated free-path sampling $p_u(s)$ is proportional to correlated transmittance $X_c(s)$.  The proportionality constant is determined by the requirement that $p_u(s)$ be a normalized distribution of path lengths.  Thus, we require that the distribution of uncorrelated free paths is the normalized correlated transmittance
	\begin{equation}
		p_u(s) = \frac{X_c(s)}{\int_0^\infty X_c(s) ds}.
	\end{equation}
	Larsen and Vasques~\shortcite{larsen11} studied this distribution, called the equilibrium spectrum of path lengths, and proved that the normalization constant is simply the \emph{mean-correlated-free-path length} $\mfp_c$
	\begin{equation}
		\int_0^\infty X_c(s) ds = \mfp_c = \int_0^\infty p_c(s) \, s \, ds.
	\end{equation}
	letting us express $p_u(s)$ as
	\begin{equation}
		p_u(s) = \frac{1 - \int_0^s p_c(s') ds'}{\int_0^\infty p_c(s') \, s' \, ds'}
	\end{equation}
	or directly from the mean-correlated-free path and macroscopic cross section~\cite{larsen11}
	\begin{equation}
		p_u(s) = \frac{1}{\mfp_c} e^{-\int_0^s \Sigma_{tc}(s') ds'}.
	\end{equation}
	The transmittance of light along a path beginning at an uncorrelated origin is given by the \emph{uncorrelated transmittance}
	\begin{equation}
		X_u(s) = 1 - \int_0^s p_u(s') ds'
	\end{equation}
	and is the quantity measured in the laboratory of the attenuation law of the random medium, provided that both the source and detector are not correlated to the medium particles.  Thus, the free-path distribution required between medium collisions for reciprocal linear transport is related to an uncorrelated measurement of the attenuation law by the second derivative,
	\begin{equation}\label{eq:xutopc}
		p_c(s) = \mfp_c \frac{\partial^2}{ \partial s^2} X_u(s).
	\end{equation}

	\subsection{\new{Relationship to Stochastic Microstructure Models}}
		\new{Our current study is agnostic to the specific model of stochastic media that leads to non-exponential free paths, yet we have arrived at a relationship previously used to study absorption in binary mixtures models.  If the mixing statistics are Markovian, a homogenization treatment gives exact results for attenuation, which remains exponential~\cite{pomraning98,dumas2000homogenization}.  For non-Markovian mixtures, homogenization is only approximate and non-exponential chord length distributions appear.}
		In the case of an isotropic two-phase random medium where phase 1 is void or non-participating and light or particles reflect from the surface of phase 2 without penetration into the interior, our formulation maps exactly to distributions studied in statistical mechanics and stereography.  Namely, $X_u(s)$ becomes the lineal-path function and $p_c(s)$ the normalized chord-length distribution of phase 1, where Eq.~\ref{eq:xutopc} is a well known relationship~\cite{torquato93}.  \new{The same relationship was also used to initialize the phase probabilities and initial chord lengths at $s = 0$ when deriving an uncorrelated transmission law for general binary mixtures of immiscible materials with non-Markovian mixing statistics using a renewal process and a master equation approach \cite{levermore88,sanchez1991statistical}.    Further work is required to investigate the accuracy of our model for multiple scattering in non-Markovian mixtures with a homogeneous single-scattering albedo.}

        \new{In the case of continuum models of quenched stochastic media where $\Sigma_t(\pos)$ is a continuous random variable~\cite{davis2011radiation,borovoi02,kostinski02}, taking on many densities in each realization as opposed to a binary set, we find that reciprocity also requires a distinction between free-paths beginning with uncorrelated origins and those that originate at previous collisions.  Intuitively, correlated free paths seem more likely to begin where the macroscopic cross section is higher than the medium average, $\Sigma_t(\pos) > \langle \Sigma_t \rangle$, because of the greater likelihood for the previous collision to end there.}

	\begin{table*}[t]
		\centering     
		\begin{tabular}{| c | l | c |} 
		
			\hline    
			\vline width0pt height2.3ex
			\textbf{Symbol} & \textbf{Description} & \textbf{Relations} \\
			\hline %
			\multicolumn{3}{|c|}{ \small \emph{medium-\textbf{c}orrelated free path origins }} \\
			\hline
			$s$ & {\small distance since last medium collision or correlated birth} & \\
			\hline
			$\Sigma_{tc}(s)$ & {\small correlated macroscopic cross section} & $\Sigma_{tc}(s) = \frac{p_c(s)}{X_c(s)}$ \\
			\hline
			$p_c(s)$ & {\small correlated free-path distribution} & $p_c(s) = \Sigma_{tc}(s) \e^{-\int_0^s \Sigma_{tc}(s') ds'} = - \frac{\partial}{ \partial s} X_c(s) = \mfp_c \frac{\partial^2}{ \partial s^2} X_u(s)$ \\
			\hline
			$X_c(s)$ & {\small correlated-origin transmittance} & $X_c(s) = 1 - \int_0^s p_c(s') ds'$ \\
			\hline
			$\mfp_c$ & {\small mean correlated free-path} & $\mfp_c = \int_0^\infty p_c(s) \, s \, ds$ \\
			\hline
			\multicolumn{3}{|c|}{ \small \emph{medium-\textbf{u}ncorrelated free path origins }} \\
			\hline
			$s$ & {\small distance since last surface/boundary or uncorrelated birth} & \\
			\hline
			$\Sigma_{tu}(s)$ & {\small uncorrelated macroscopic cross section} & $\Sigma_{tu}(s) = \frac{p_u(s)}{X_u(s)}$ \\
			\hline
			$p_u(s)$ & {\small uncorrelated (equilibrium) free-path distribution} & $p_u(s) = \Sigma_{tu}(s) \e^{-\int_0^s \Sigma_{tu}(s') ds'} = - \frac{\partial}{ \partial s} X_u(s) = \frac{X_c(s)}{\mfp_c}$ \\
			\hline
			$X_u(s)$ & {\small uncorrelated-origin transmittance} & $X_u(s) = 1 - \int_0^s p_u(s') ds'$ \\
			\hline
			$\mfp_u$ & {\small mean uncorrelated free-path} & $\mfp_u = \int_0^\infty p_u(s) \, s \, ds$ \\
			\hline

		\end{tabular}    
		\caption{Summary of our notation and relationships between quantities in our formalism.}
	\end{table*}

	\subsection{Path-integral formulation}
	  It is instructive to analyze the weak violation of reciprocity using the path-integral formulation of the light transport paths as defined by Veach~\shortcite{veach97}.  We refer the reader to the excellent review by Novak et al.	\shortcite{novak18} for complete details and speak only briefly about several terms in the path measurement contributions for the single-scattering subpaths inside the medium in our thought experiment (Figure~\ref{fig-ss}).  In this formulation, the measurement contribution (throughput) for the subpath inside the medium $\bar{\pos} = \pos_1 \pos_2 \pos_3$ (ignoring the vertex contributions at the boundary) is
	  \begin{equation*}
	  	f_j(\pos_1 \pos_2 \pos_3) = T(\pos_1,\pos_2) G(\pos_1,\pos_2) f_s(\pos1,\pos2,\pos3) T(\pos_2,\pos_3) G(\pos_2,\pos_3)
	  \end{equation*}
	  in the forward direction and
	  \begin{equation*}
	  	f_j(\pos_3 \pos_2 \pos_1) = T(\pos_3,\pos_2) G(\pos_3,\pos_2) f_s(\pos3,\pos2,\pos1) T(\pos_2,\pos_1) G(\pos_2,\pos_1)
	  \end{equation*}
	  in the adjoint direction.  The geometry terms $G$ are reciprocal $G(\pos_i,\pos_j) = G(\pos_j,\pos_i)$ but the other terms are not.  The transmittance terms $T$ remain dimensionless free-flight probabilities in the non-exponential case,
	  \begin{align*}
	  	&T(\pos_1,\pos_2) = X_u(a) \\
	  	&T(\pos_2,\pos_1) = X_c(a) \\
	  	&T(\pos_2,\pos_3) = X_c(b) \\
	  	&T(\pos_3,\pos_2) = X_u(b)
	  \end{align*}
	  and since $\pos_2$ is a correlated medium event, edges containing $\pos_2$ do not have reciprocal measurement contributions $T(\pos_1,\pos_2) \neq T(\pos_2,\pos_1)$ in the non-exponential case.  Similarly, for the measurement contribution at the scattering vertex $\pos_2$,
	  \begin{align*}
	  	&f_s(\pos_1,\pos_2,\pos_3) = \Sigma_{tu}(a) \, c \, P(\dir_i,\dir_o) \\
	  	&f_s(\pos_3,\pos_2,\pos_1) = \Sigma_{tu}(b) \, c \, P(\dir_o,\dir_i)
	  \end{align*}
	  we see a different probability density $\Sigma_{tu}(s)$ to collide at the vertex $\pos_2$ in the case that $a \neq b$ because of the statistical correlations in the medium.
	  However, since $p_u(s) = X_u(s) \Sigma_{tu}(s)$, weak reciprocity is attained over the entire subpath
	  \begin{equation}
	  	f_j(\pos_1 \pos_2 \pos_3) = f_j(\pos_3 \pos_2 \pos_1).
	  \end{equation}

	  \begin{figure}
		  \centering
		  \includegraphics[width=.99\linewidth]{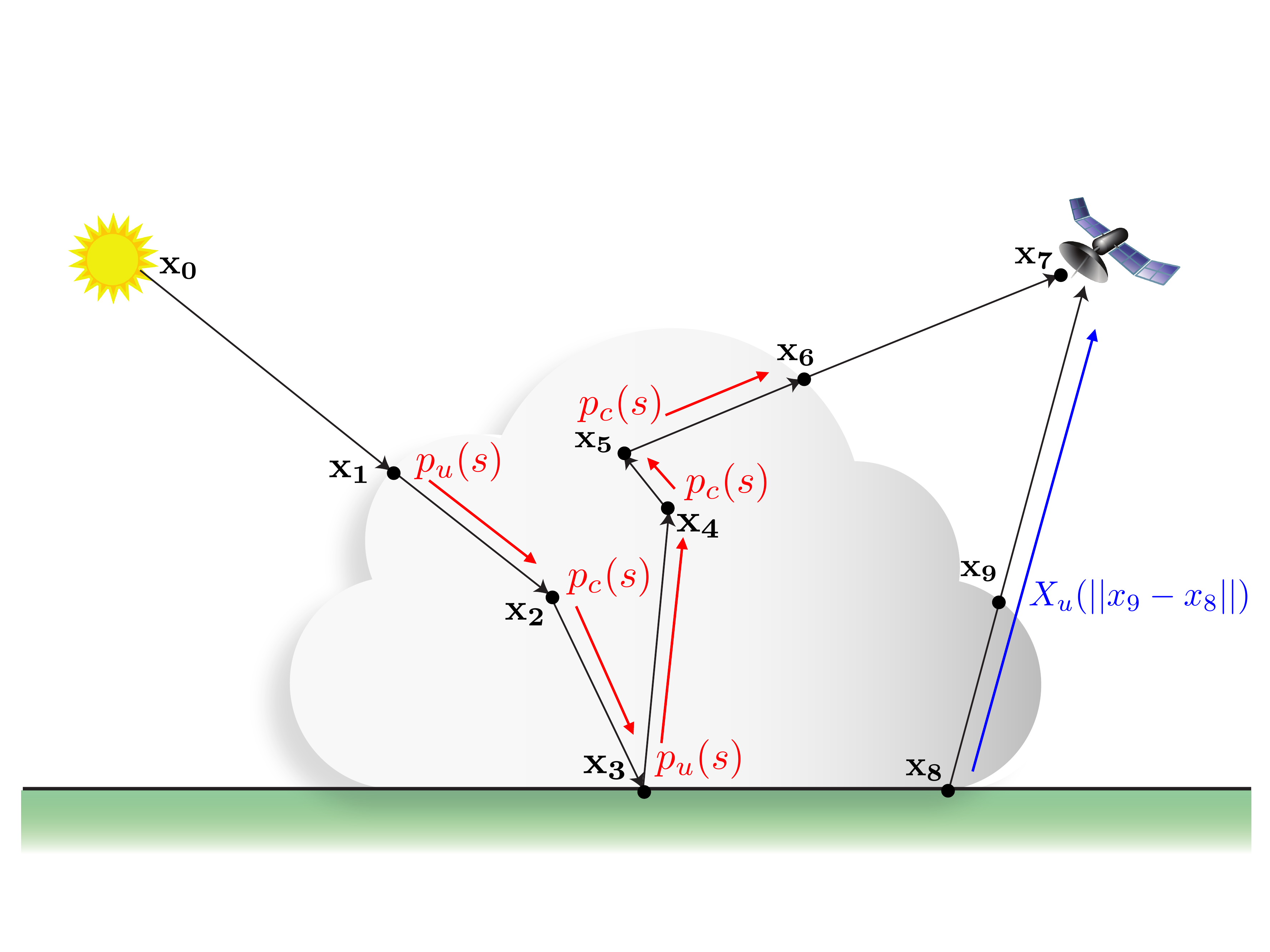}
		  \caption{Application of the correlated-origin $p_c(s)$ and uncorrelated-origin $p_u(s)$ free-path distributions in the forward analog sampling of a path through a homogeneous cloud.  Arriving at the cloud interface $\pos_1$ and reflecting off of the ground $\pos_3$ (positions that are statistically uncorrelated from the cloud particles) we use $p_u(s)$ to sample subsequent free-paths.  After scattering inside the medium at $\left\{\pos_2, \pos_4, \pos_5 \right\}$, $p_c(s)$ is used to preserve the ensemble-average free-path lengths between medium collisions.  Attenuation of the radiance leaving the ground $\pos_8$ on the way to the detector at $\pos_7$ uses the uncorrelated transmittance for the medium, $X_u(s)$.}
		  \label{fig-satellite} 
	    \end{figure}

	\subsection{Discussion} We first note that our proposal includes classical transport as a special case and, further, that exponential free-paths statistics
	\begin{align}
		&p_c(s) = \Sigma_t e^{-\Sigma_t s} \\
		&X_c(s) = e^{-\Sigma_t s} \\
		&p_u(s) = \frac{X_c(s)}{\int_0^\infty X_c(s) ds} = \frac{e^{-\Sigma_t s}}{1 / \Sigma_t} = p_c(s)
	\end{align}
	is the unique scenario where the same free-path statistics can be used for both boundary-medium and medium-medium segments and produce reciprocal transport.  Also immediately apparent is that, in needing $\mfp_c$ to define $p_u(s)$ as a normalization, this approach cannot technically be applied in fractal random media when the free paths between scatterers are best described by a heavy-tailed distribution with an unbounded mean.  In practice, however, free paths of arbitrary length in bounded media are always interrupted by collision with a medium boundary or inclusion, in which case clamping the tail of a heavy-tailed distribution would be a possible practical work-around.

	\subsection{General Extension - A Sketch}
	  The above derivation was based on only a single medium interaction, yet produces a form of $p_u(s)$ that exhibits weak-reciprocity generally over longer paths with multiple collisions inside the medium with the throughput equality condition always reducing to $p_u(s_j) X_c(s_k) = p_u(s_k) X_c(s_j)$ for the two correlated-uncorrelated segments of lengths $s_j$ and $s_k$, given that the correlated-correlated length-statistics $p_c(s)$ are reciprocal.  Further, nothing specific about this half-space analysis limits the scope of its result: we are free to replace either end point in our thought experiment with uncorrelated emitters, detectors or objects imbedded in the medium in a manner that is uncorrelated to the scattering particles and arrive at the same conclusions.  

	  Extending the GLBE in this way requires a number of changes.  Correlated birth behaves as previously, but light born at uncorrelated sources inside of correlated random media sample free-paths using $p_u(s)$ and has a distinct macroscopic cross section that can be determined from~\cite{larsen11}
		\begin{equation}
			\Sigma_{tu}(s) = \frac{p_u(s)}{X_u(s)}
		\end{equation}
		requiring the distinction between correlated and uncorrelated radiances $L_c(\pos,\dir,s)$ and $L_u(\pos,\dir,s)$ respectively with the total radiance formed via
		\begin{equation}
			L(\pos,\dir) = \int_0^\infty \left( L_c(\pos,\dir,s) + L_u(\pos,\dir,s) \right) ds.
		\end{equation}
		We denote the two sources $Q_c(\pos,\dir)$ and $Q_u(\pos,\dir)$.  The balance equation for $L_c$ includes new collision contributions from both radiances with their respective macroscopic cross sections
		{\small \begin{multline*}
		  	\frac{\partial L_c}{\partial s}(\pos,\dir,s) + \dir \cdot \nabla L_c(\pos,\dir,s) + \Sigma_{tc}(s) L_c(\pos,\dir,s) = \delta(s) \frac{Q_c(\pos)}{4 \pi} + \\ \delta(s) \, c \, \int_{4\pi} \int_0^\infty P(\dir_i \cdot \dir_o) \left( \Sigma_{tc}(s') L_c(\pos,\dir',s') + \Sigma_{tu}(s') L_u(\pos,\dir',s') \right) ds' d\dir' 
		  \end{multline*} }
		  whereas $L_u$ has no contributions from medium collisions,
		  \begin{equation*}
		  	\frac{\partial L_u}{\partial s}(\pos,\dir,s) + \dir \cdot \nabla L_u(\pos,\dir,s) + \Sigma_{tu}(s) L_u(\pos,\dir,s) =  \frac{\delta(s)}{4 \pi} Q_u(\pos).
		  \end{equation*}
		  New boundary conditions are required that integrate over both varieties of incident light, evaluating \new{bidirectional surface-scattering distribution functions} (BSDFs)~\cite{pharr10} and emitting only uncorrelated radiance together with any uncorrelated source entering from outside.  Finally, radiance $L_u$ is streamed uncollided with attenuation $X_u(s)$ between any two uncorrelated adjacent path space vertices.  \new{Figure~\ref{fig-satellite} illustrates an example application of the formalism to a multiple scattering problem in piecewise homogeneous media.}

		  \new{Concurrently, Jarabo et al.~\shortcite{jarabo18} have proposed a similar extension of the GLBE, going further to include path-length-dependent single-scattering albedo and phase functions, and also consider a multitude of different volumetric sources that can have independent correlation statistics to the scattering centers in the medium.  Heterogeneous density in stochastic random media (\cite{camminady17}) and correlation across boundaries between two medium regions with different compositions is also discussed.  Reciprocity and determinstic relationships between the various distributions have not been studied in these cases and remain an important area of future work.}

	  We devote the remainder of this paper to motivating our proposal using only basic transport scenarios.  In the follow-up papers, we present a complete transport formalism for application to general complex transport problems complete with discussion of collision, track-length and expected-value estimators, density estimation, bidirectional estimators, diffusion \new{bidirectional scattering-surface reflectance-distribution functions} (BSSRDFs) and other variance reduction methods typically used for transport in classical participating media.  The much more complicated relationship between radiance and collision rate density requires revisiting many of these approaches in detail.  The notion of correlation across boundaries and losing the power of invariant imbedding, adding/doubling and delta-tracking will also be discussed.

\section{Monte Carlo Validation}\label{sec:MC}

\begin{figure}
	  \centering
	  \subfigure[independent scattering centers]{\includegraphics[width=.49\linewidth]{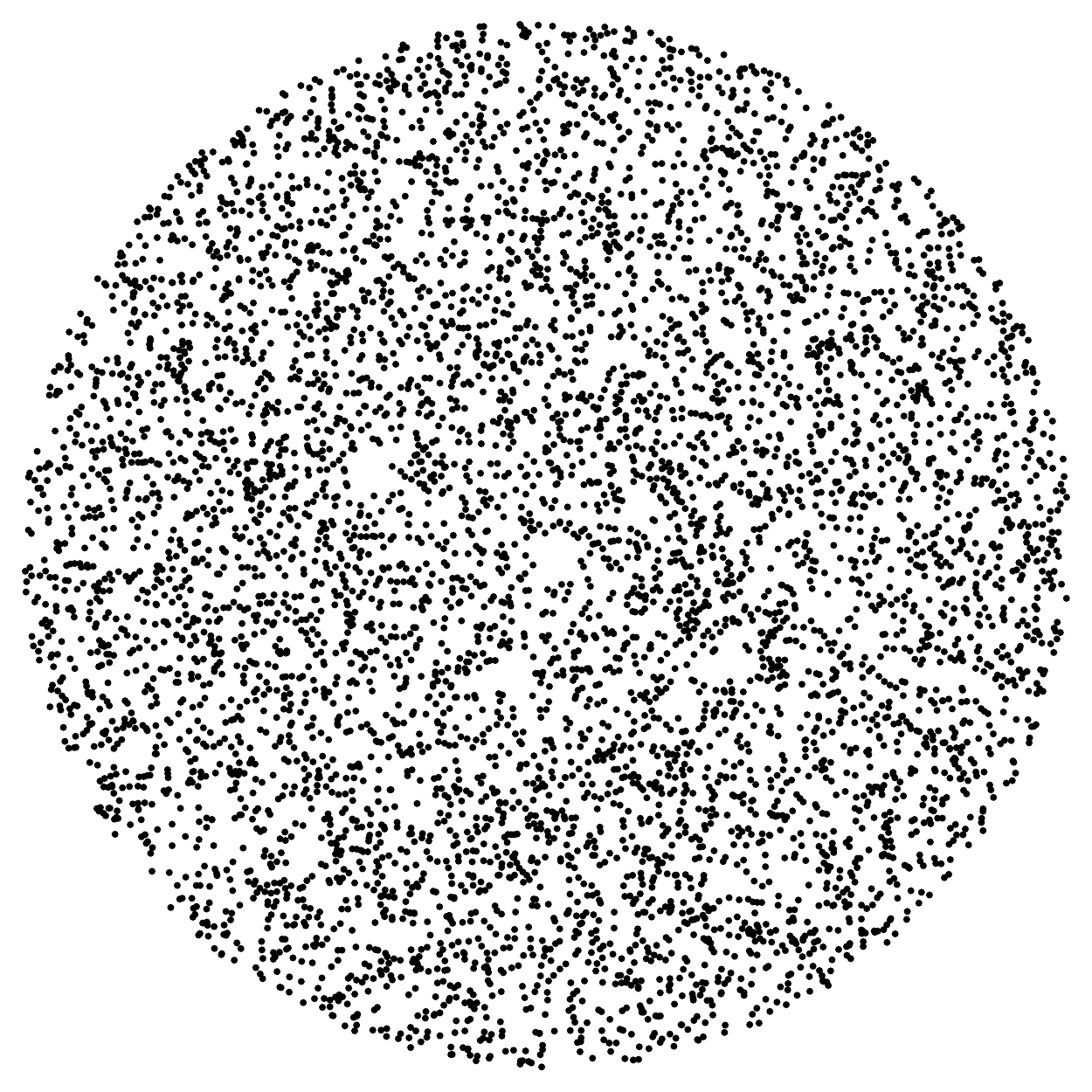}}
	  \subfigure[blue noise, $\smin = 0.1$]{\includegraphics[width=.49\linewidth]{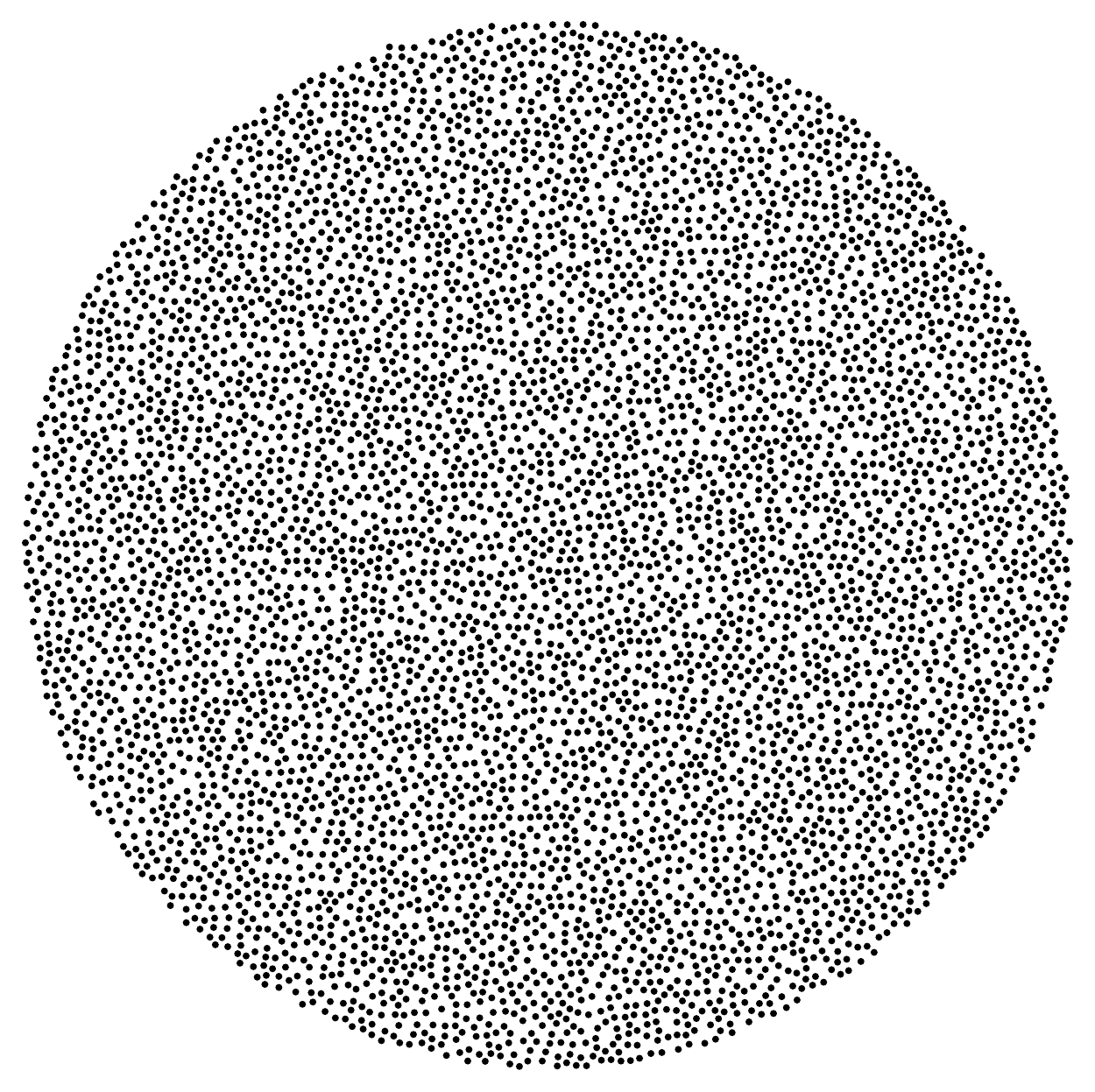}}
	  \caption{Two random realizations in flatland of scattering particles in a disk of radius $R = 6$, with particle radius $r = 0.04$.  Both realizations have a number density $\rho = 180 / \pi$.}
	  \label{fig-flatlanddisks} 
	\end{figure}

  To test the utility of our application of the GLBE in bounded media, we study bulk scattering by random realizations of particles distributed by minimum-distance Poisson-disk sampling~\cite{cook86,lagae08} with blue noise properties \new{(Figure~\ref{fig-flatlanddisks})}.  We chose this form of correlated random media because
  \begin{compactitem}
  	\item it admits a trivial sampling procedure, enabling a large number of Monte Carlo simulations for validating our statistical modeling
  	\item it seems physically relevant in that physical scattering and absorbing particles cannot self-intersect (the concept of hard spheres/disks in fluids)
  	\item the minimum separation length $\smin$ of the correlation requires that light must fly unattenuated for approximately this distance before the next medium interaction, which is helpful for thought experiments that distinguish generalized transport theory from classical.
  \end{compactitem}
  More specifically, we study linear transport (under the assumptions of geometrical optics) within random realizations of radially-symmetric particles with blue noise distribution in their relative positions generated by a simple dart-throwing process with rejection for any particle whose center is closer than some threshold $\smin$ to any other accepted center in the current realization.  \new{To simplify the study,} the scattering particles in each realization are assumed to be static over time-scales close to those of the mean time of flight (quenched disorder).  We currently restrict our Monte Carlo investigations to a two-dimensional ``Flatland'' domain for computational simplicity.  Efficient Poisson-disk sampling methods are known in higher dimensions~\cite{metropolis53,ebeida12}.  
  
  The transport medium is thus characterized by
    \begin{compactitem}
      \item The number density of particles $\rho$
      \item The particle cross section $\sigma = 2 r$ (for Flatland, twice the particle radius)
      \item The minimum separation distance between any two scattering centers $\smin$
      \item Absorption process and phase function (if applicable).
    \end{compactitem}

\begin{figure}
		  \centering
		  \subfigure[standard dart throwing]{\includegraphics[width=.49\linewidth]{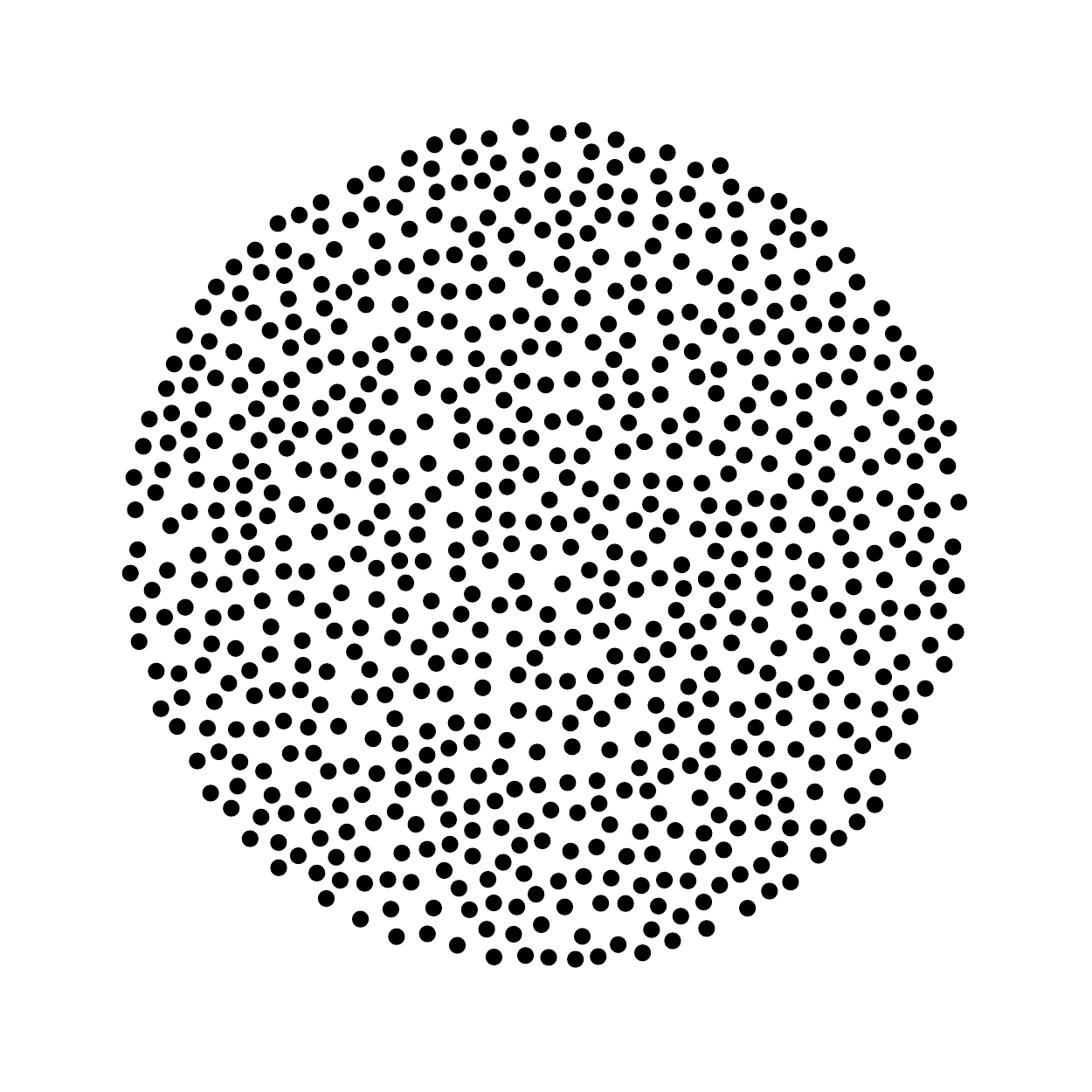}}
		  \subfigure[extended-boundary dart throwing]{\includegraphics[width=.49\linewidth]{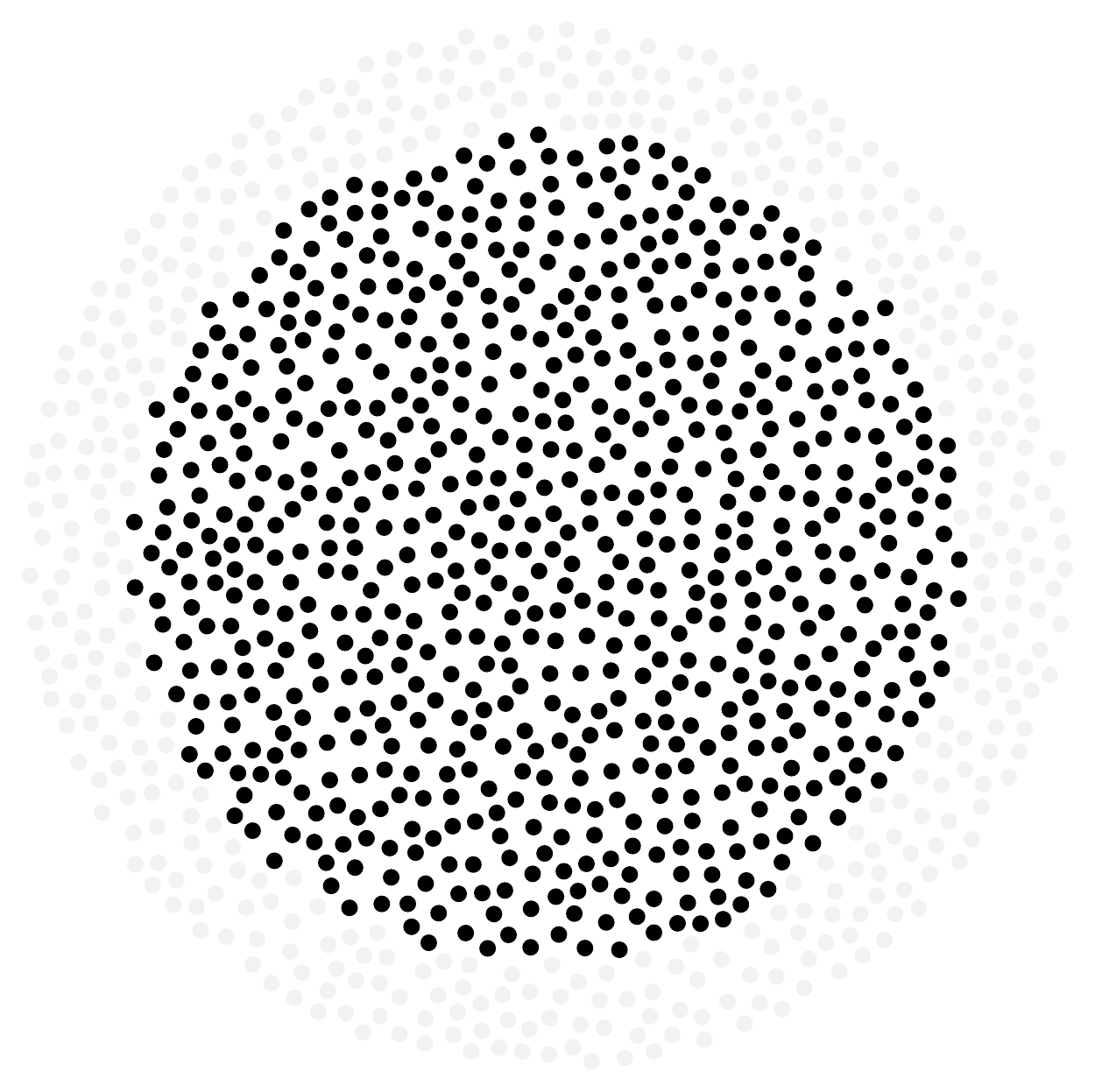}}
		  \caption{Extended-boundary vs non-extended-boundary minimum-distance Poisson-disk sampling.  (b) Ghost particles generated in the extended region, discarded after sampling terminates, change the distribution characteristics at the boundary producing random media more representative of eroded materials.}
		  \label{fig-extendedsampling} 
	  \end{figure}

  \paragraph{Dart-throwing in the presence of boundaries}\label{sec:dart}
    We compare two distinct forms of our medium sampling in the presence of boundaries (illustrated in Figure~\ref{fig-extendedsampling})
    \begin{compactitem}
      \item simple minimum-distance Poisson-disk dart-throwing~\cite{cook86} where particle center proposals are generated uniformly within the specified volume with rejection parametrized by $\smin$.
      \item \emph{extended-region dart-throwing} where dart throwing begins within a region whose boundaries are extended outward by at least several times $\smin$ and, after the requested density of particles is attained in the interior region, all particles who centers lie outside of the original region are discarded.  In doing so, extended sampling preserves the blue noise distribution right up to boundary edges.
    \end{compactitem}

  \paragraph{Collisions and Birth}

    In constructing free paths and light transport paths we consider two distinct forms of collisions with the disks in Flatland of finite radius $r$:
    \begin{compactitem}
      \item \emph{edge-collision} where the distance to the edge of the disk is returned
      \item \emph{center-warp} where, upon colliding with a disk, the collision point is warped to the center of the disk and a new distance calculated for free-path statistics, and this center is the start of the consequent free path (if any).
    \end{compactitem}

    In some simulations scattering particles may be uncorrelated to the extent that they overlap in space.  Further, some sampling procedures begin with starting locations chosen uniformly at random with no consideration of the occupancy of space at a given location.  In both of the scenarios it is important to note that our Monte Carlo tracing ignores testing against any particle already surrounding the start location (as if it does not exist).  We intentionally choose such simplifying assumptions and neglect the specifics of what process is appropriate for scattering from the finite sized scatterers, such as surface reflection off of small mirror spheres imbedded in paint layers, or classical volumetric scattering within the interior of the spheres.  The comparisons will reflect this intentional negligence with small errors on the order of the particle radius.  \new{For the rest of the paper, edge-collision is used with the exception of the single and double scattering half-space experiments in Section~\ref{sec:half-space}, where we choose to study isotropic scattering for comparison to known analytic solutions for the exponential case.  In 3D, specular reflection from spheres could be used to produce far-field isotropic scattering, which corresponds to the billiard transport assumptions in the study of Lorentz gas dynamics~\cite{golse12}.  However, far-field reflectance from spheres in $d$ dimensions is only isotropic for $d = 3$, so we warp to the center and sample an isotropic deflection before the next free-path to simulate energy leaving from the center of each particle, on average, and to avoid numerical precision issues leading to undesired double collision with the same particle.}

   \section{Minimum-Poisson-Disk Blue-noise Transport}\label{sec:blue}

     In this section we apply the Monte Carlo sampling methods previously described to compute a variety of statistics for transport in volumes with correlated scattering centers.  We begin by estimating the two free-path distributions required by our transport formalism for the blue noise media we sample and find approximate models for both distributions for making predictions using our model.  Specifically, we measure and propose analytic approximations for
	\begin{compactitem}
	  \item $p_u(s)$ - the free-path distribution in a blue noise medium when starting from a random position and random direction in the medium \emph{uncorrelated} from the positions of the scatterers
	  \item $p_c(s)$ - the free-path distribution in a blue noise medium when starting from the center of one of the particles of the realization, chosen uniformly at random and with a random direction (thus, a \emph{correlated free-path})
	\end{compactitem}

	  \subsection{Uncorrelated Minimum-Poisson-Disk free paths}\label{sec:bluepu}

\begin{figure*}
		  \centering
		  \subfigure[independent scattering centers, $\smin = 0$]{\includegraphics[width=.49\linewidth]{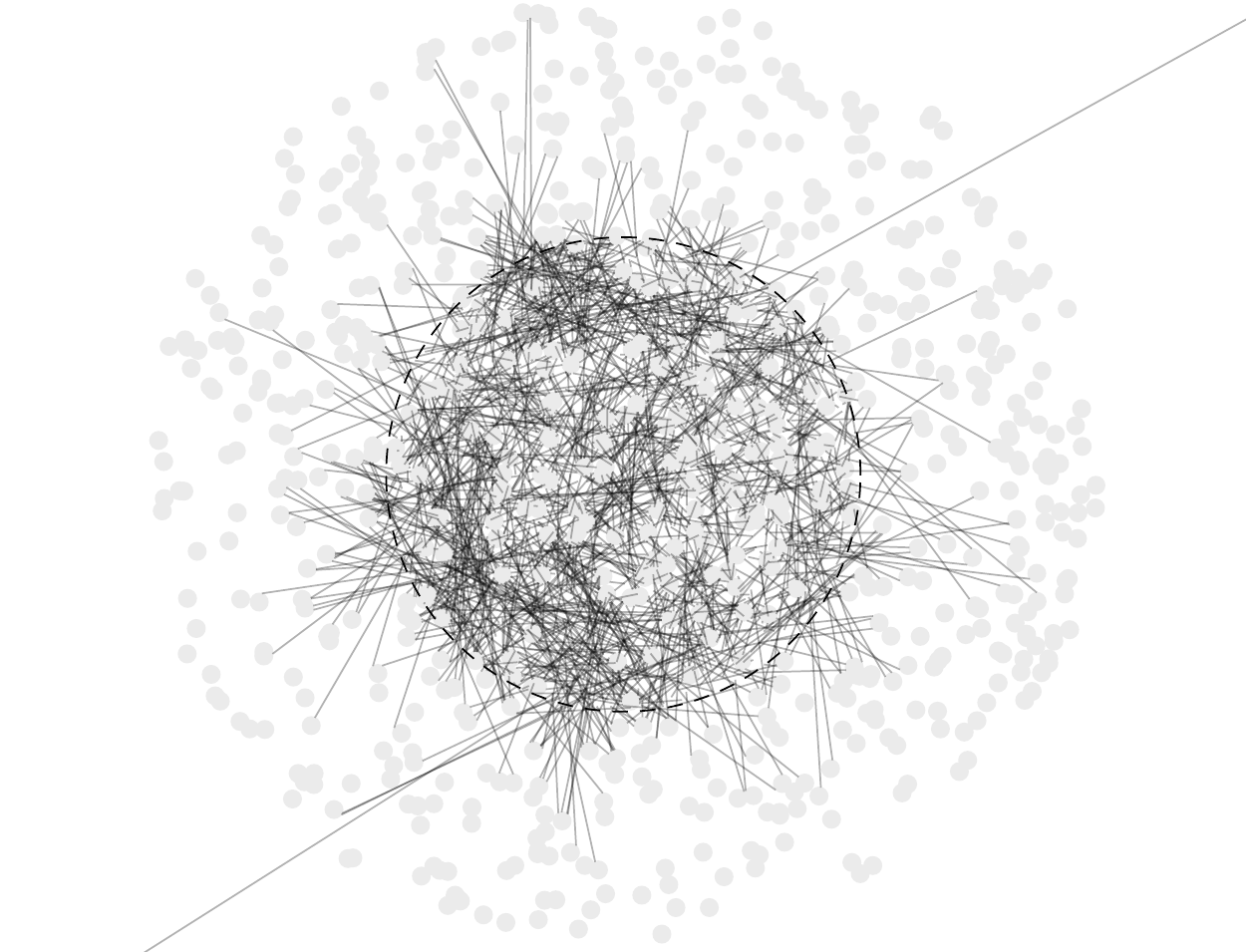}}
		  \subfigure[blue noise, $\smin = 0.1$]{\includegraphics[width=.49\linewidth]{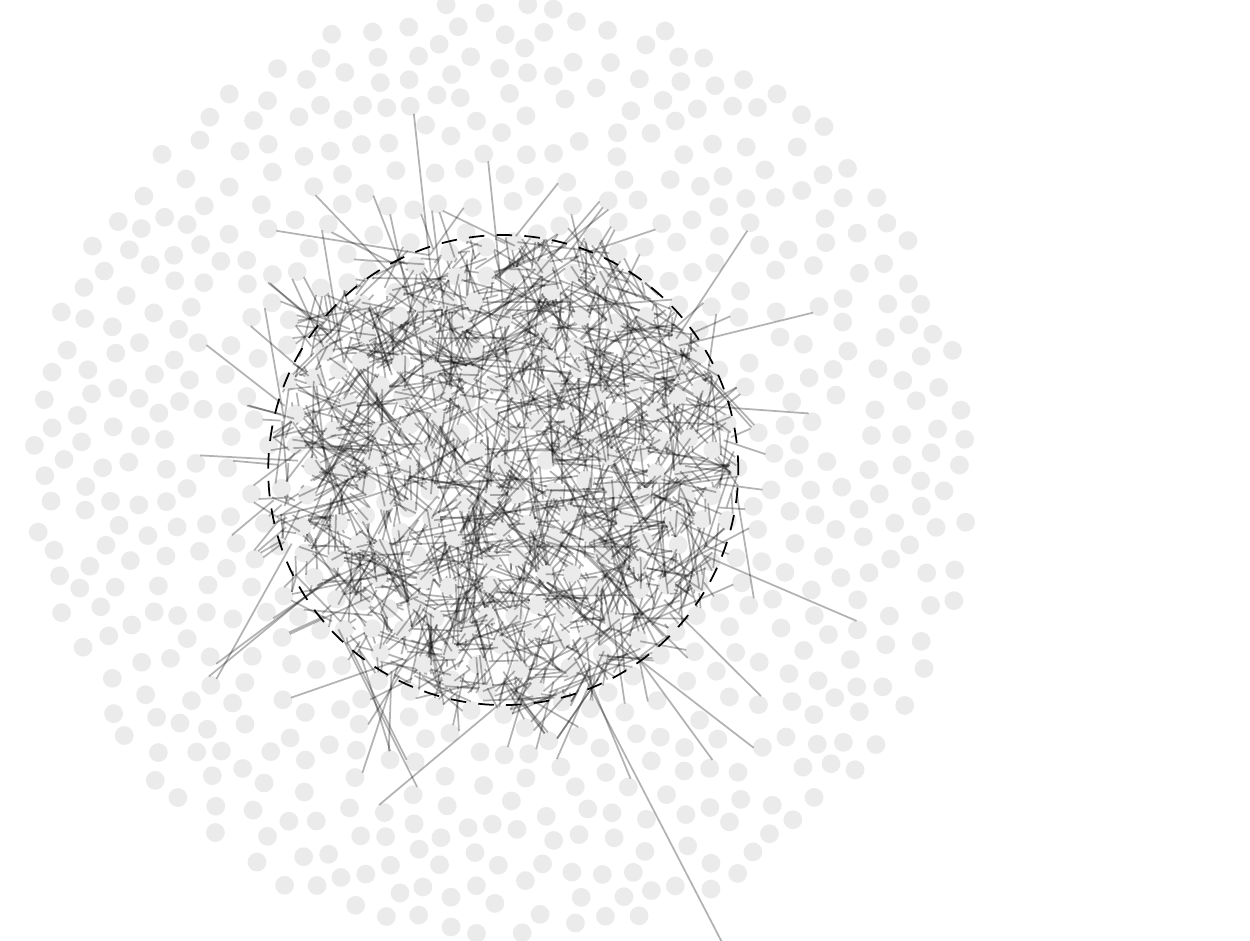}}
		  \caption{Illustration of our uncorrelated free-path sampling procedure within two random realizations of radius $R = 2$, number density $\rho = 180 / \pi$, and particle radius $r = 0.04$.  2000 random paths are shown per figure.  The unit disk of initial position is shown with a dashed boundary for reference.  Note the longer average free path lengths in the uncorrelated exponential medium in (a) compared to the blue noise scatterers (b).  \remove{Figures (c) and (d) show a more expensive sampling initialization that rejects starting from occupied space.  We performed a number of comparisons between these two forms of uncorrelated path sampling and found negligible difference in the overall statistics in most cases.}}
		  \label{fig-freepathtraces} 
	  \end{figure*}

      We measured the free-path statistics with uncorrelated starting positions in blue noise Flatland media via the following sampling procedure:
      \begin{compactitem}
      	\item Each random realization populated a disk of radius $R$ until number density $\rho$ was achieved using simple dart throwing and rejection based on minimum separation length $\smin$
      	\item A number of random samples, $N_R$, per realization were traced and the free path length histogram updated after each, and then a new realization was sampled
      	\item A starting location for each ray was chosen uniformly at random within the disk of unit radius
      	\item The ray orientation was chosen uniformly at random
      	\item The free path length returned was the minimum of the distance to the outer surface of all particles in the medium with intersection distances $t > 0$ where all particles overlapping the starting location ignored
      \end{compactitem}
      Examples of 2000 random paths generated via this procedure within two different realizations are illustrated in Figure~\ref{fig-freepathtraces}.  We see that as the particles repel and correlation increases, fewer large gaps appear and the mean free path decreases \new{(by about $20\%$ in this case)}.

	  \begin{figure*}
		  \centering
		  \subfigure[$\sigma = 0.08, \rho = \frac{180}{\pi}$]{\includegraphics[width=.3\linewidth]{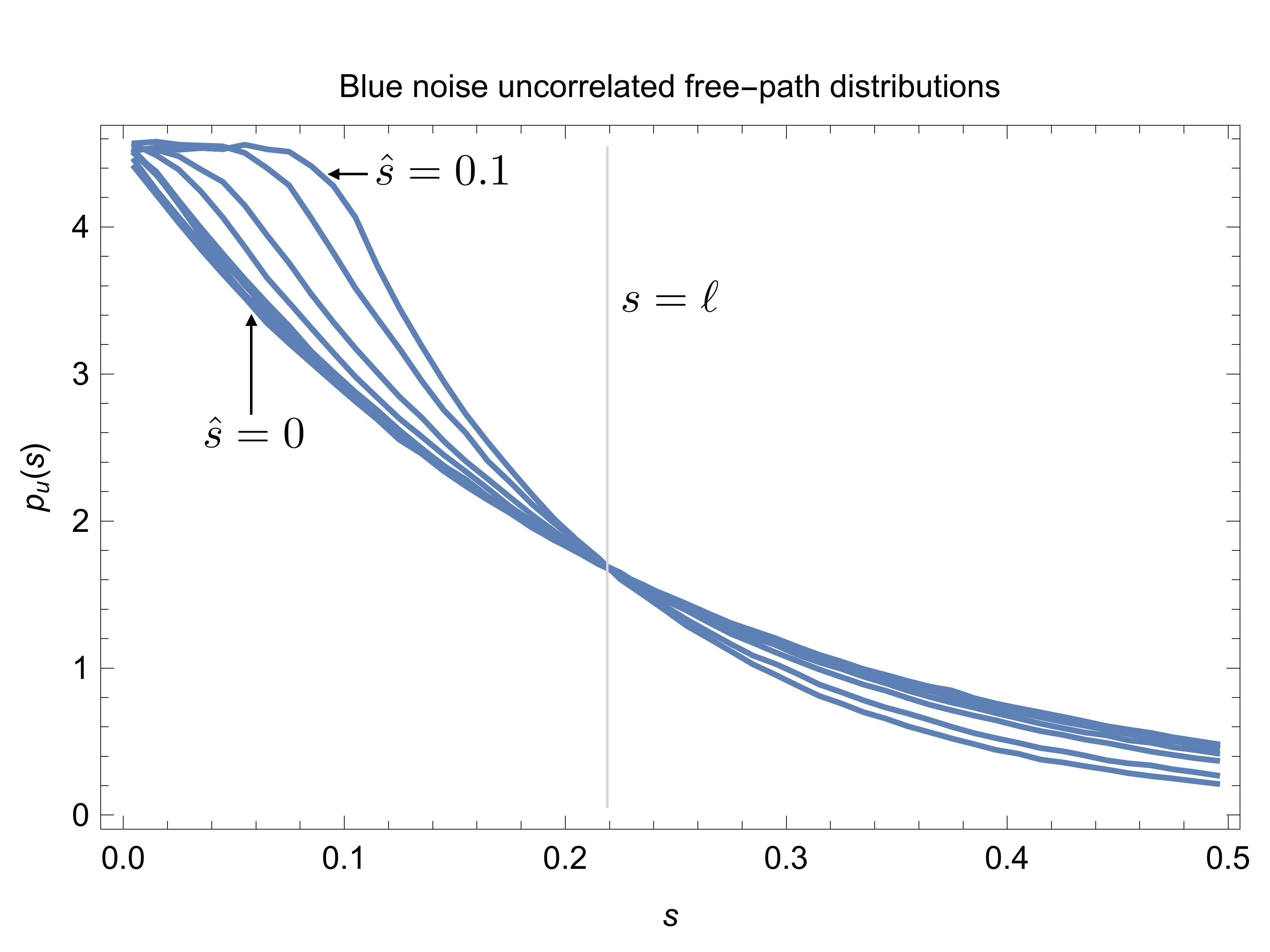}}
		  \subfigure[Semilog plot of (a)]{\includegraphics[width=.3\linewidth]{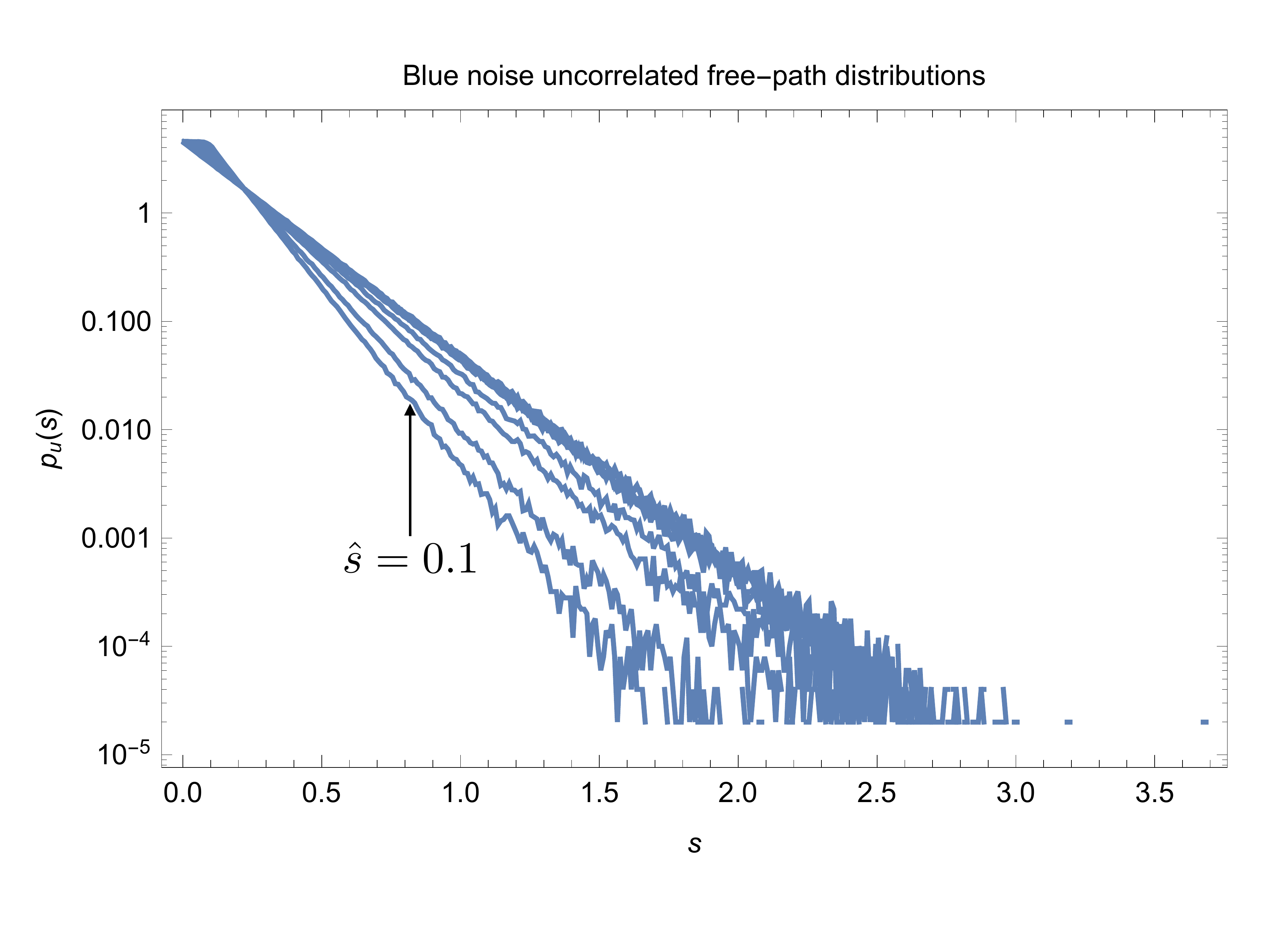}}
		  \caption{Monte Carlo uncorrelated free-path distributions in Flatland with Poisson disk sampled scatterers.  We note the distributions are relatively constant from $s = 0$ to $s = \smin$, the tails are exponential, and the distributions are equal at $s = \ell$.  Each plot shows 9 Monte Carlo simulations with 5000000 random free paths, $R = 6$, and $\smin \in \{ {0, 0.005, 0.01, 0.015, 0.02, 0.035, 0.05, 0.075, 0.1} \}$.}
		  \label{fig-puMC} 
		\end{figure*}

	As an initial validation we verified that our Flatland implementation of this sampling procedure in the classical case of uncorrelated scatterers ($\smin = 0$) was well modeled by the predictions of classical transport theory with \emph{classical mean free path}
	\begin{equation}
    	\ell = \frac{1}{\rho \sigma}
    \end{equation}
    and an exponential free-path length distribution
    \begin{equation}
    	p_u(s) = \frac{1}{\ell} e^{-s/\ell}.
    \end{equation}
    We then performed a number of blue noise Monte Carlo simulations for the uncorrelated free path lengths in a disk medium of radius $R = 6.0$.  The number density $\rho$ of the medium was fixed at $180 / \pi$ while the particle radius $r$ and minimum separation length $\smin$ were varied.  Figure~\ref{fig-puMC} illustrates the observed behaviours, from which we observed three trends:
	\begin{compactitem}
		\item The distributions have exponential tails
		\item The distributions are roughly constant from $s = 0$ to $s = \smin$
		\item Each family of distributions over $\smin$ intersect at the classical mean free path length $s = \ell$.
	\end{compactitem}
	Chord length and free path distributions in hard disk packings are a well studied problem~\cite{torquato93,olson08}, and these results are not surprising and serve to validate our Monte Carlo implementation.  While more accurate approximations are known, for simplicity, we propose an easy-to-sample, first-order approximation for blue noise free-path distributions with these observed properties given by
	\begin{equation}\label{eq:publue}
		p_u(s) = \frac{1}{\ell} \begin{cases}
		1 & 0 \le s < \smin\\
		e^{-\frac{s-\smin}{\ell-\smin}} & s \ge \smin
		\end{cases},
    \end{equation}
    a normalized pdf
    \begin{equation}
    	\int_0^\infty p_u(s) ds = 1
    \end{equation}
    with mean free path
    \begin{equation}
    	\mfp = \int_0^\infty p_u(s) \, s \, ds = \ell - \smin + \frac{\smin^2}{2 \ell}
    \end{equation}
    and mean square free path
    \begin{equation}
    	\mfpsqr = \int_0^\infty p_u(s) \, s^2 \, ds = 2 \ell^2-\frac{2 \smin^3}{3 \ell}-4 \ell \smin+3
   \smin^2
    \end{equation}
    As desired, $p_u(s)$ reduces to the classical case of exponential free-paths as $\smin \rightarrow 0$.
    This family of free-path distributions has an uncorrelated transmittance
    \begin{equation}
    	X_u(s) = 1 - \int_0^s p_u(s') ds' = \begin{cases}
		1 - \frac{s}{\ell} & 0 \le s < \smin\\
		\frac{(\ell-\smin) e^{-\frac{s-\smin}{\ell-\smin}}}{\ell} & s \ge \smin
		\end{cases}.
    \end{equation}
    \begin{figure}
		  \centering
		  \includegraphics[width=.99\linewidth]{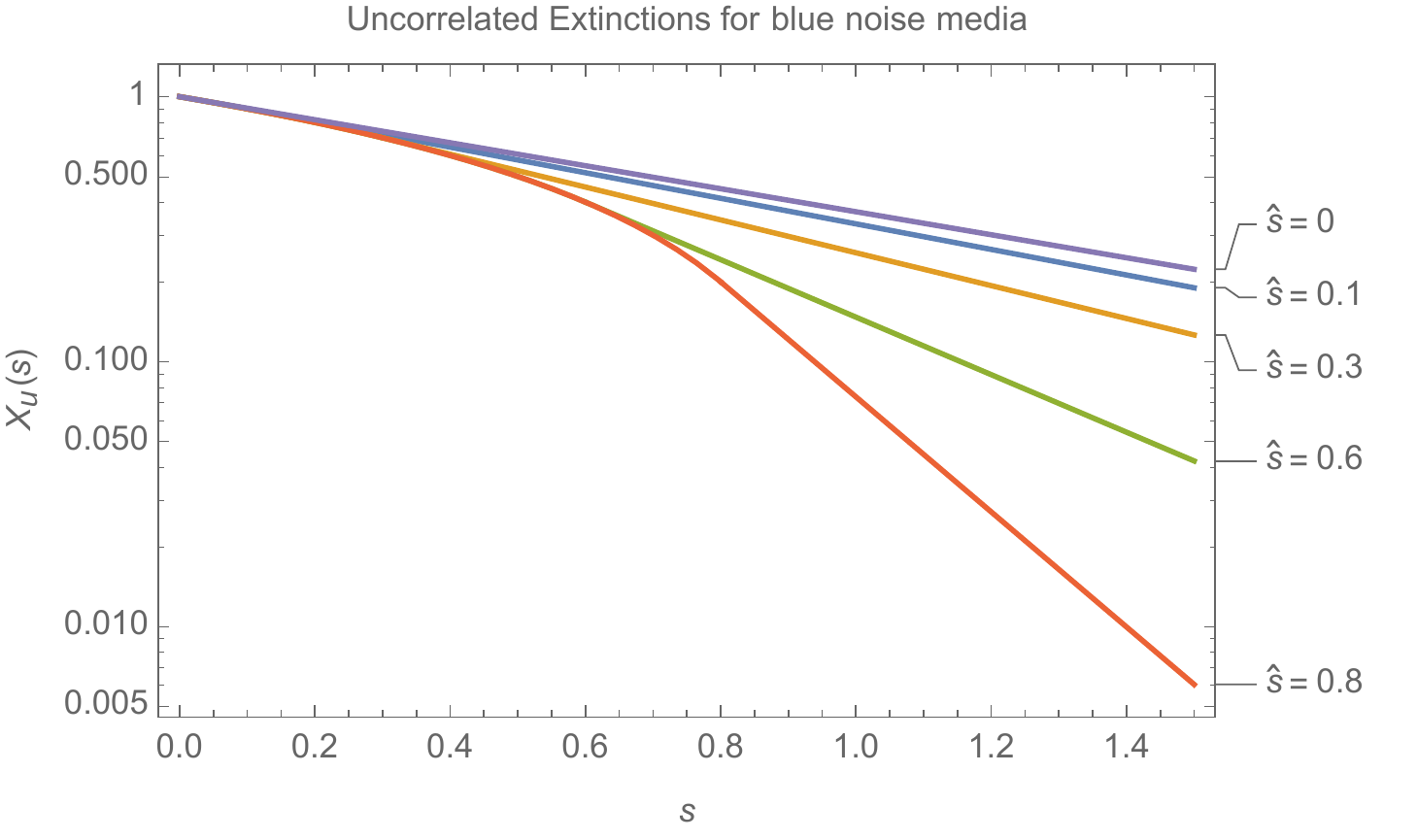}
		  \caption{Family of transmittance profiles for uncorrelated paths in minimal Poisson Disk blue noise scattering media.  Each plot has the classical mean free path $\ell = 1$, with classical exponential transmittance shown when the minimum separation length $\smin = 0$.}
		  \label{fig-Xu} 
	\end{figure}
    
    \new{The deviation of this family of transmittance functions from the classical exponential case is shown in Figure~\ref{fig-Xu}.}
    Uncorrelated free paths are easily sampled from a single uniform random variable $\xi \in [0,1]$ via
    \begin{equation}
    	s = \begin{cases}
		\ell \xi & \xi < \frac{\smin}{\ell} \\
		\smin-(\ell-\smin) \log
   \left(\frac{\xi-\frac{\smin}{\ell}}{1-\frac{\smin}{\ell}}\right) & \text{else}
		\end{cases}.
    \end{equation}
    \begin{figure*}
	  \centering
	  \subfigure[]{\includegraphics[width=.33\linewidth]{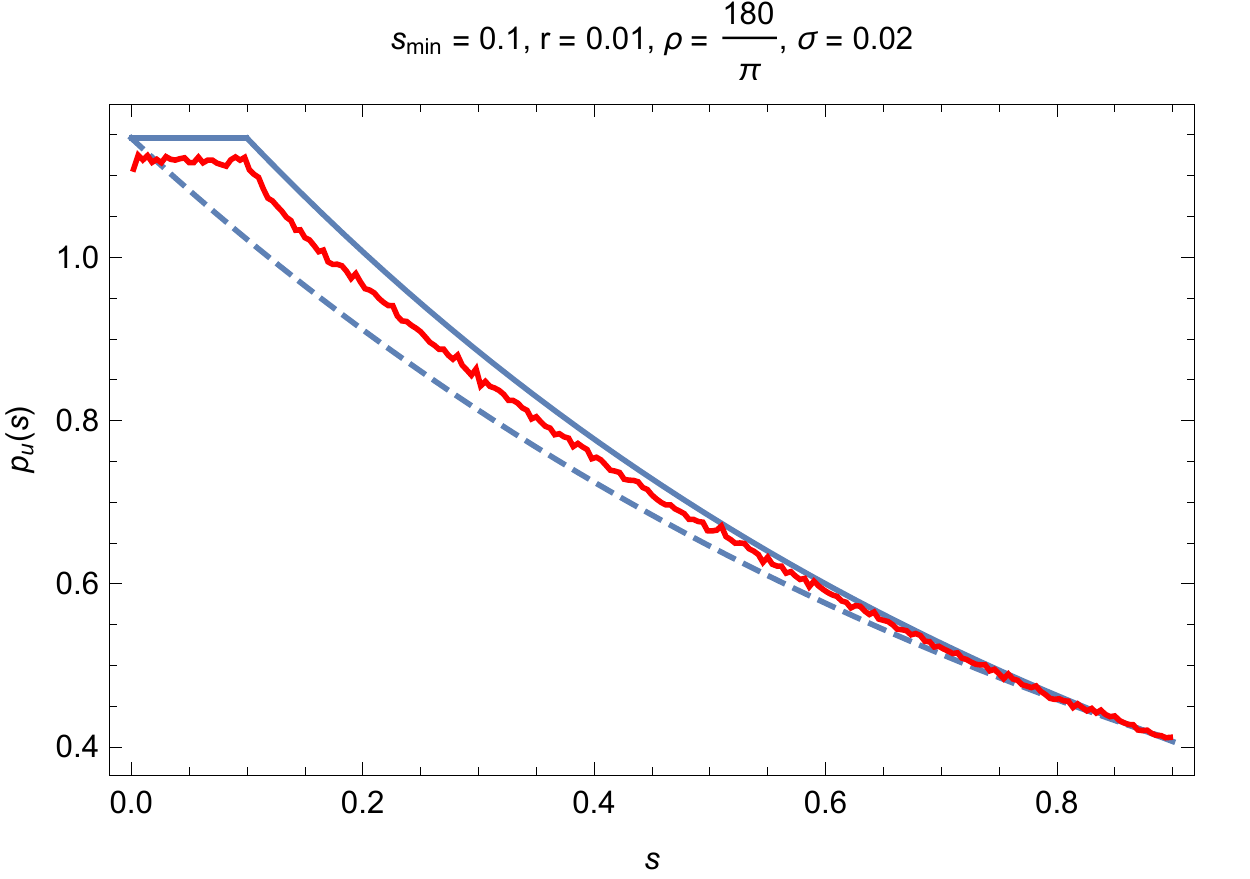}}
	  \subfigure[]{\includegraphics[width=.33\linewidth]{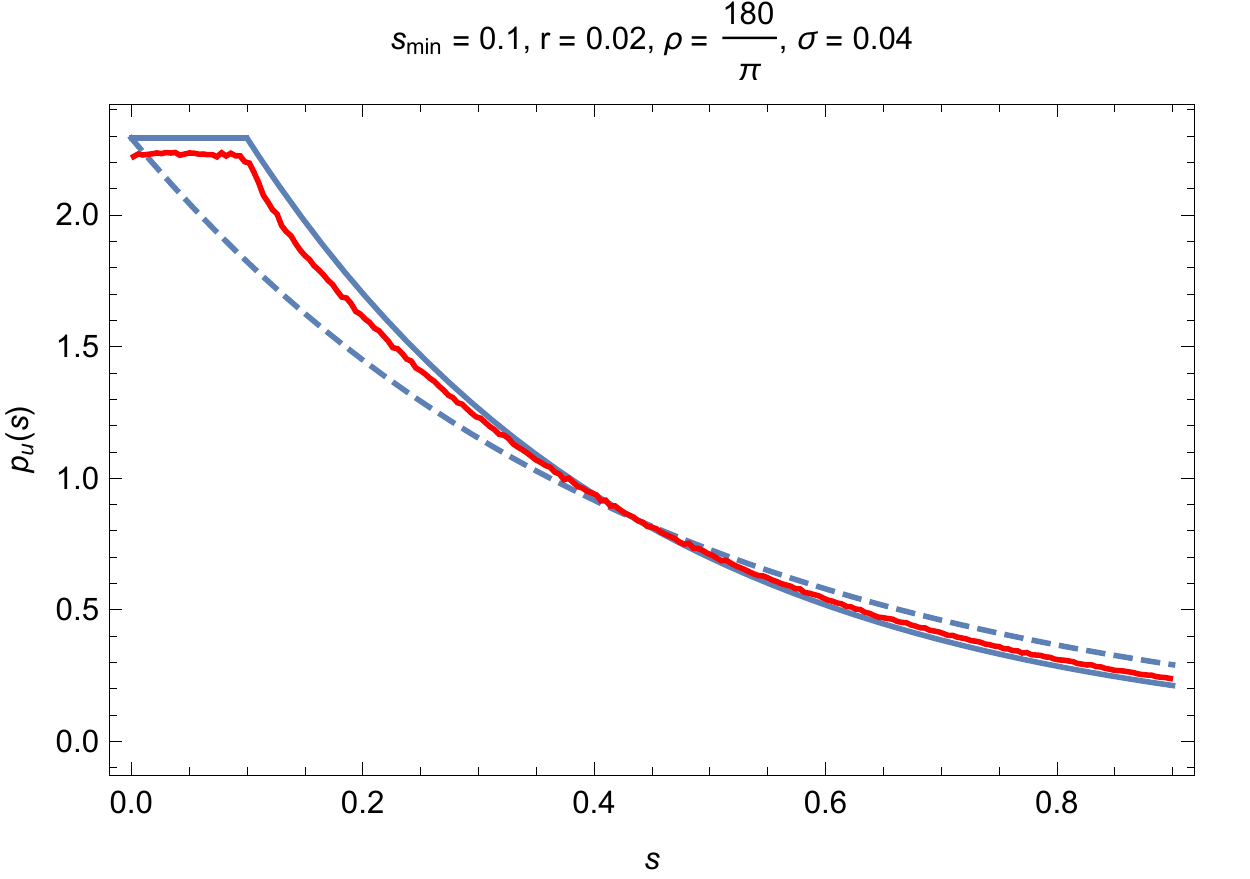}}
	  \subfigure[]{\includegraphics[width=.33\linewidth]{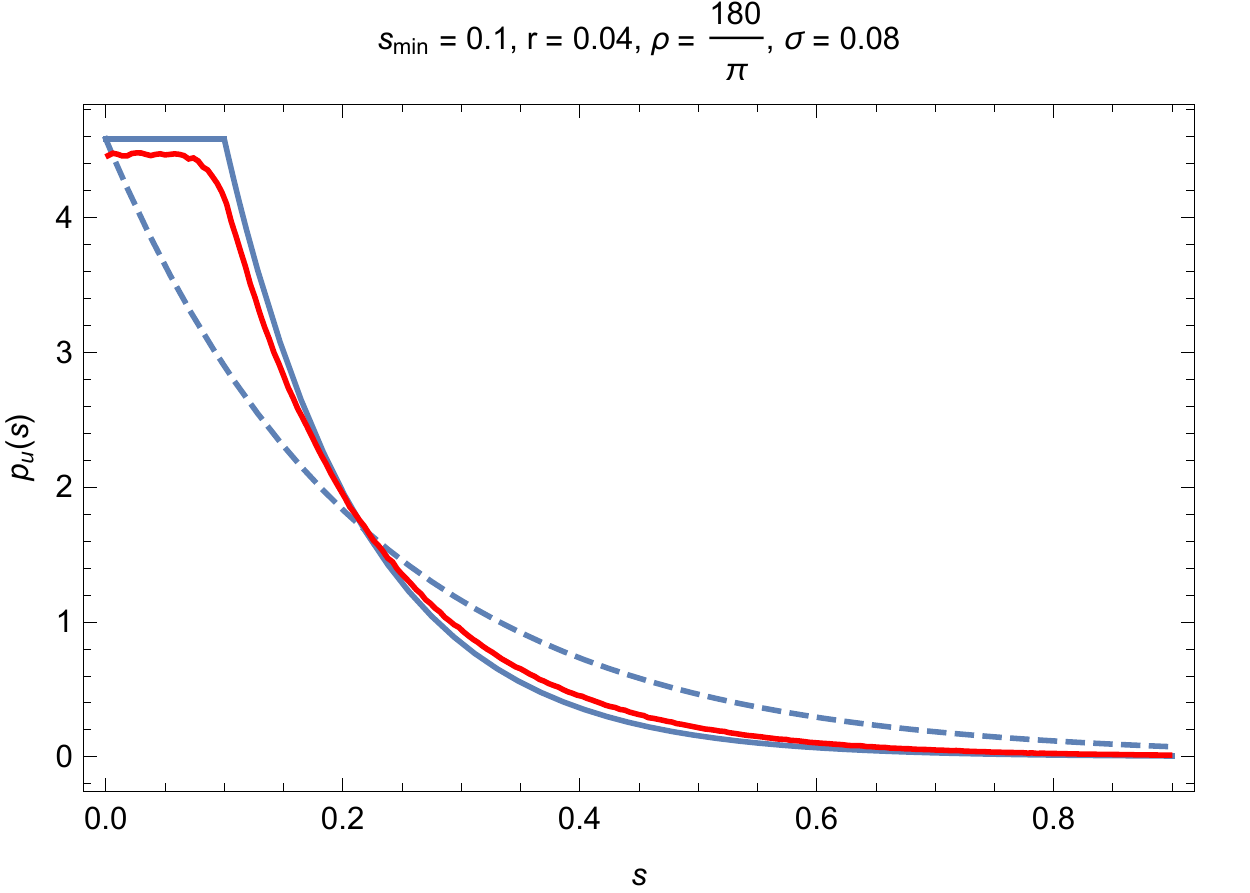}}
	  \subfigure[]{\includegraphics[width=.33\linewidth]{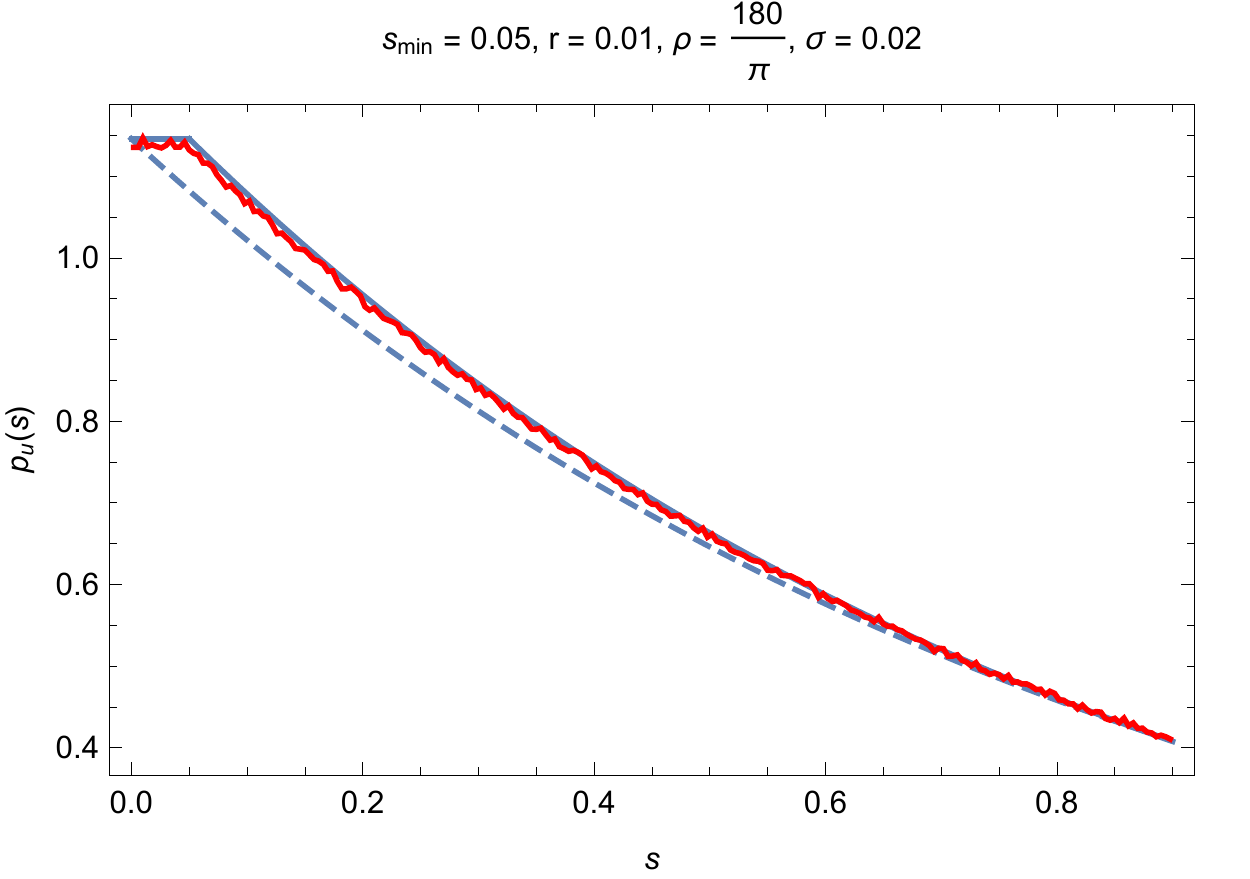}}
	  \subfigure[]{\includegraphics[width=.33\linewidth]{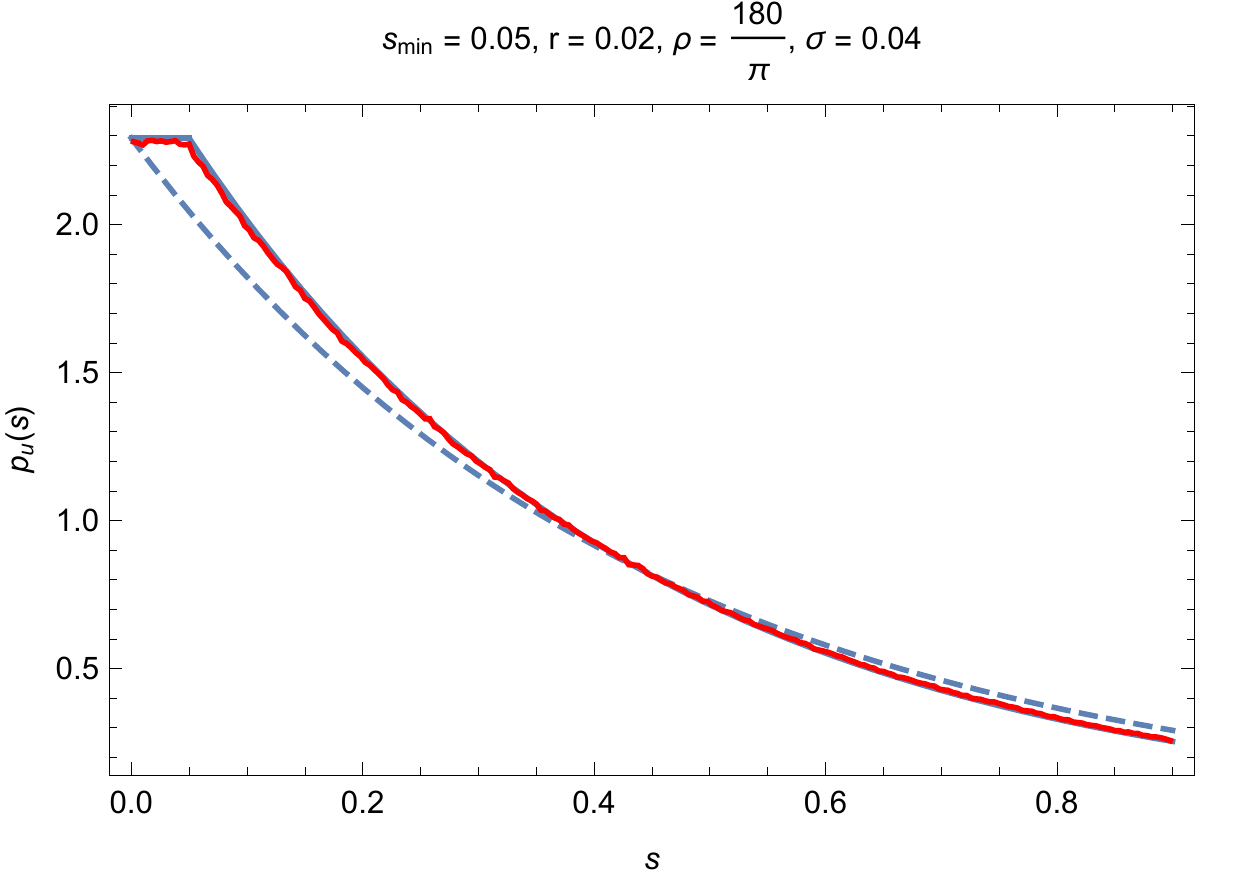}}
	  \subfigure[]{\includegraphics[width=.33\linewidth]{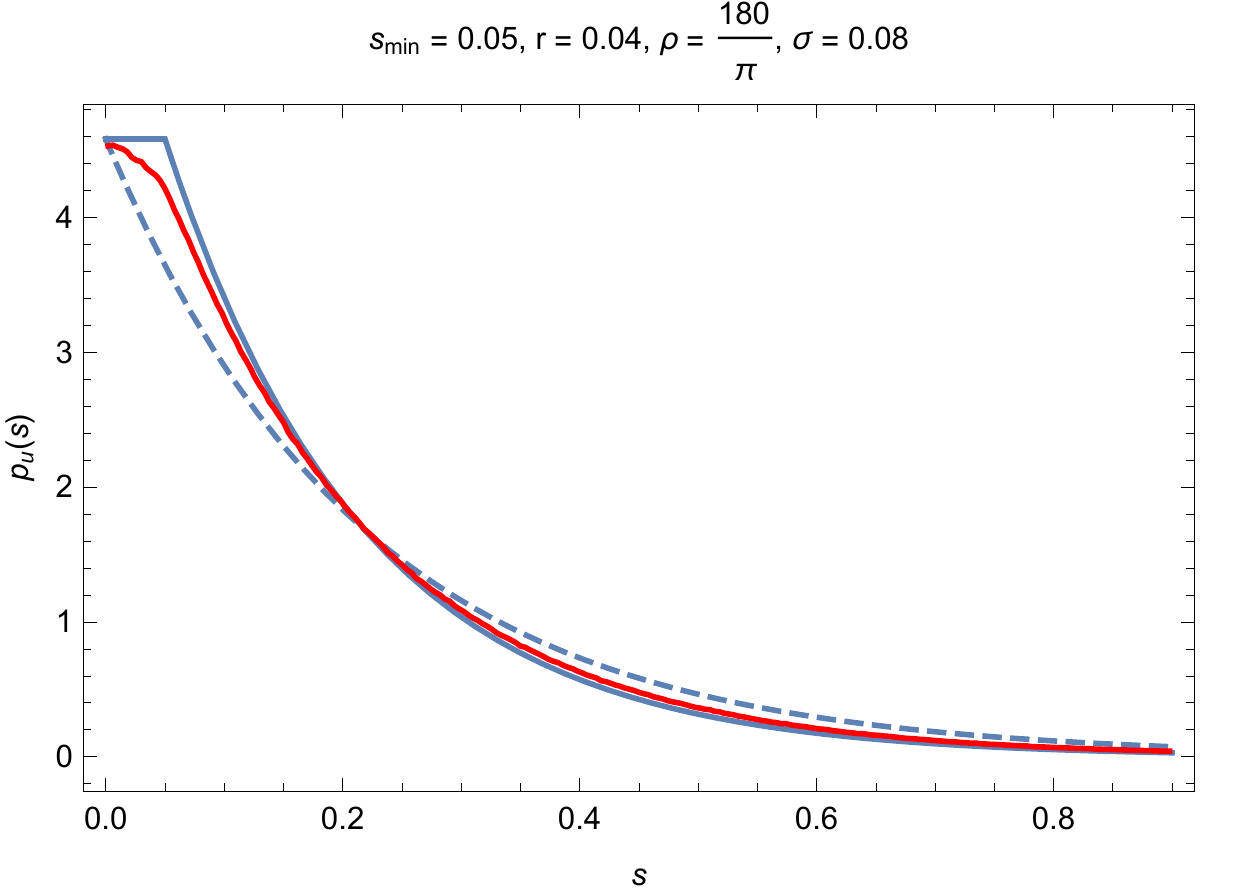}}
	  \subfigure[]{\includegraphics[width=.33\linewidth]{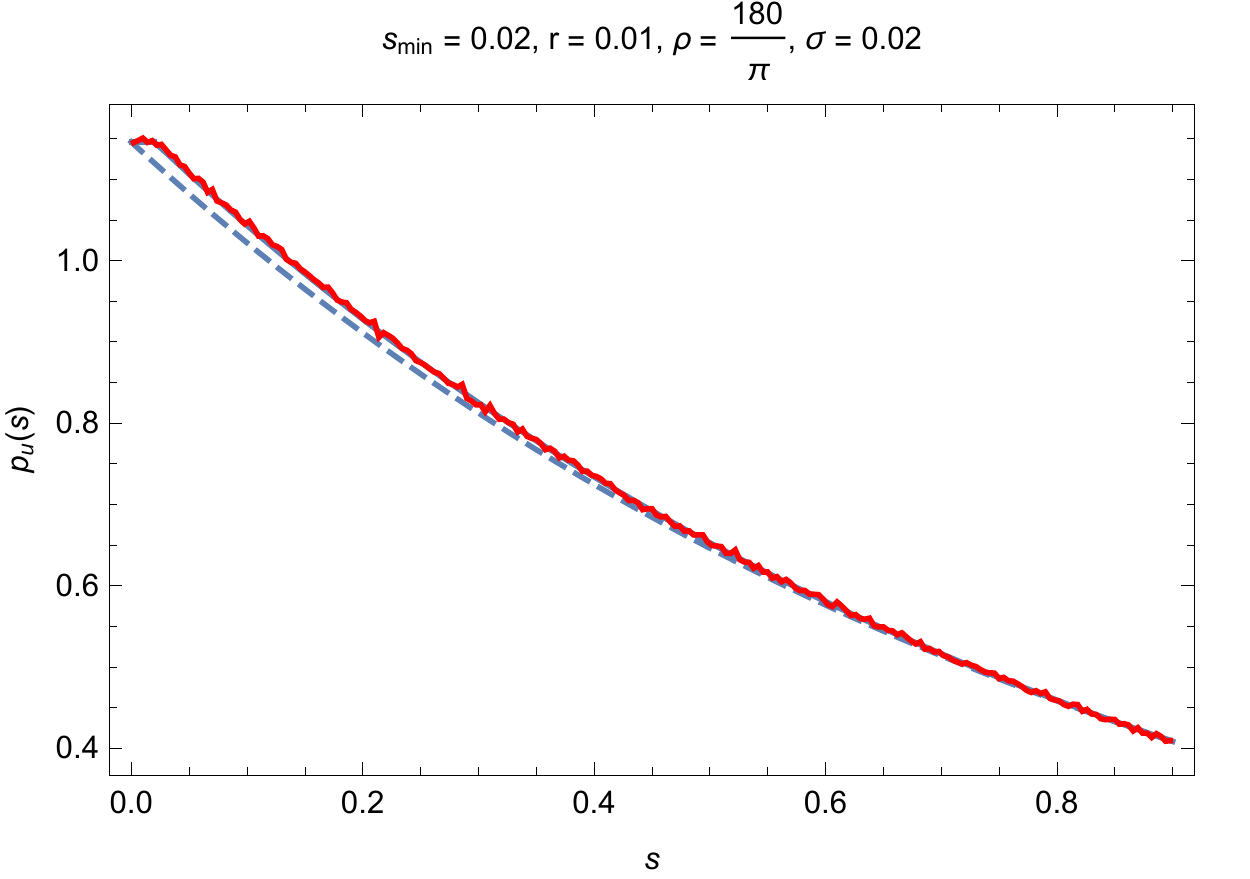}}
	  \subfigure[]{\includegraphics[width=.33\linewidth]{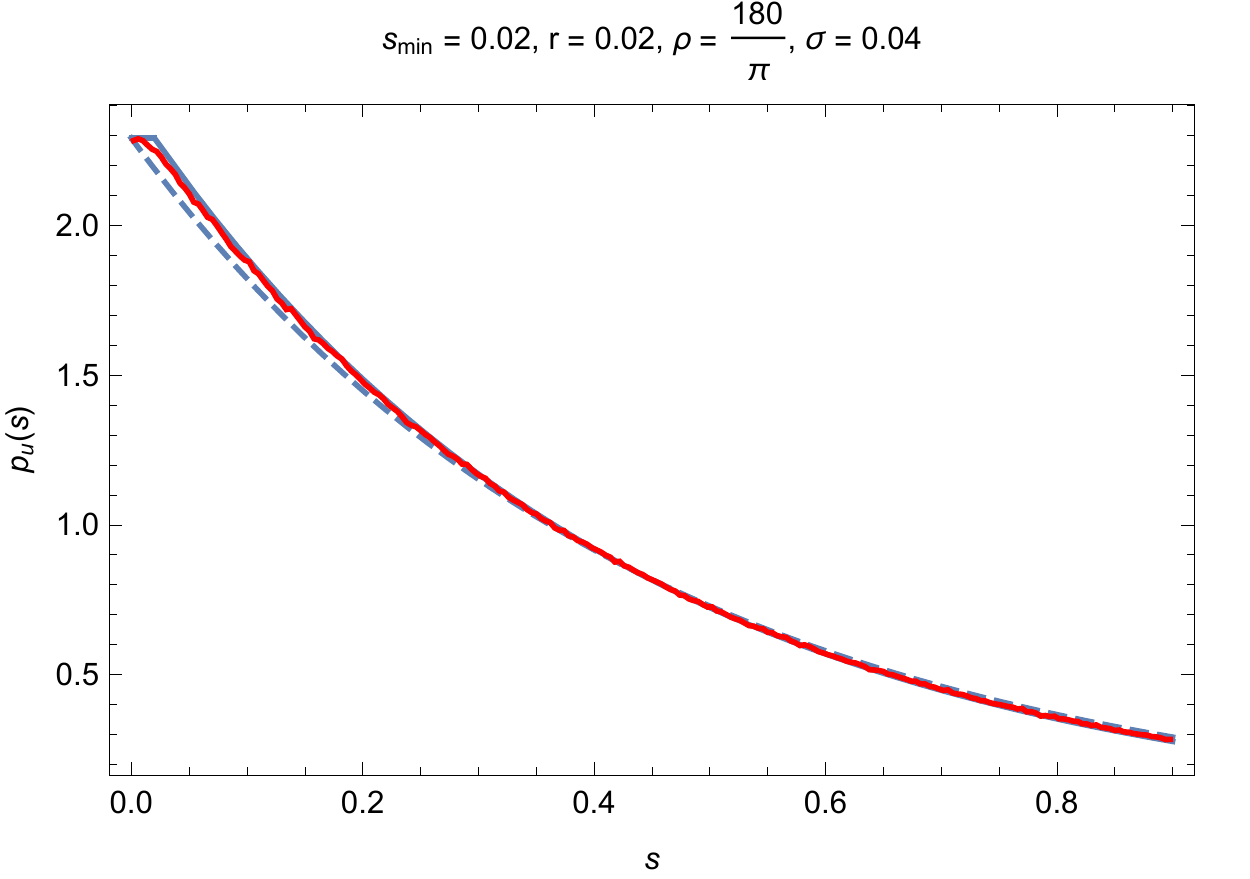}}
	  \subfigure[]{\includegraphics[width=.33\linewidth]{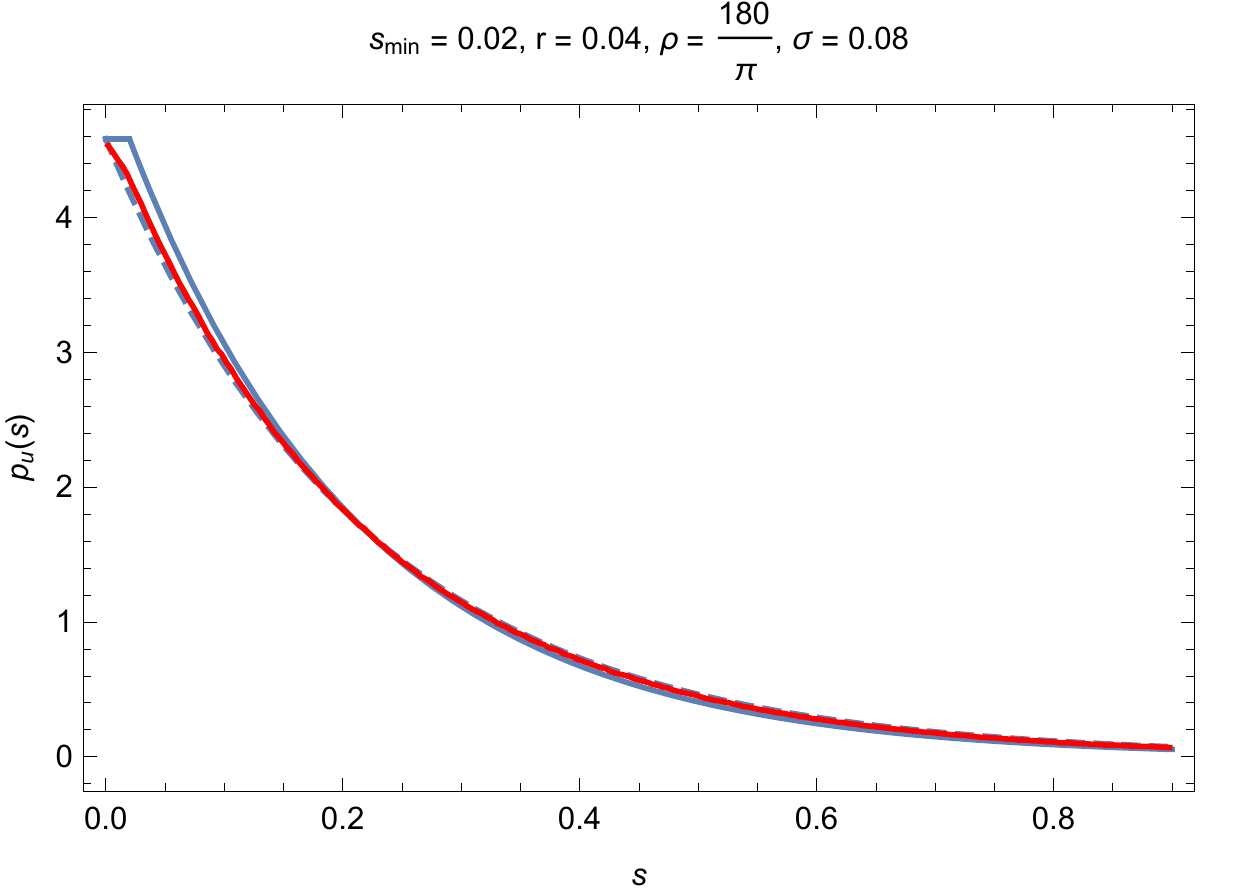}}
	  \subfigure[]{\includegraphics[width=.33\linewidth]{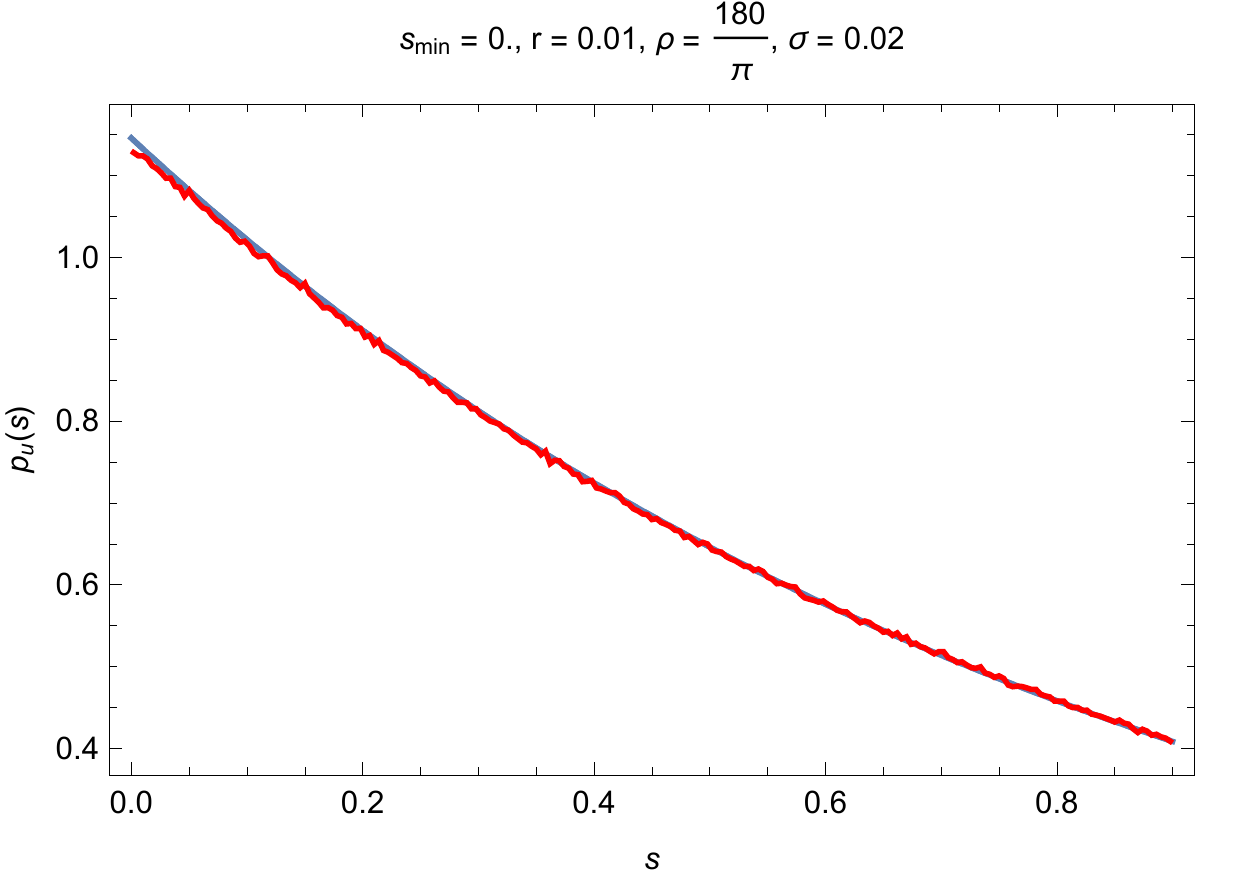}}
	  \subfigure[]{\includegraphics[width=.33\linewidth]{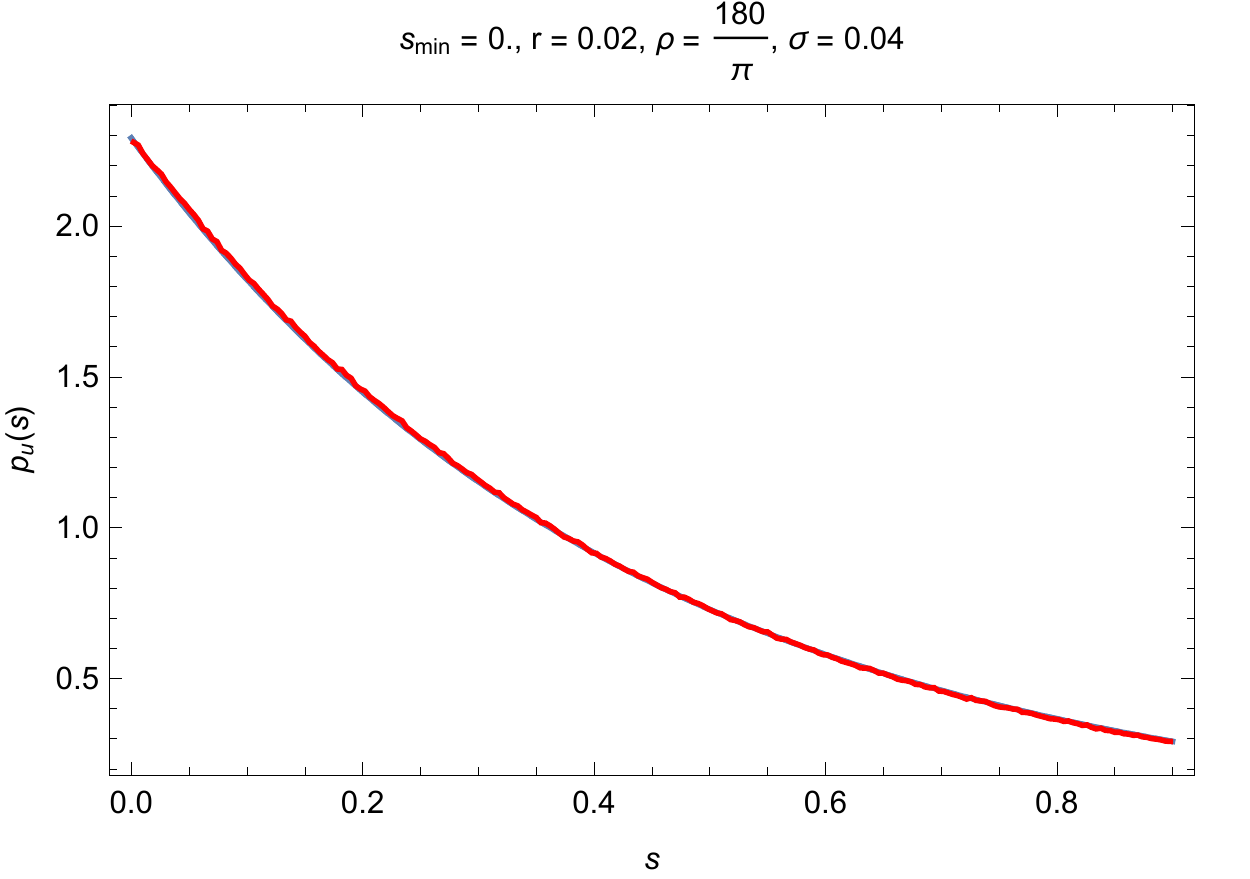}}
	  \subfigure[]{\includegraphics[width=.33\linewidth]{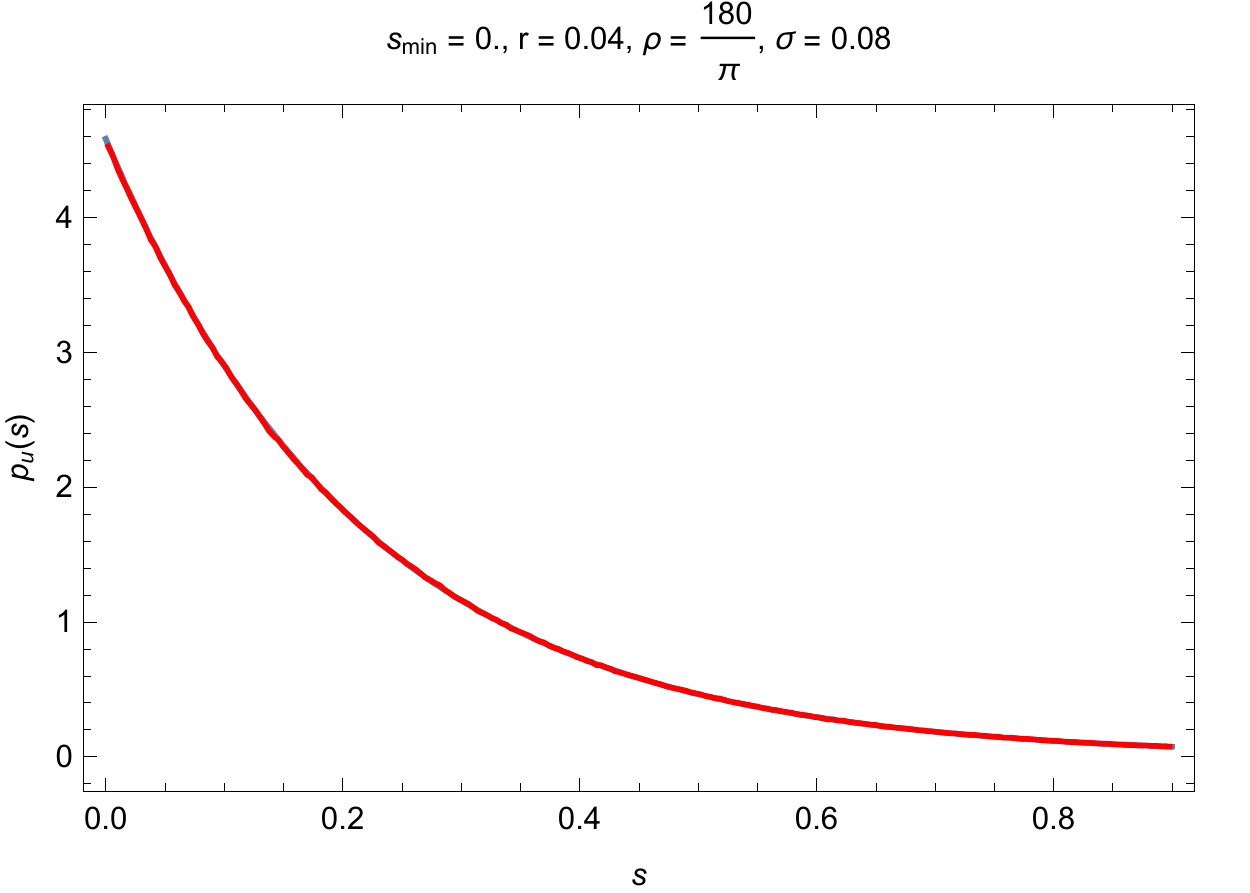}}
	  \caption{Uncorrelated free-path distributions for blue noise scattering.  Our proposed simple analytic form $p_u(s)$ \new{(blue continuous)} vs Monte Carlo (red) and classical exponential free-path (dashed).}
	  \label{fig-freepaths} 
	\end{figure*}
	Figure~\ref{fig-freepaths} illustrates the accuracy of this new simple form of blue noise free-path distributions over a variety of cross sections and separation lengths.  

\begin{figure}
		  \centering
		  \includegraphics[width=.8\linewidth]{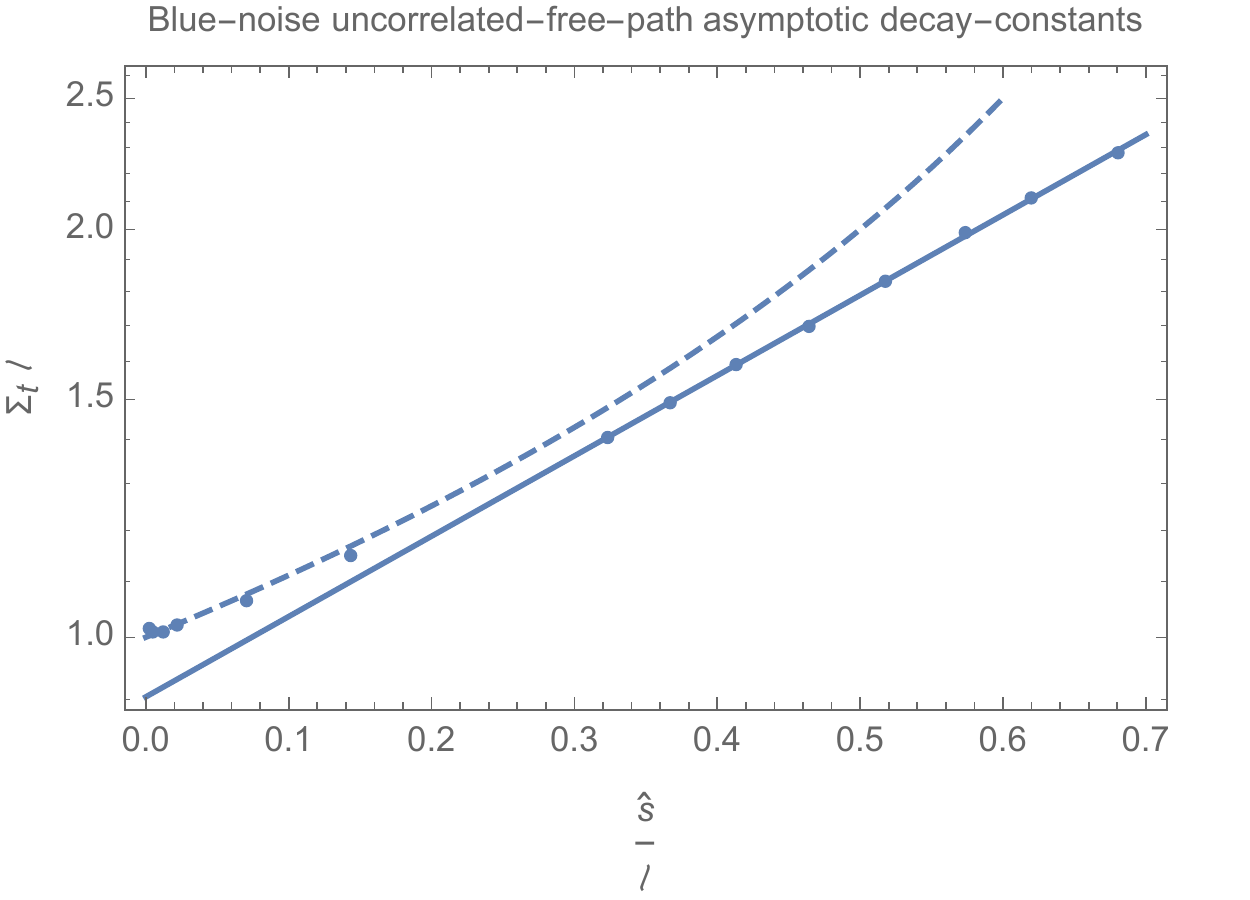}
		  \caption{For dilute blue noise scatters with a small ratio of minimum separation length $\smin$ to classical mean-free-path $\ell$, we find the ratio of classical decay rate to correlated decay rate $\Sigma_t \ell$ well approximated by our analytic model \new{(dashed)}, but observe an exponential trend \new{(continuous blue curve)} for highly occupied volumes and recommend an interpolation of the two.}
		  \label{fig-poorasymp} 
	\end{figure}

	\paragraph{Increased-accuracy asymptotics}
	  We found that the asymptotics of the uncorrelated free paths were well approximated by exponentials but note that we did not test ranges past $18$ classical mean free paths.  For shielding calculations and other applications where the asymptotic attenuation is required with great accuracy we note that our exponential tails $e^{-\Sigma_t s}$ have poor decay-constants $\Sigma_t$ as the ratio of separation length to classical mean free path ($\smin / \ell$) gets large.  \new{We observed up to $25\%$ error between our simple model and MC simulations in the range $ 0.3 < \smin / \ell < 0.65$}.   Figure~\ref{fig-poorasymp} shows the observed decay-constant ratios of classical to correlated $\Sigma_t \ell$ as a function of the ratio of separation length to classical mean free path for the special case of hard disks ($\smin = 2 r$).  For less dilute volumes we observe the decay-rate ratio to be better approximated by an exponential of $\smin / \ell$,
	  \begin{equation}
	  	\Sigma_t \ell \approx 0.903537 e^{1.36543 \hat{s} / \ell}
	  \end{equation}
	  as an improvement to our simpler model's
	  \begin{equation}
	  	\Sigma_t \ell = \frac{1}{1 - \frac{\smin}{\ell}},
	  \end{equation}
	  \new{which was found by a least squares fit to an exponential form from 8 measured data points with $ 0.3 < \smin / \ell < 0.65$}.

  \subsection{Correlated Minimum-Poisson-Disk free paths}
    We measured the free-path statistics for random paths with correlated starting positions in blue noise Flatland media by using the sampling procedure described in Section~\ref{sec:bluepu} with the modification that the initial position of the path is chosen by randomly selecting one of the medium particles inside the unit source disk and starting the ray at that particle's center.
    \begin{figure*}
	  \centering
	  \subfigure[]{\includegraphics[width=.33\linewidth]{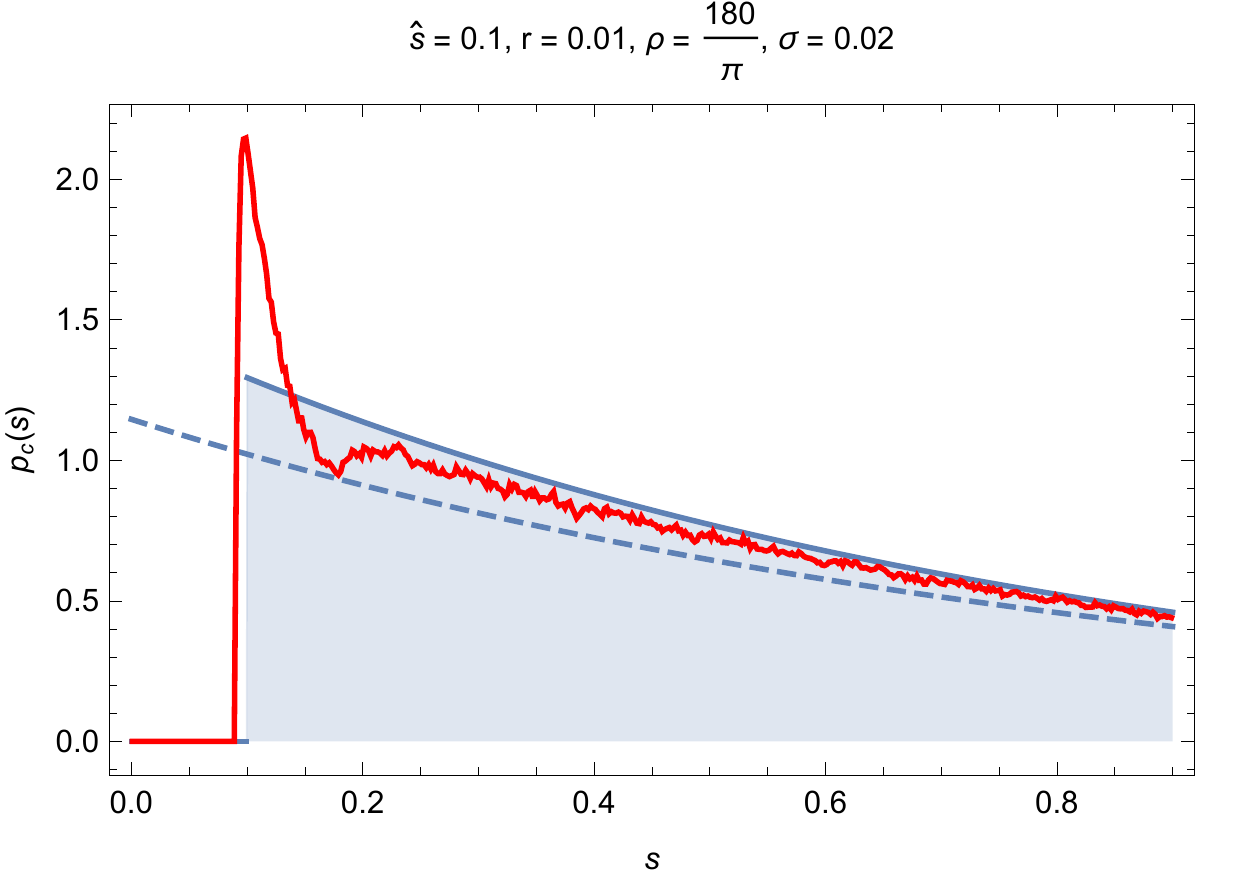}}
	  \subfigure[]{\includegraphics[width=.33\linewidth]{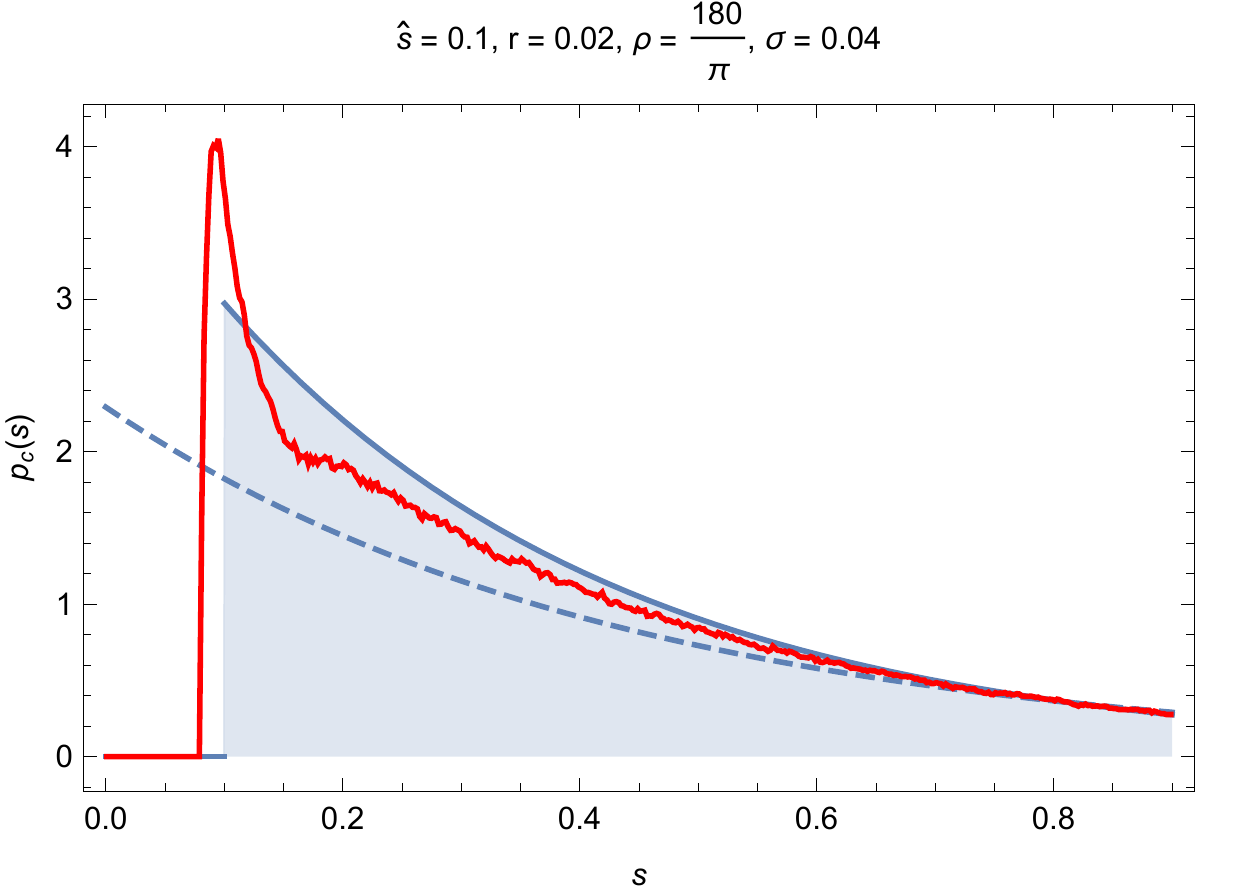}}
	  \subfigure[]{\includegraphics[width=.33\linewidth]{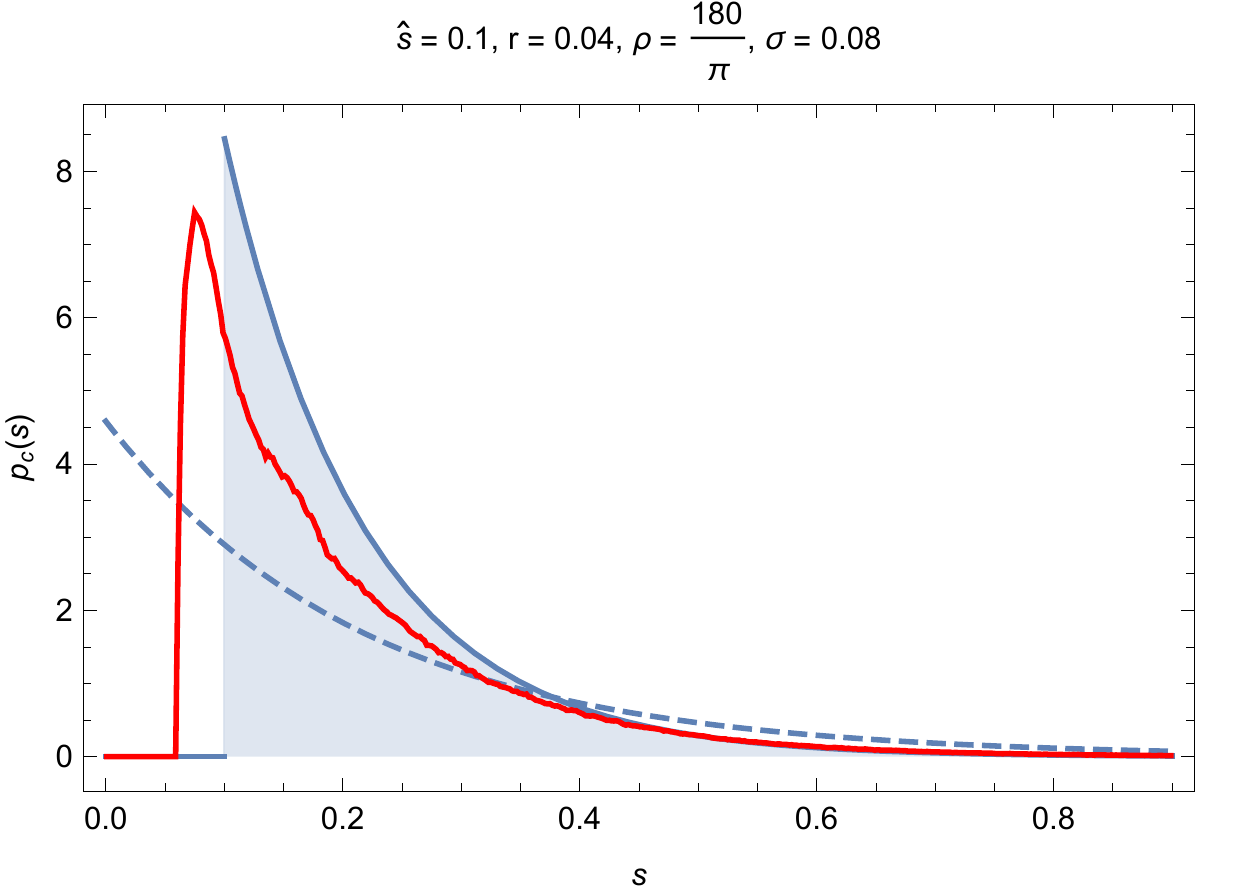}}
	  \subfigure[]{\includegraphics[width=.33\linewidth]{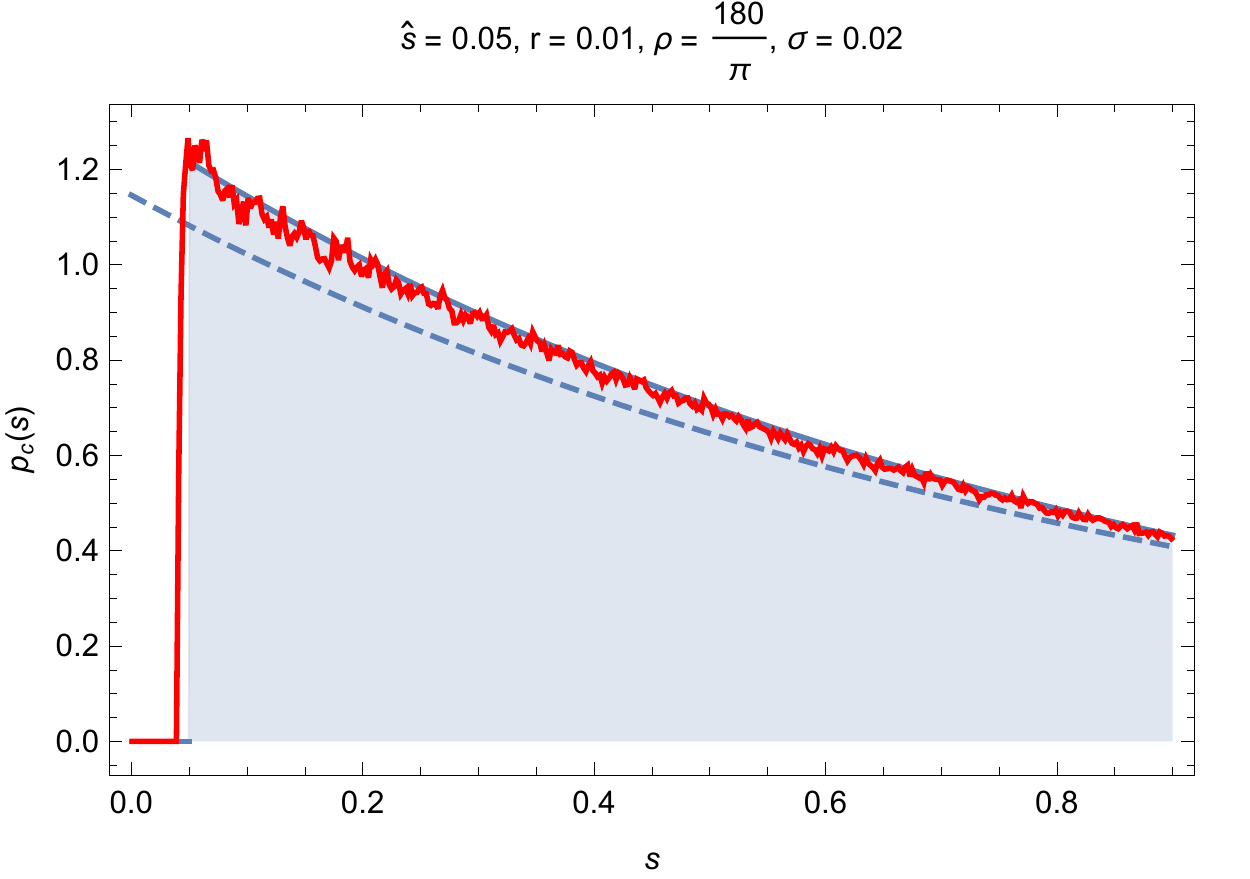}}
	  \subfigure[]{\includegraphics[width=.33\linewidth]{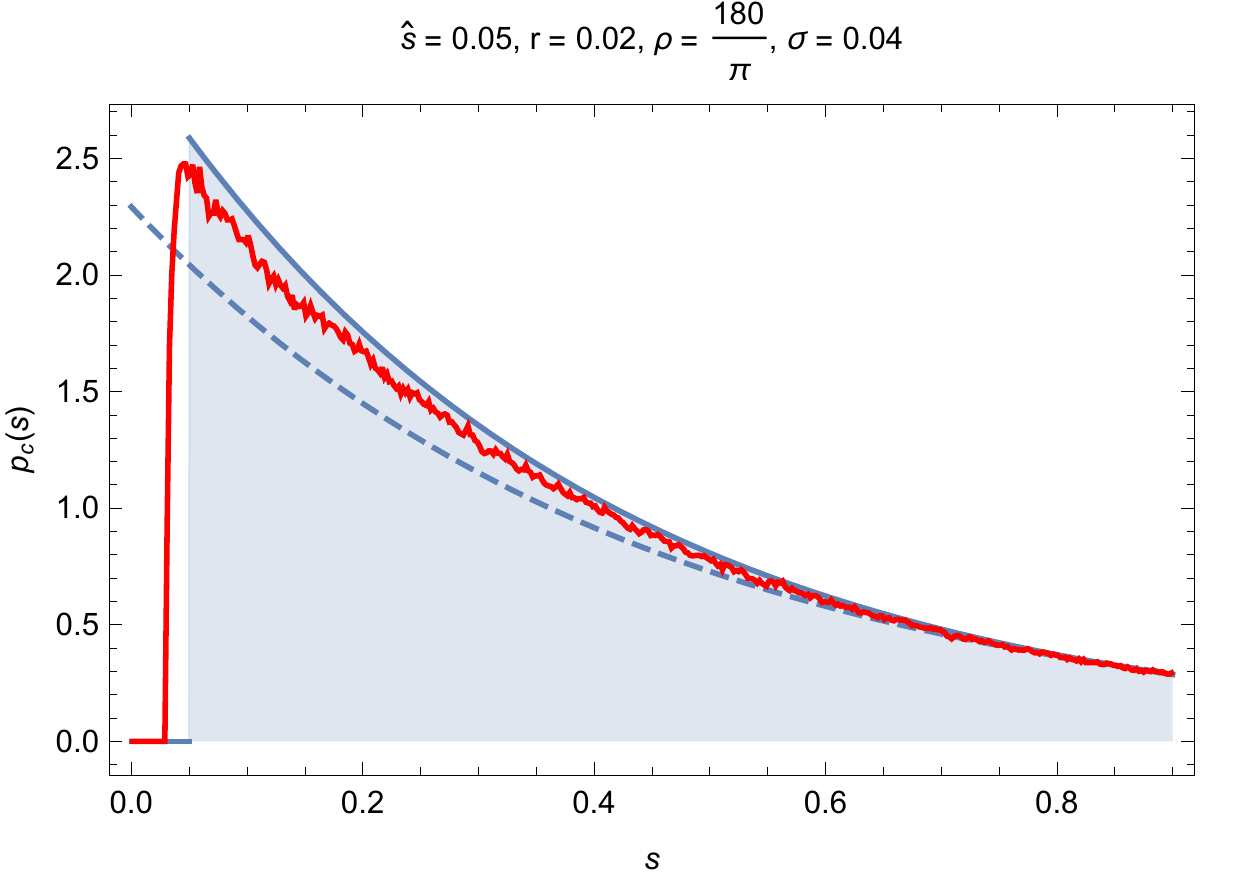}}
	  \subfigure[]{\includegraphics[width=.33\linewidth]{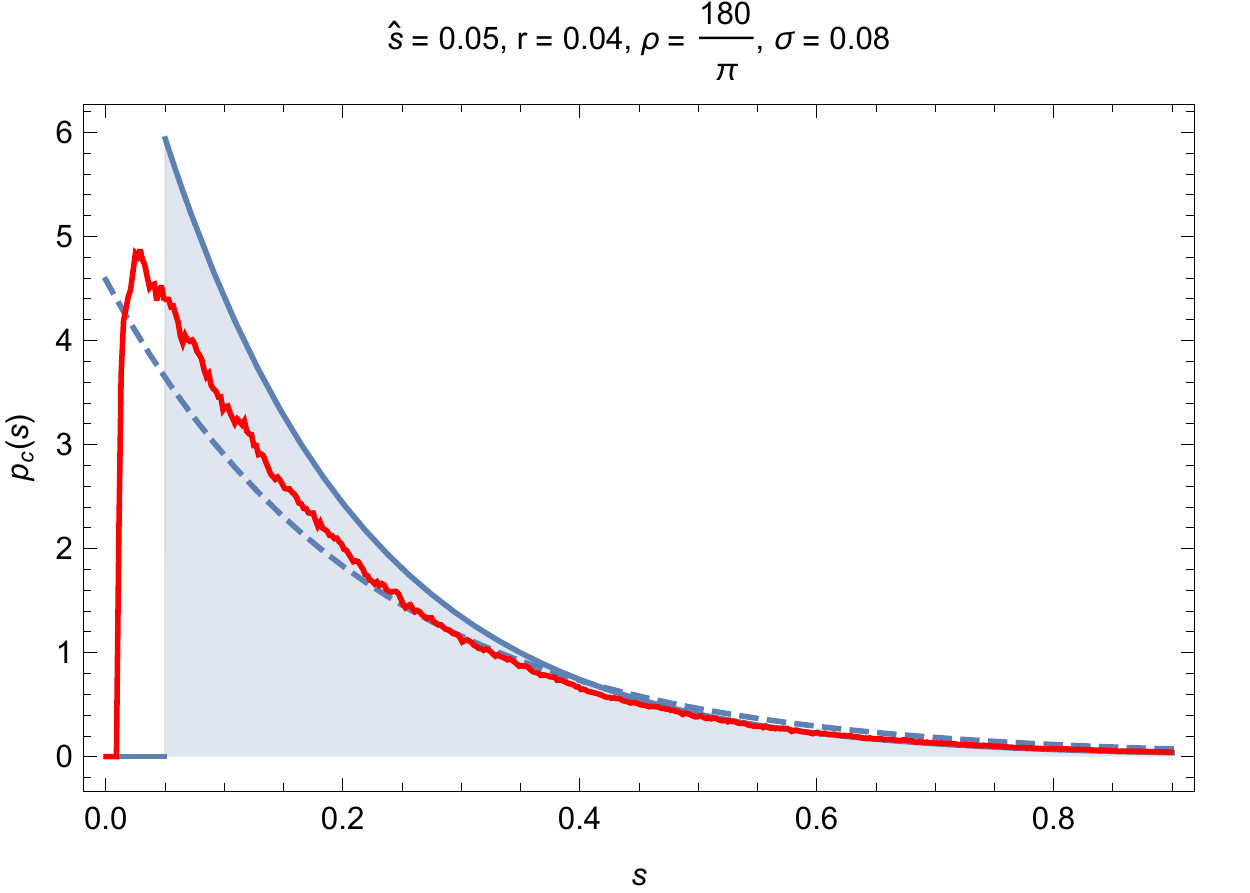}}
	  \subfigure[]{\includegraphics[width=.33\linewidth]{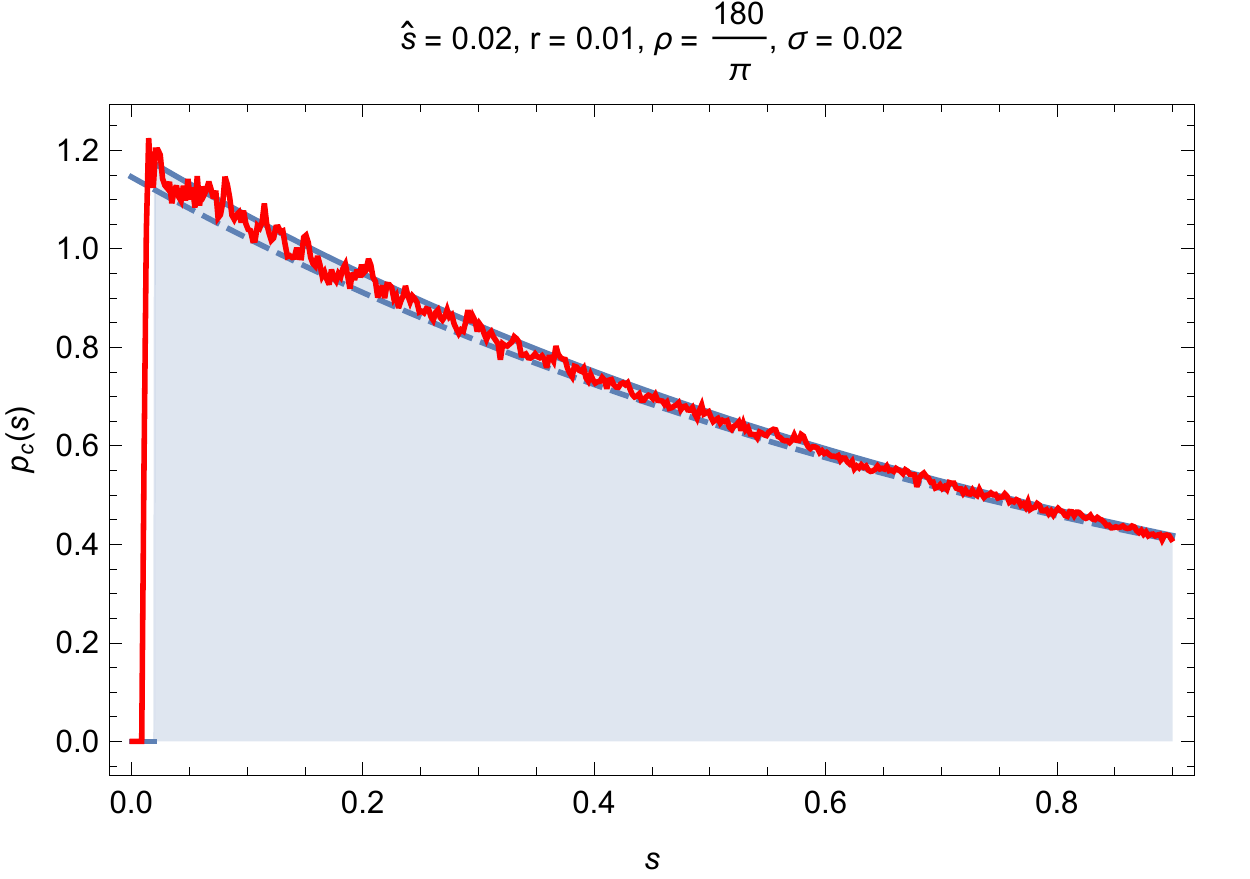}}
	  \subfigure[]{\includegraphics[width=.33\linewidth]{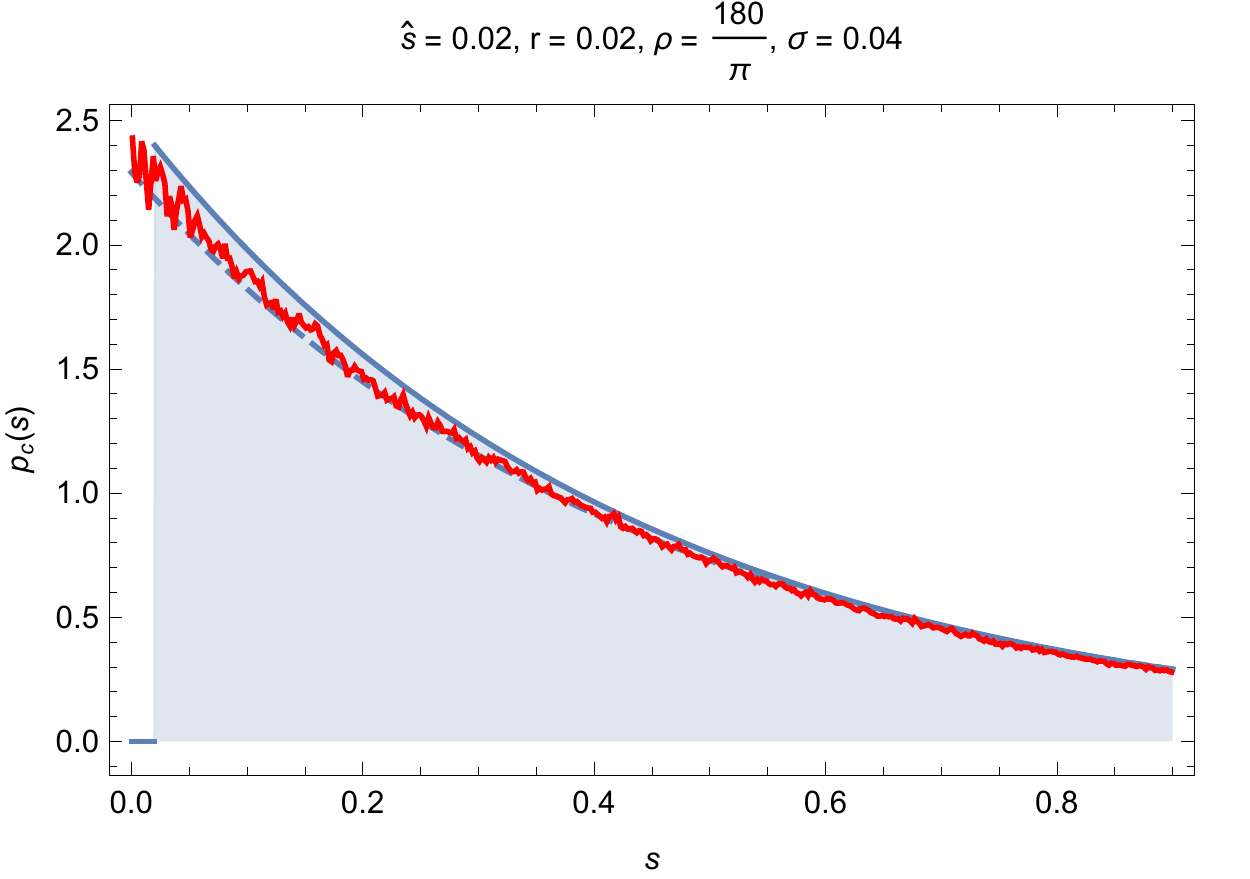}}
	  \subfigure[]{\includegraphics[width=.33\linewidth]{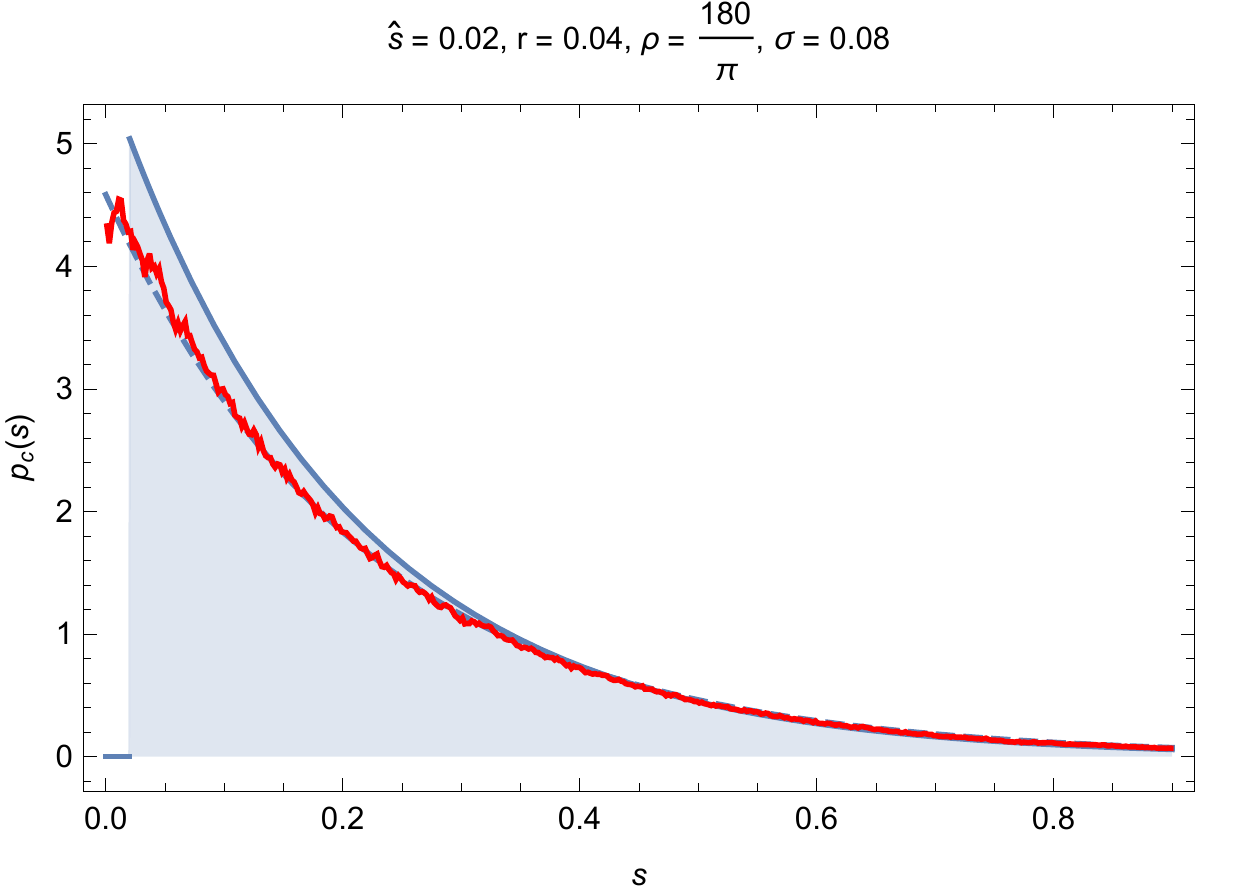}}
	  \caption{Correlated free-path distributions for blue noise scattering.  Our proposed simple analytic form $p_c(s)$ (filled) vs Monte Carlo (red) vs classical exponential free-path (dashed).}
	  \label{fig-corrfreepaths} 
	\end{figure*}  
	A selection of results are summarized in Figure~\ref{fig-corrfreepaths}.  As the separation length increases we see that correlated free paths are significantly different for small $s$ than the uncorrelated case of the previous section.  The absence of any collisions below the separation length is clearly apparent, as well as the ringing behaviour seen in Percus-Yevick distributions~\cite{smith70b,henderson09}.  Because our estimator returns the path length to the edge of finite-sized disks, the distributions rise sooner than $s = \smin$, but rather at $s = \smin - r$.

	To avoid the complexity of defining $\Sigma_{tc}(s)$ in terms of Percus-Yevick solutions, and to choose a form consistent with the uncorrelated case, we propose using a simple model for the correlated free path found by setting the macroscopic cross section to
	\begin{equation}
		\Sigma_{tc}(s) = \begin{cases}
		0 & 0 \le s < \smin\\
		\frac{1}{\ell-\smin} & s \ge \smin
		\end{cases}
	\end{equation}
	which, by Equation~\ref{eq:pc} produces a correlated free-path distribution
	\begin{equation}
		p_c(s) = \begin{cases}
		0 & 0 \le s < \smin\\
		\frac{e^{-\frac{s-\smin}{\ell-\smin}}}{
   \ell-\smin} & s \ge \smin
		\end{cases}.
	\end{equation}
	This distribution has a mean free path $\mfp_c = \ell$ and mean square free path $\mfp_c^2 = 2 \ell^2-2 \ell \smin+\smin^2$.
	The correlated transmittance is then
	\begin{equation}
		X_c(s) = \ell \, p_u(s)
	\end{equation}
	where $p_u(s)$ is defined by Eq.~\ref{eq:publue}.
	Correlated free-path lengths are easily sampled from a single uniform random variable $\xi \in [0,1]$ via
    \begin{equation}
    	s = \smin - (\ell - \smin) \log \xi.
    \end{equation}
	These distributions are compared to Monte Carlo in Figure~\ref{fig-corrfreepaths} and perform reasonably well for dilute media.

  \subsection{Correlated vs uncorrelated free paths origins}

    We have just seen that as the correlation in the medium increases the free path statistics differ significantly for paths beginning at a medium particle vs at an uncorrelated random starting location.  The form of correlation we have chosen presents the most significant differences for short paths and makes clear the importance to distinguish between the two types of free path sampling that might need to be applied.  

\begin{figure*}
	  \centering
	  \subfigure[]{\includegraphics[width=.24\linewidth]{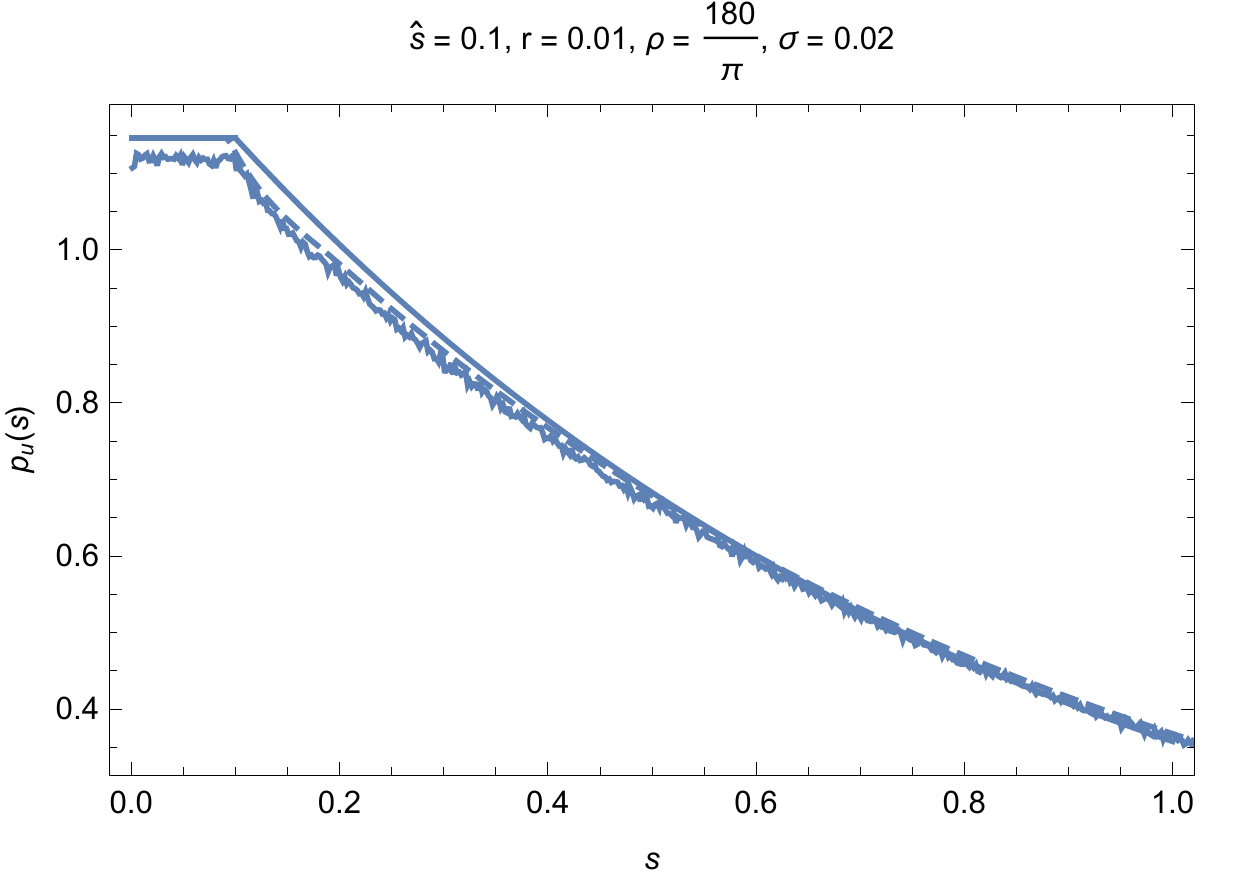}}
	  \subfigure[]{\includegraphics[width=.24\linewidth]{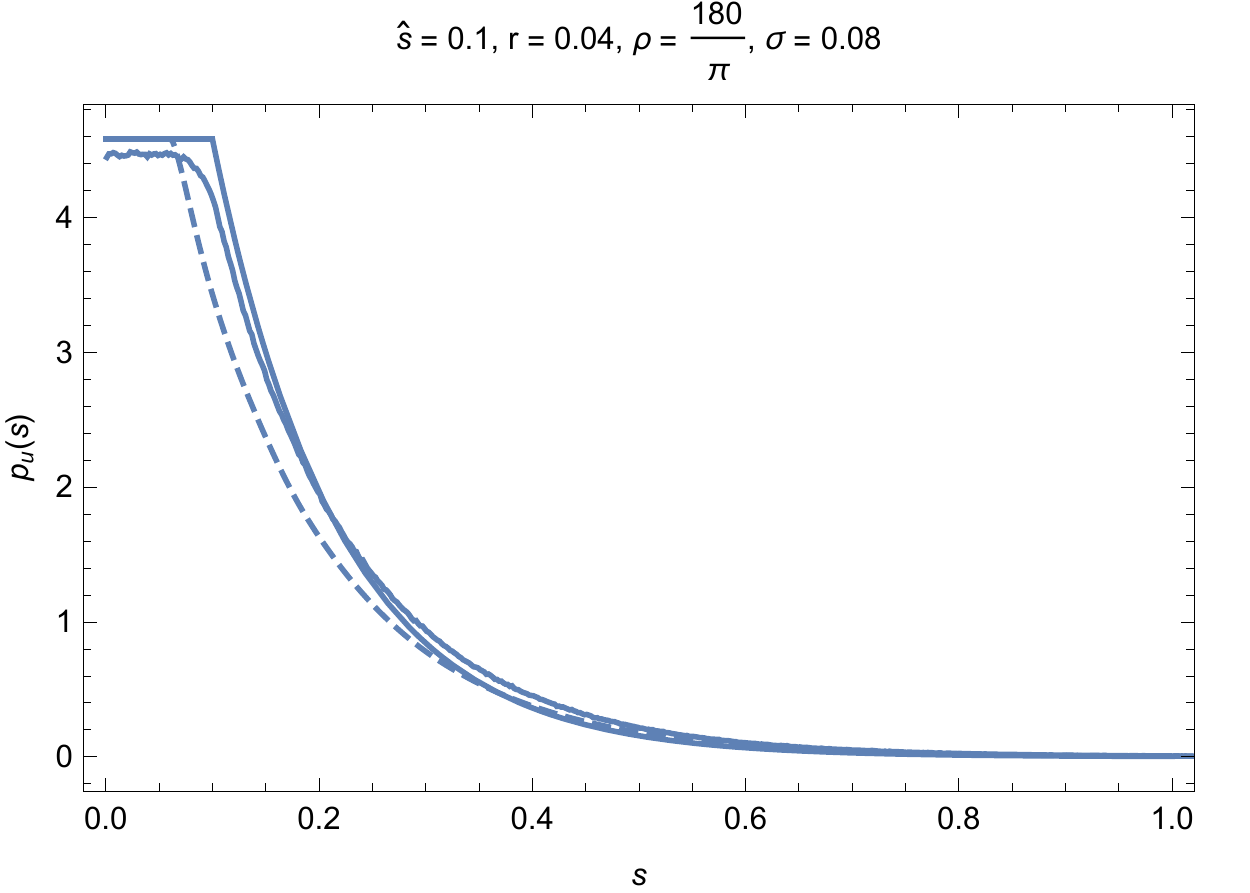}}
	  \subfigure[]{\includegraphics[width=.24\linewidth]{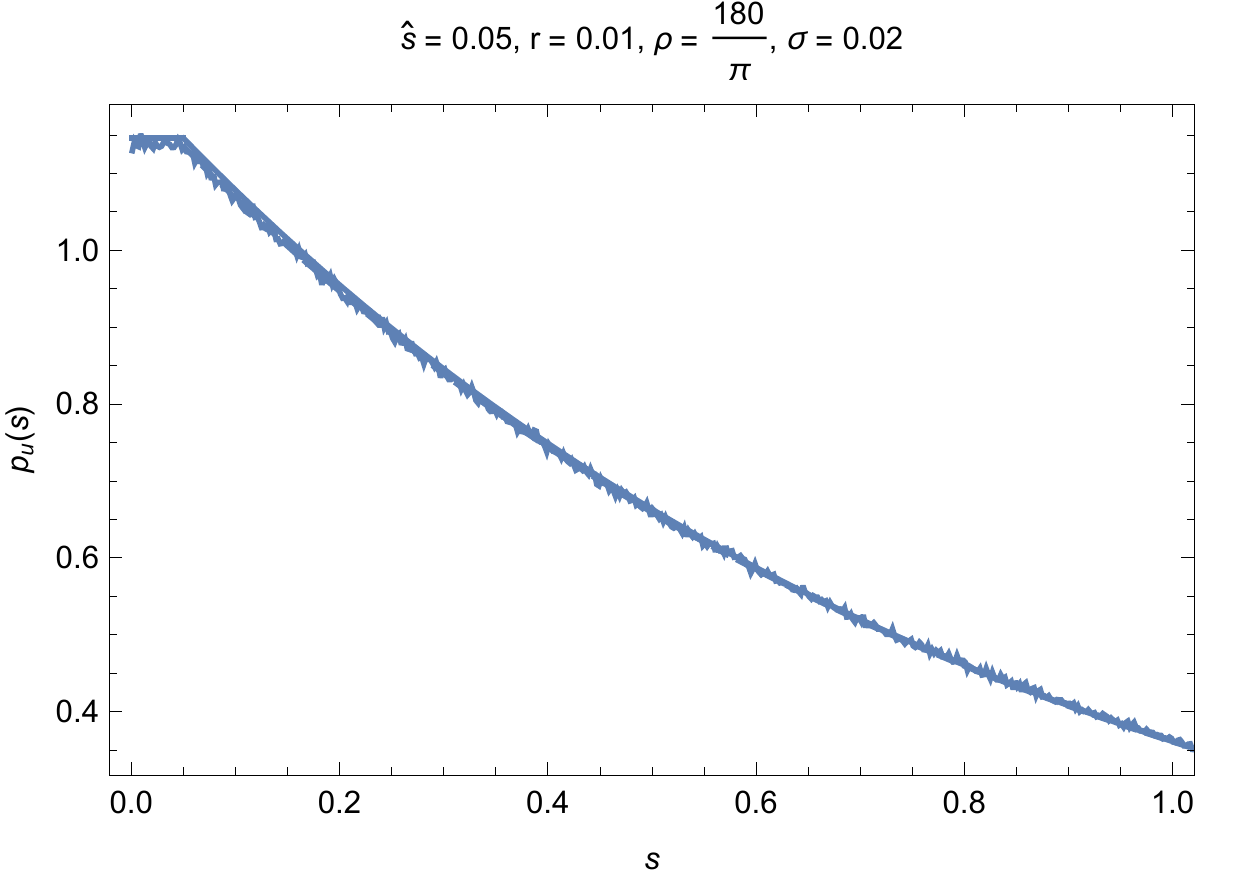}}
	  \subfigure[]{\includegraphics[width=.24\linewidth]{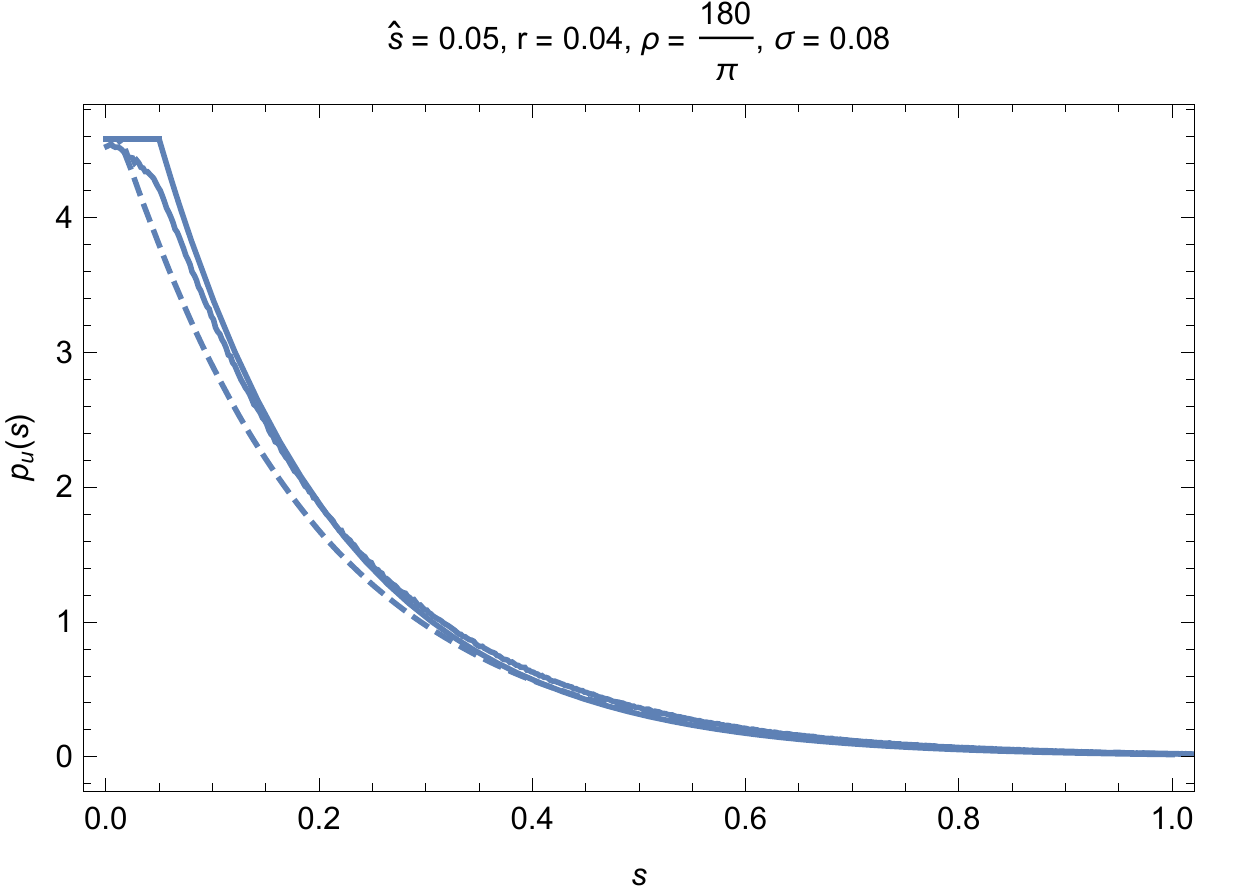}}
	  \caption{Uncorrelated free-path distributions in blue noise infinite media as measured by Monte Carlo (noisy curve) vs as predicted from the correlated free-path distribution by re-normalizing its transmittance (dashed) vs our analytic approximation (smooth curve).}
	  \label{fig-uvsc} 
	\end{figure*}

    If we define the construction of bounded random correlated media by the process of carving finite chunks out of infinite random volumes such that none of the interior particles are correlated to the boundary in any way then it seems intuitive to select the uncorrelated free-path distribution when sampling entry paths into the medium for light arriving at the boundary.  
    We found earlier that weak reciprocity will only be exhibited if the entry sampling is exactly the renormalized correlated transmittance inside the medium and, indeed, we see this relationship appear in our Monte Carlo results, providing additional motivation for our proposal.  In Figure~\ref{fig-uvsc} we show the uncorrelated free-path distributions as measured by our Monte Carlo process compared against the renormalized correlated transmittance as measured by Monte Carlo (and include our analytic proposal for reference).

    \new{We note that the use of $p_u(s)$ for sampling all free-path lengths (including those between two medium collisions) would lead to a transport model with a single attenuation law and might be tempting, in order to simplify the implementation of Monte Carlo methods and the transport equation.}  However, as we have seen, this must lead to non-reciprocal transport for non-exponential media.  Additionally, this will degrade low-range low-order scattering accuracy, especially in the case of strong near-field correlations like those studied in this section.  Finally, the long-range multiple-scattering asymptotics approach a diffusion limit determined by the mean free path and mean square free path~\cite{larsen11} and thus, bulk scattering accuracy will degrade by making this approximation, as these moments can be significantly different in the uncorrelated and correlated cases.
 	 
\newpage
  \subsection{Angular memory}

    We now briefly consider the angular aspect of transport memory in random collections of slow moving or static scatterers, as we expect coherent backscattering to appear in our simulations and not in our transport model.  We would hope to find that any errors in our transport model due to coherent backscattering to be no greater in magnitude than in the case of uncorrelated random media.  To include the opposition or hot-spot effect in a transport formalism, Myneni et al.~\shortcite{myneni91} proposed generalizing the concept of macroscopic cross section to depend on the location of the last two scattering events.  This permits allowing a higher probability to return uncollided along a previously traversed path in the case that a near-back scattering event is sampled.  This is closely related to the three-point probability function in two-phase random media.

\begin{figure*}
	  \centering
	  \subfigure[$\smin = 0$]{\includegraphics[width=.33\linewidth]{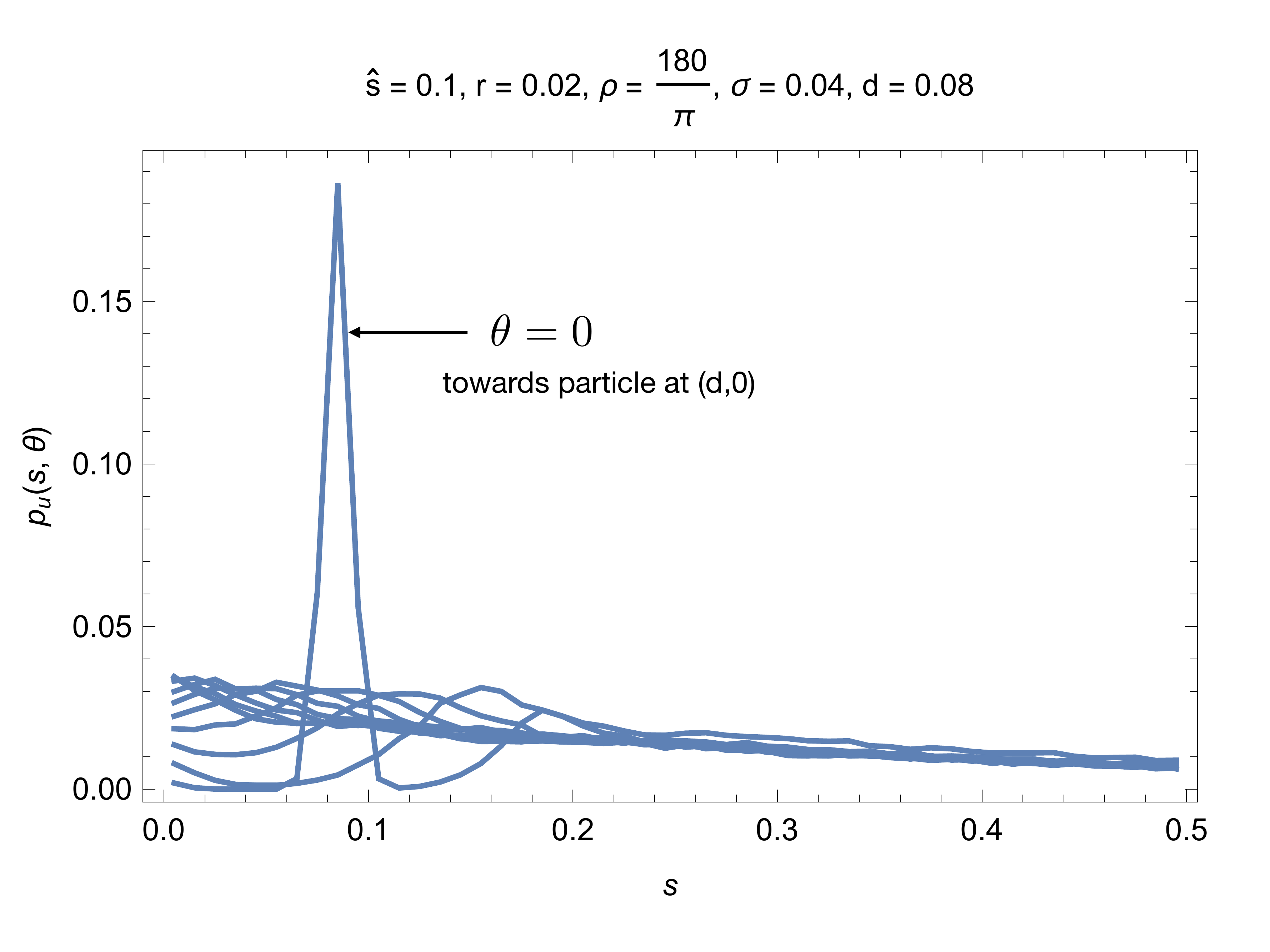}}
	  \subfigure[$\smin = 0$]{\includegraphics[width=.33\linewidth]{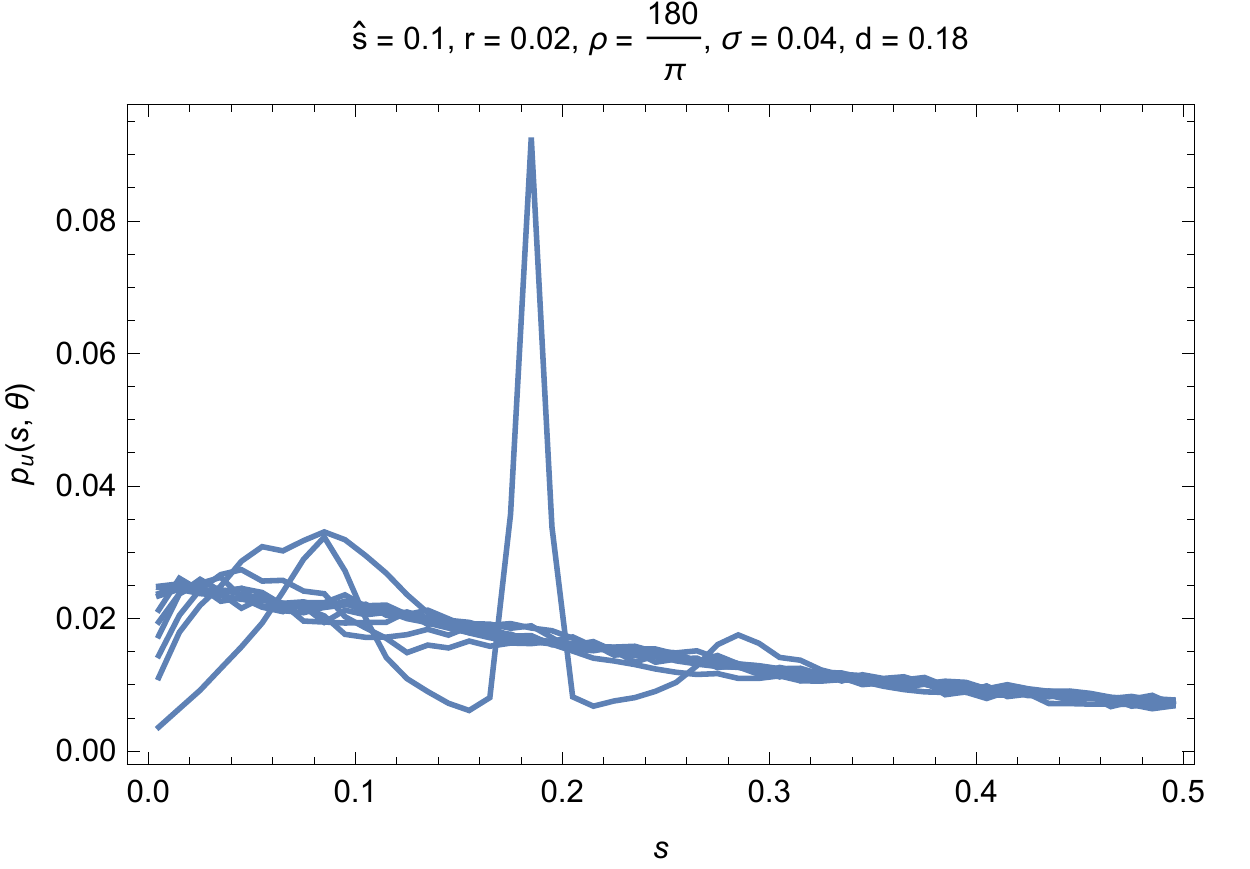}}
	  \subfigure[$\smin = 0$]{\includegraphics[width=.33\linewidth]{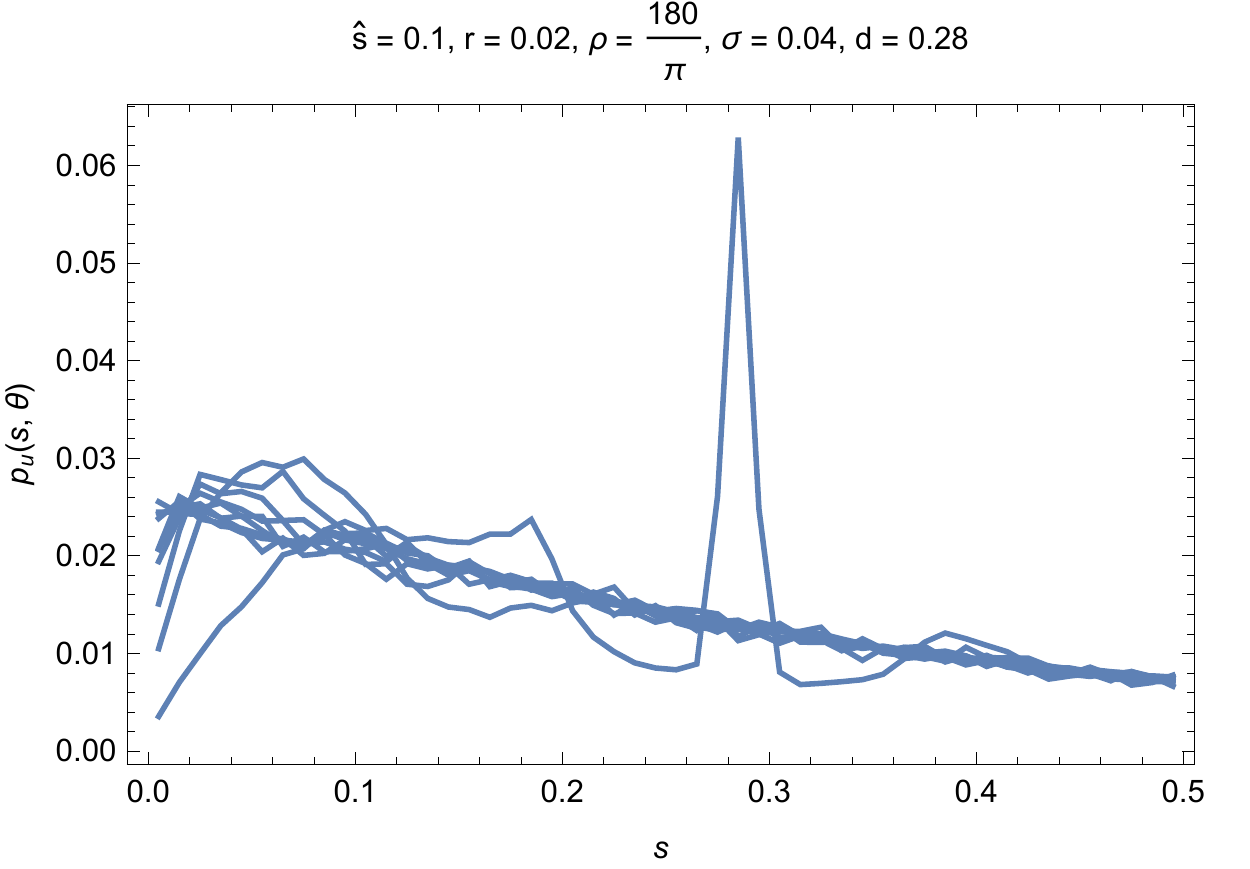}}
	  \subfigure[$\smin = 0.1$]{\includegraphics[width=.33\linewidth]{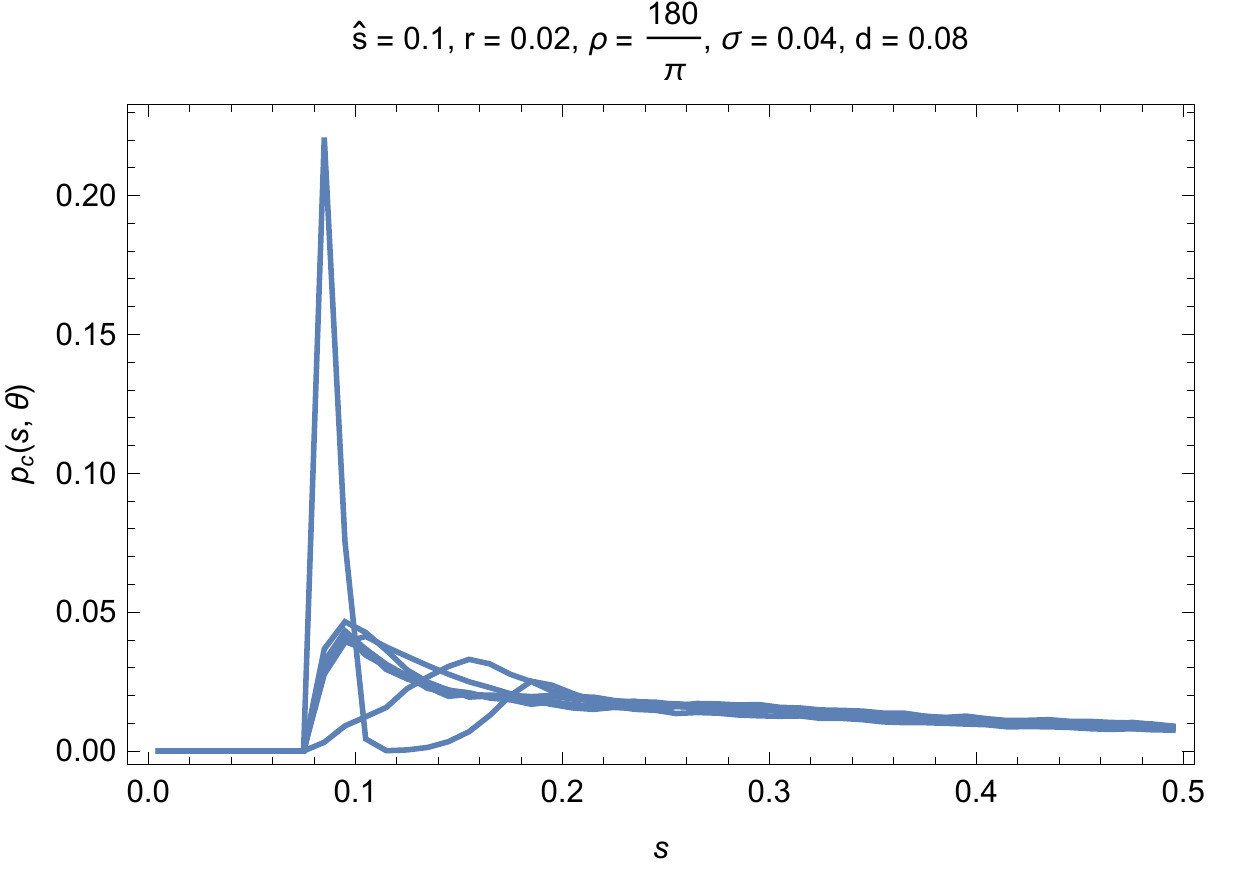}}
	  \subfigure[$\smin = 0.1$]{\includegraphics[width=.33\linewidth]{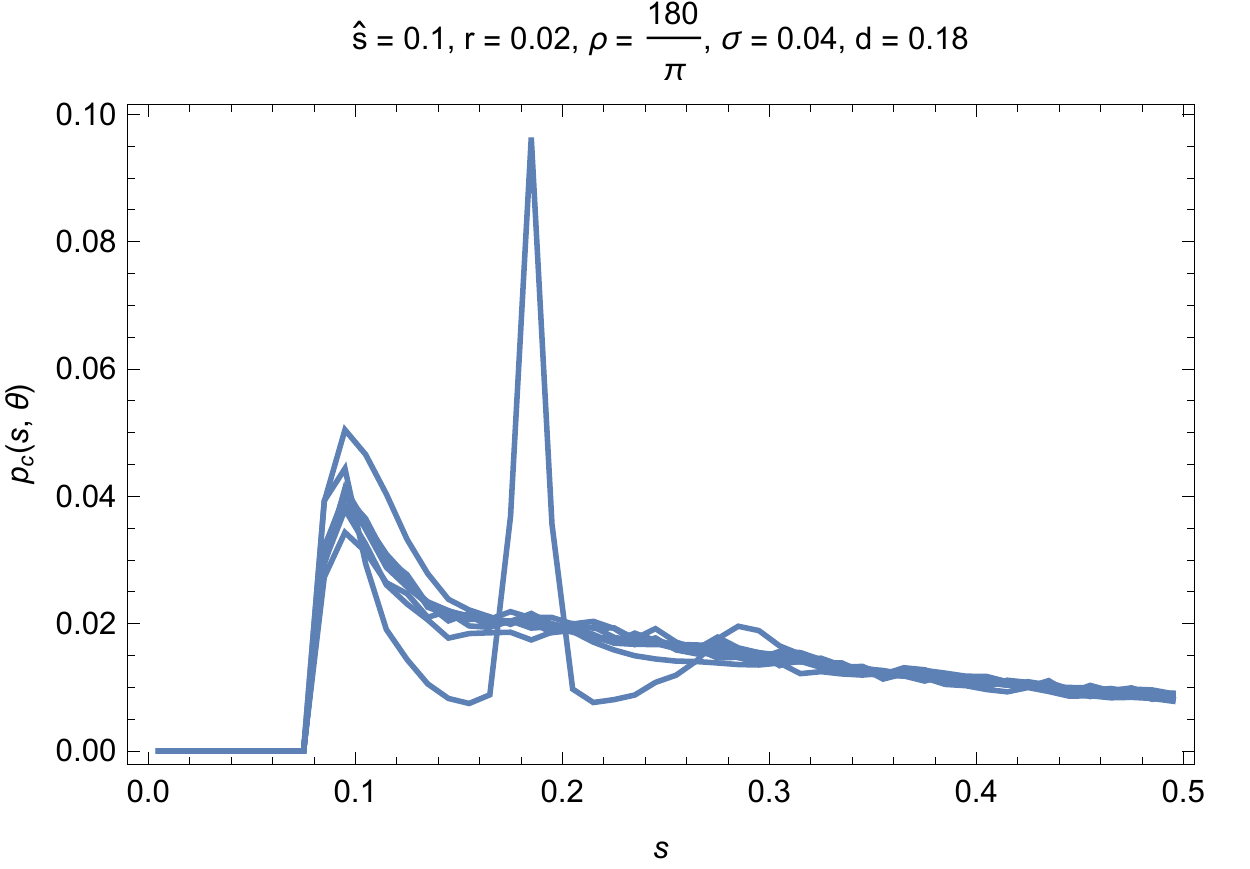}}
	  \subfigure[$\smin = 0.1$]{\includegraphics[width=.33\linewidth]{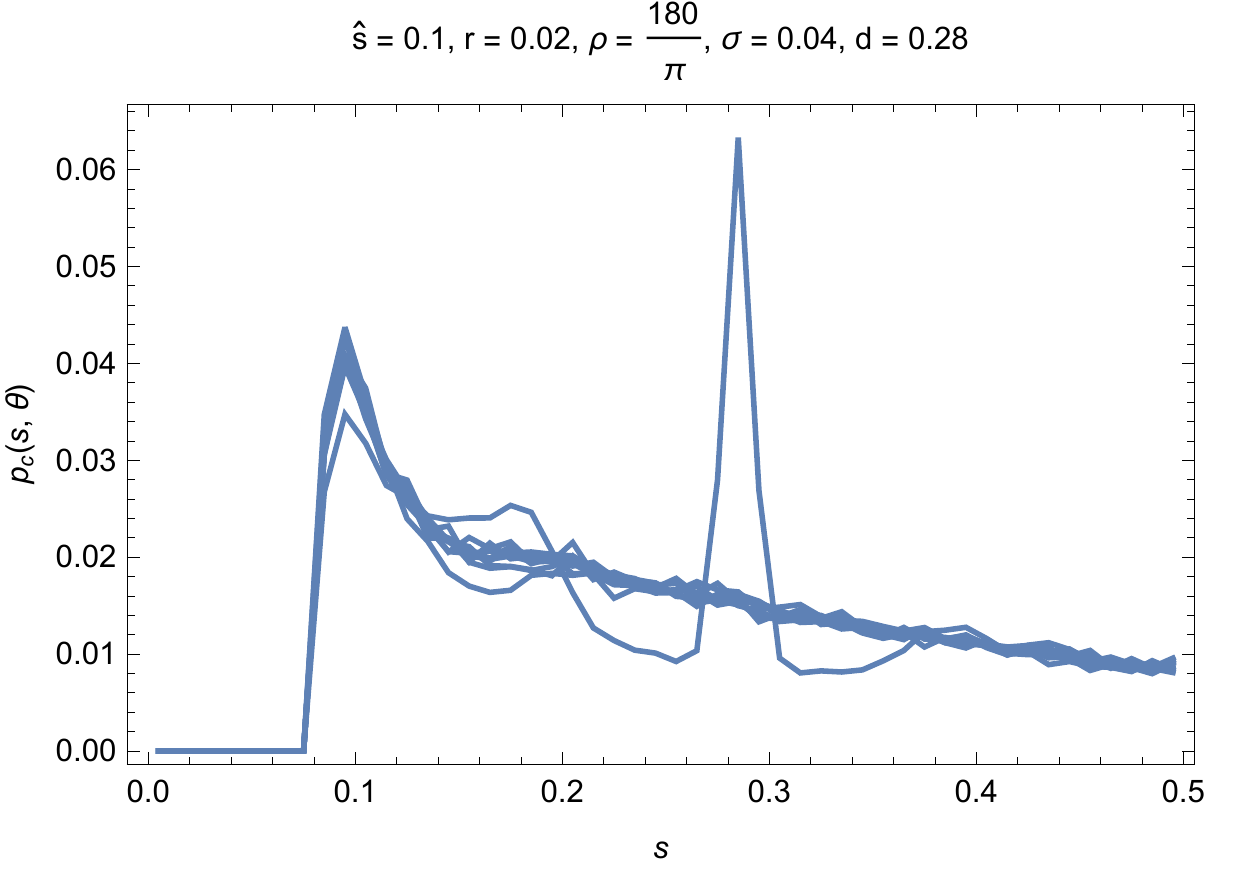}}
	  \subfigure[density plot $\smin = 0$]{\includegraphics[width=.33\linewidth]{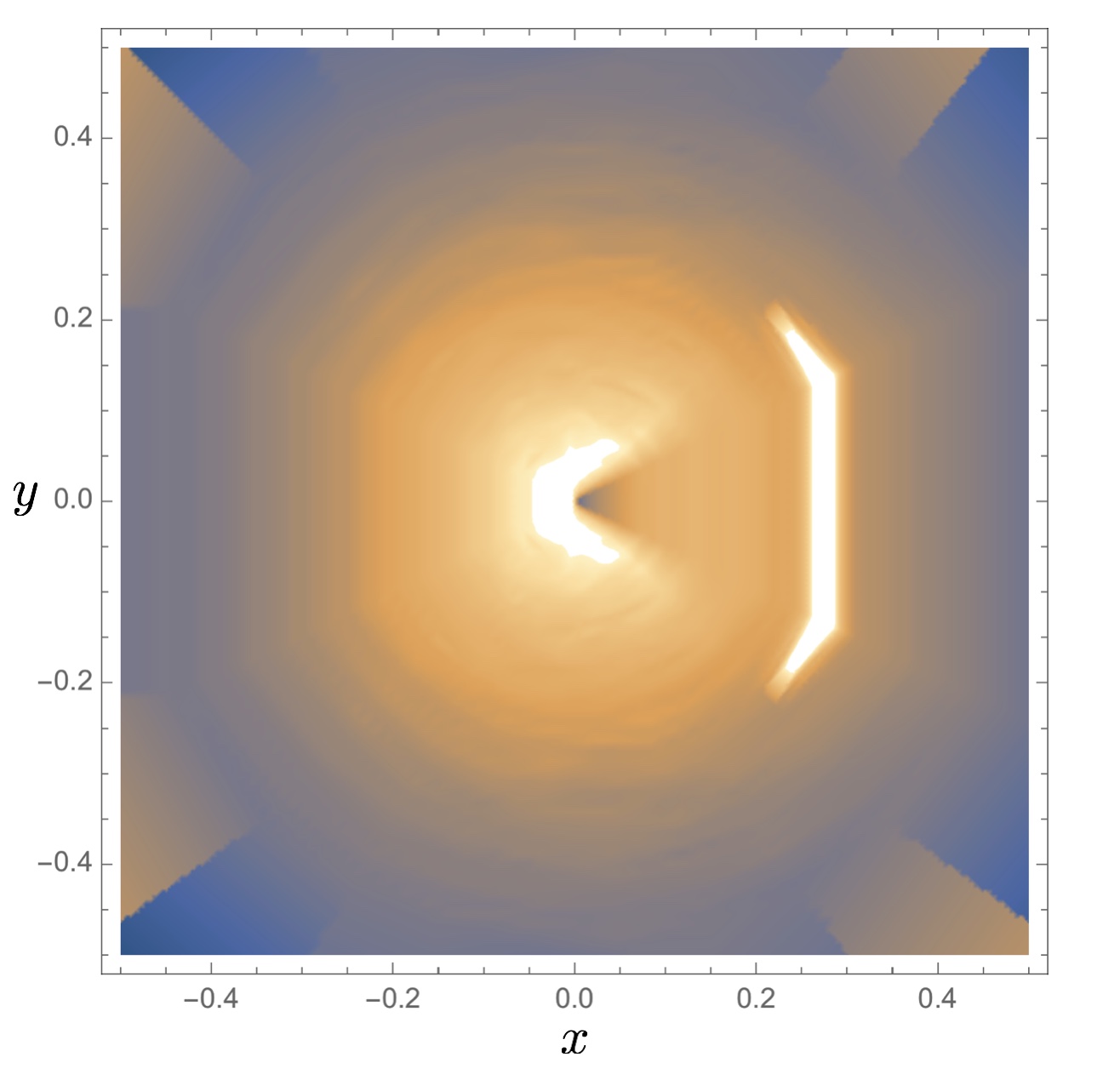}}
	  \subfigure[density plot $\smin = 0.1$]{\includegraphics[width=.33\linewidth]{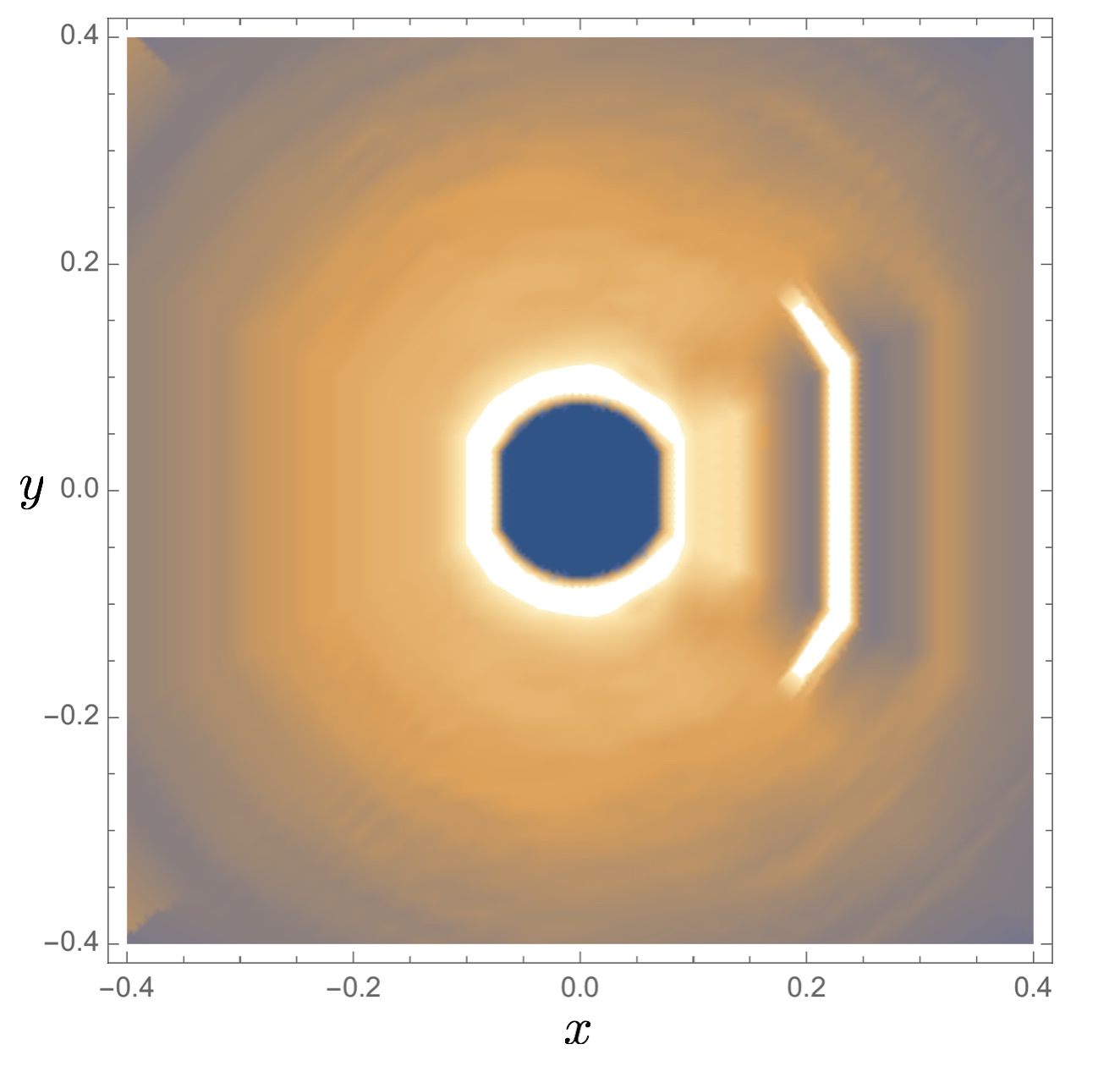}}
	  \caption{Free-path distributions conditioned on the last scattering event occurring a distance $d$ to the right of the current location.  The minimum separation distance $\smin$ creates gaps around the current scattered position as well as the prior.}
	  \label{fig-angular} 
	\end{figure*}

    To measure how the correlation of scattering centers influences the complexity of the \emph{three-point macroscopic cross section} $\Sigma_{tc}(s,-\dir,\pos_1,\pos_2)$, we performed Monte Carlo simulations in blue noise media to accumulate angularly-resolved free path statistics.  Each sample was rotated and accumulated such that the statistics were conditioned on the presence of a particle in the medium directly to the right, a distance $d$ away from the starting location.
    In Figure~\ref{fig-angular} we show the measured angularly-resolved free path distributions for a variety of distances $d$.  For each plot the angular distributions are overlayed, with the majority of them---moving away from the last scattering particle---showing identical statistics to the previously measured unconditioned case.   For the directions heading back towards the last scattering particle, a large spike of collision probability at the distance $d$ of separation is evident.  Proximity and particle radius have relatively intuitive influence on the results.  Several results are also visualized as 2D density plots, which better illustrate the observed anisotropy of $\Sigma_{tc}(s,-\dir,\pos_1,\pos_2)$ when the previous collision location is remembered.
    We note nothing surprising in the case of minimum Poisson disk free paths, with no collisions occurring near the current location as well as a valley of low probability around the last scatterer a distance $d$ away in direction $\dir$.  Future work is required to propose a practical analytic form of $\Sigma_{tc}(s,-\dir,\pos_1,\pos_2)$ that would describe these behaviours well in most cases and add coherent backscattering to GRT.

  \subsection{Purely-absorbing slabs}

    Our first bounded Monte Carlo experiment considers purely-absorbing blue noise slabs with vacuum boundaries.  In measuring penetration depth statistics at various angles of illumination of a half-space, we effectively measure the transmission probabilities for slabs of arbitrary thickness.  Our sampling procedure was to randomly populate optically thick and optically wide rectangles with blue noise using the standard- or extended-dart-throwing method described in \new{S}ection~\ref{sec:dart}.  \new{1000} distinct realizations were sampled and \new{5000} paths were traced in each, beginning at a location near the center of one of the slab surfaces with the starting location pulled back outside of the slab to ensure collision with particles that protrude \new{(by up to $r$)} outside of the sampled boundary due to their finite radius.  

\begin{figure}
		  \centering
		  \includegraphics[width=.8\linewidth]{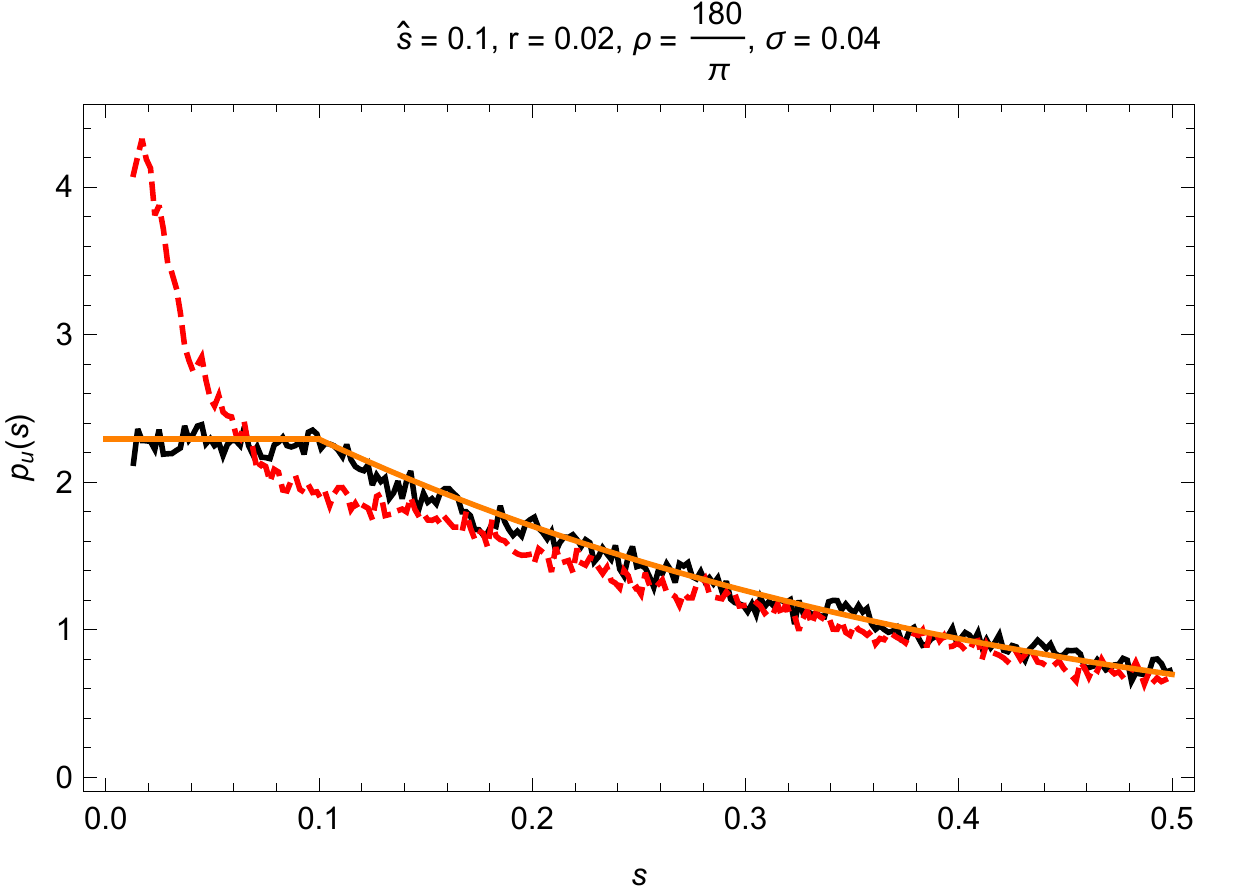}
		  \caption{Uncorrelated free-path statistics for light entering at the boundary of a blue-noise medium with standard dart-throwing (dashed) vs extended-dart-throwing (black) vs our analytic approximation (orange).}
		  \label{fig:penetration} 
	\end{figure}

    For extended-sampled blue noise we found \remove{the exact same} \new{similar (identical apart from MC noise)} free-path statistics for penetration depth, regardless of incoming angle, as we saw in the infinite-medium uncorrelated measurements above.  This was not the case, however, for standard sampling.  The lack of particles outside the boundaries from which to negatively correlate to changes the distribution of the particles near the slab boundaries in a way that was clearly evident in the free path distributions for entry.  Figure~\ref{fig:penetration} shows the difference in statistics for the two varieties of sampled medium with normal incidence.  This highlights the importance of understanding the physical formation of correlated medium particles in the problem being analyzed and to only use a homogeneous free-path distribution if the correlation is completely preserved up to the boundary.  \new{The need for this precaution has been previously noted in the study of stacked pebbles in pebble bed reactors by Vasques and Larsen~\shortcite{vasques13b} and citations within.  For the case of edge-effects for binary stochastic mixtures, see also~\cite{griesheimer11}}.  The remainder of this paper considers only extended-sampling to ensure homogeneous correlation up to boundary edges.

  \subsection{Single and double scattering from a half space}\label{sec:half-space}

    We compared the predictions of our transport model to Monte Carlo simulation of low order scattering from a Flatland half space with vacuum boundary conditions and isotropic scattering.  The \remove{full} analytic solution for this problem in the case of exponential \new{free-path lengths} is known~\cite{deon18}.  Due to the computational limitations of sampling high numbers of random realizations, we restrict this analysis to single and double scattering only.  The sampling procedure was analogous to previous problems except center-warp collision was performed after the first and (optionally) second scattering.  

    \paragraph{Uncorrelated exponential case}
    We first verified our implementation for uncorrelated disks and found good agreement to the \newtwo{bidirectional reflectance distribution functions} (BRDFs) for single and double scattering $f_1(\mu_i,\mu_o)$ and $f_2(\mu_i,\mu_o)$, respectively, given by
    \begin{equation}
    	f_1(\mu_i,\mu_o) = \frac{c}{2 \pi} \frac{1}{\mu_i + \mu_o}
    \end{equation}
    and
    \begin{equation}
    	f_2(\mu_i,\mu_o) = \frac{c^2}{2 \pi^2} \frac{\frac{\sec^{-1} \mu_i}{\sqrt{1-\frac{1}{\mu_i^2}}}+\frac{\sec^{-1} \mu_o}{\sqrt{1-\frac{1}{\mu_o^2}}}}{\mu_i + \mu_o}
    \end{equation}
    where $\mu_i$ and $\mu_o$ are the direction cosines of the incident and exitant directions, respectively.

\begin{figure*}
	  \centering
	  \subfigure[classical]{\includegraphics[width=.33\linewidth]{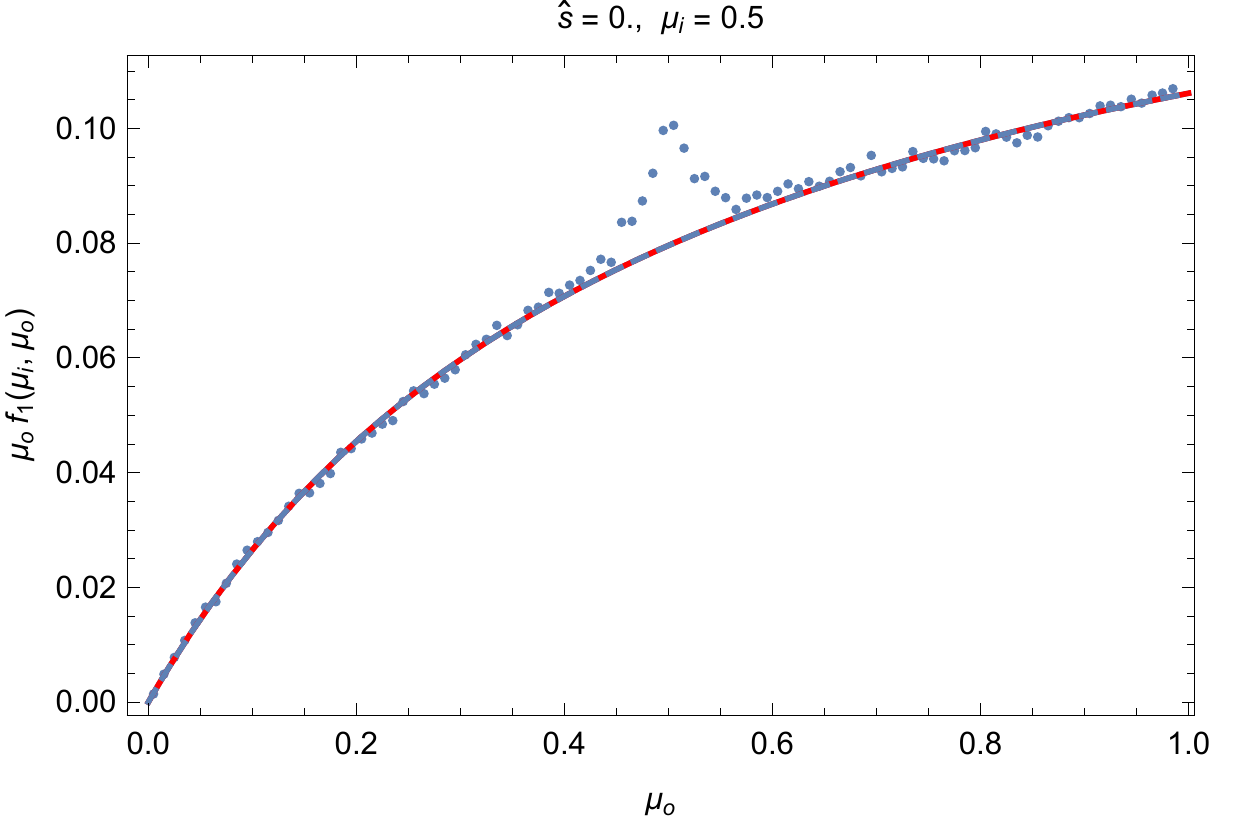}}
	  \subfigure[blue]{\includegraphics[width=.33\linewidth]{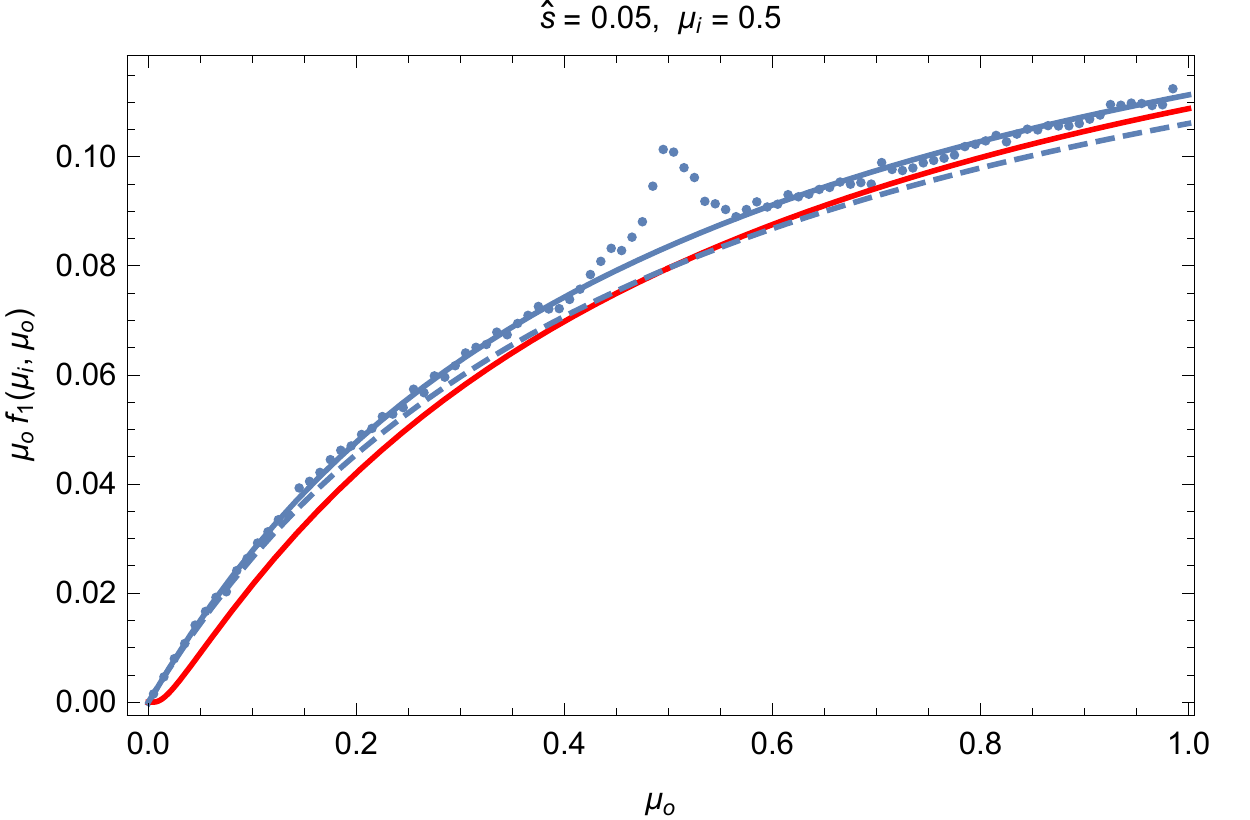}}
	  \subfigure[blue]{\includegraphics[width=.33\linewidth]{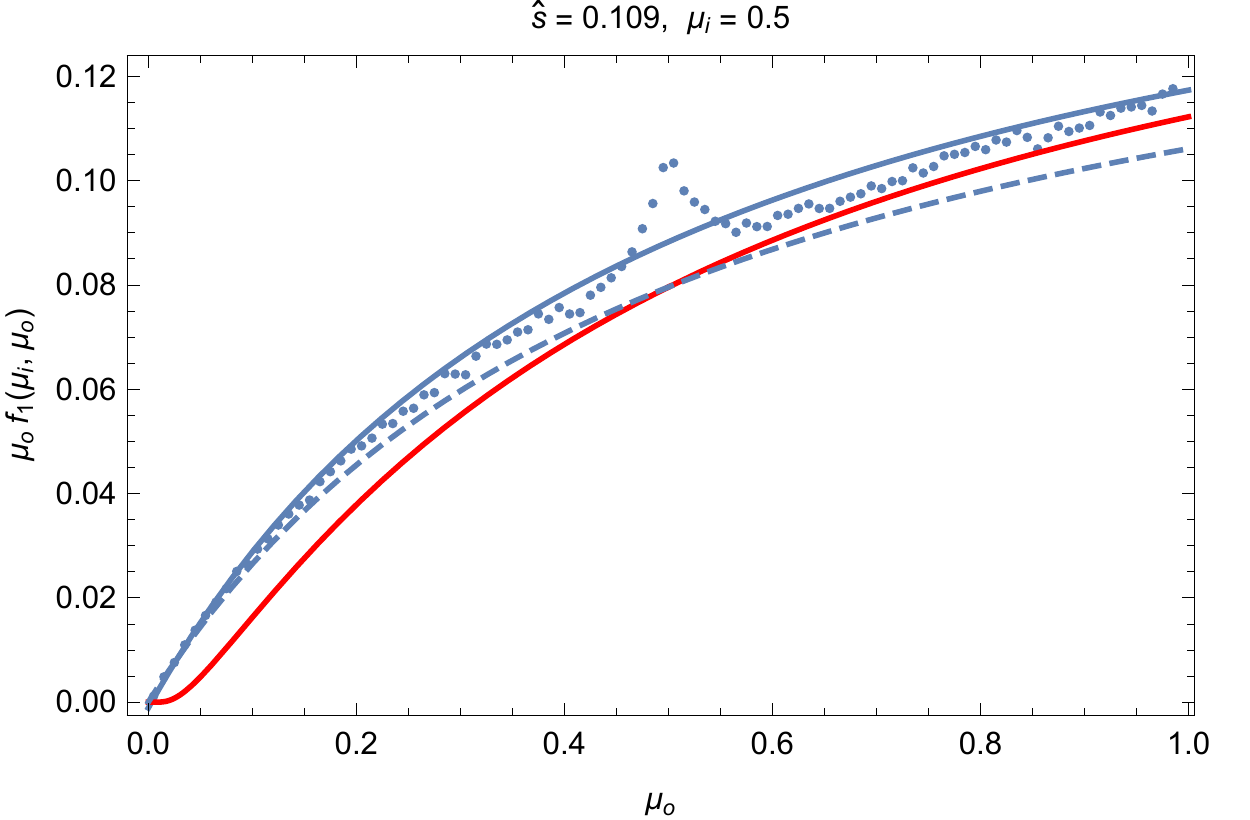}}
	  \subfigure[classical]{\includegraphics[width=.33\linewidth]{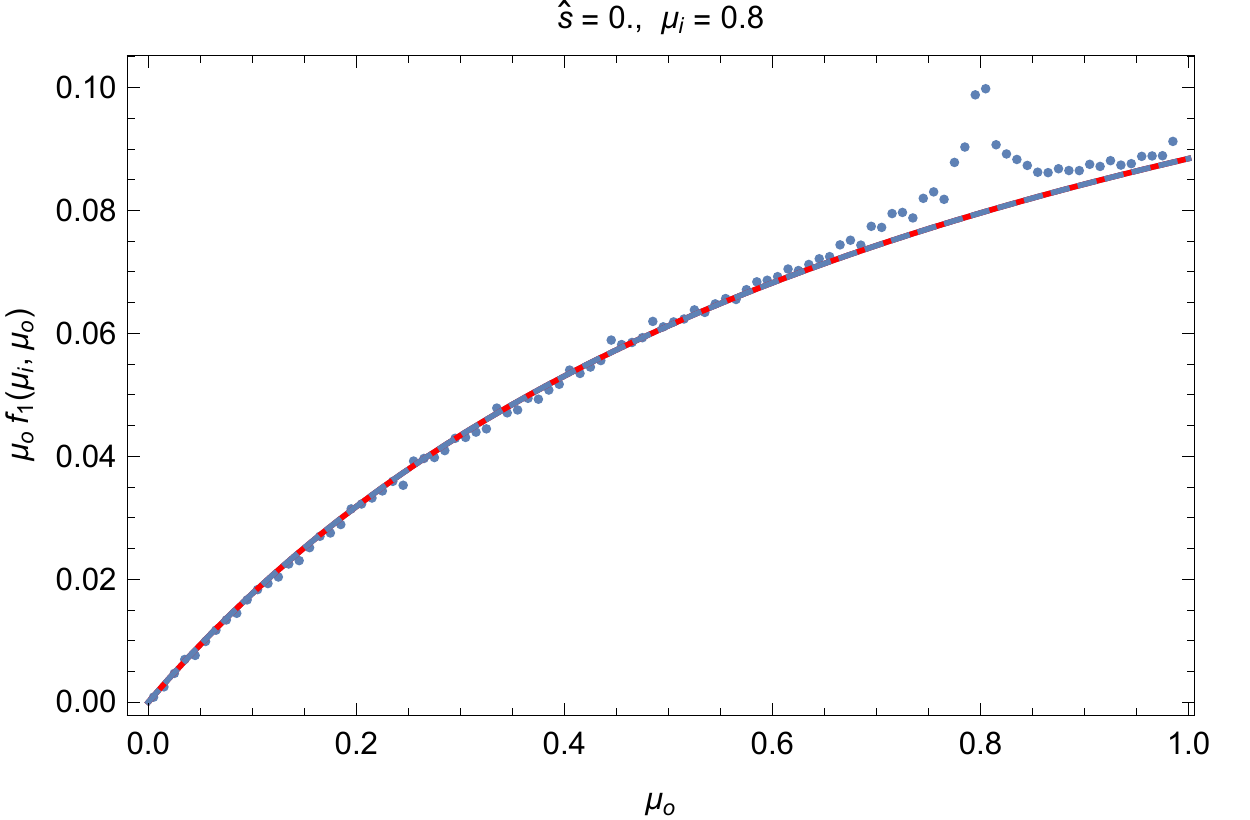}}
	  \subfigure[blue]{\includegraphics[width=.33\linewidth]{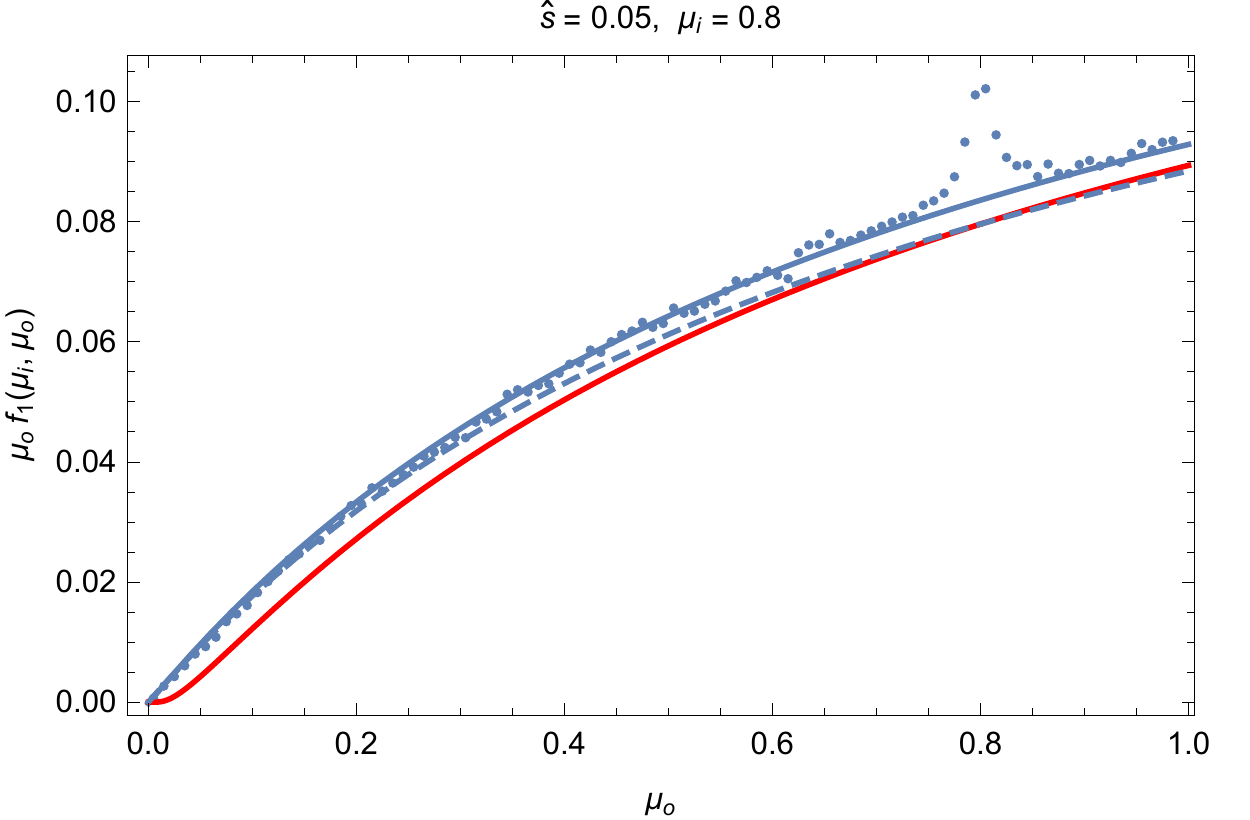}}
	  \subfigure[blue]{\includegraphics[width=.33\linewidth]{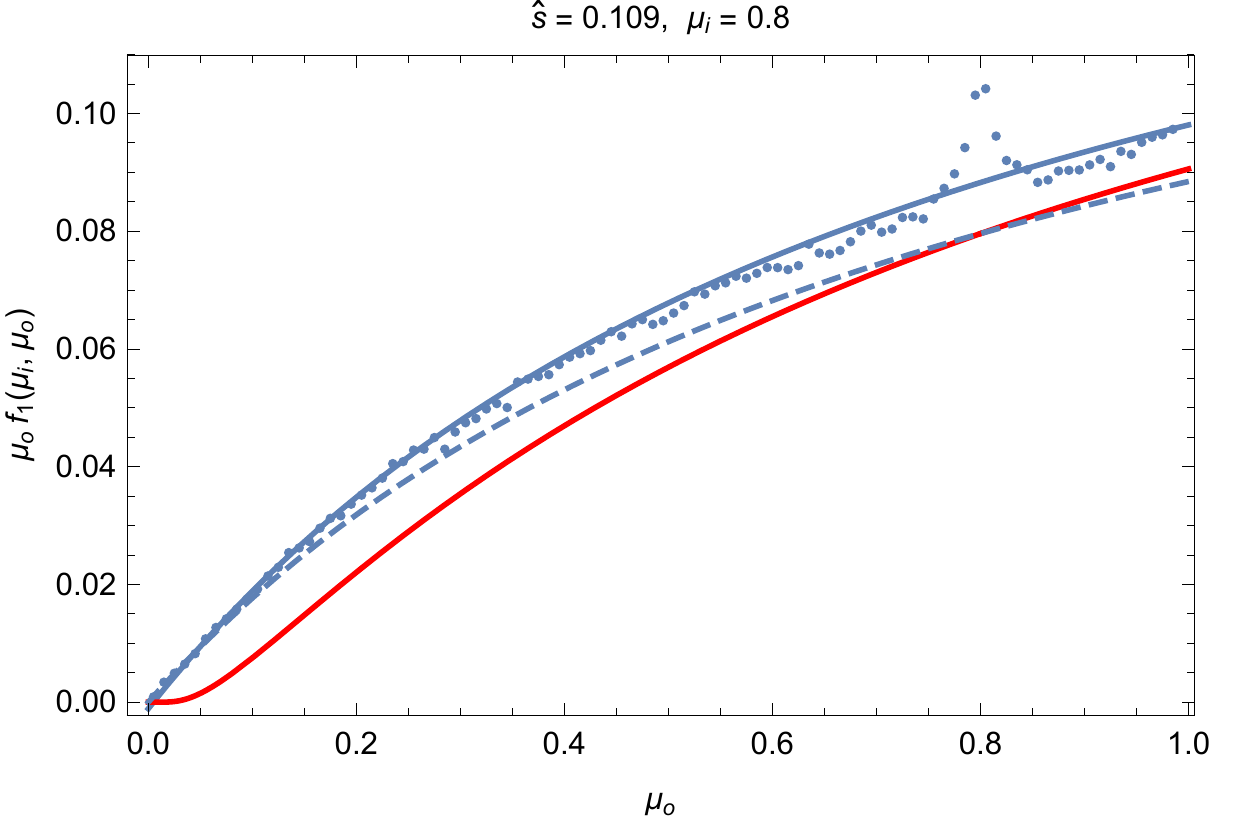}}
	  \caption{Single-scattering from a purely-scattering Flatland half-space with vacuum boundary conditions.  Monte Carlo (dots) vs classical exponential media (dashed) vs our reciprocal model (\new{blue} continuous curve) vs a non-reciprocal model (red).}
	  \label{fig-ssMC} 
	\end{figure*}

    \paragraph{Blue-noise case}
      In our reciprocal formulation, scattering events in the medium are sampled at distances $p_u(s)$ along the incident ray and escape attenuated by $X_c(s)$.  Thus, the single-scattered BRDF can be written
      \begin{equation}\label{eq:ssgeneral}
      	f_1(\mu_i,\mu_o) = \frac{1}{\mu_i \mu_o} \frac{c}{2 \pi} \frac{1}{\mfp_c} \int_0^\infty  X_c \left(\frac{z}{\mu_i}\right) X_c \left(\frac{z}{\mu_o}\right) dz.
      \end{equation}
      Using our proposal for $p_u(s)$ in the blue noise medium we find (for $\mu_i < \mu_o$)
      \begin{equation}
      	f_1(\mu_i,\mu_o) = \frac{(\frac{\smin}{\ell}-1) \mu_i e^{\frac{\frac{\smin}{\ell}
   (\mu_o-\mu_i)}{(\frac{\smin}{\ell}-1) \mu_i}}+\mu_i+\mu_o}{2 \pi 
   \mu_o (\mu_i+\mu_o)}.
      \end{equation}
      In Figure~\ref{fig-ssMC} we compare this to Monte Carlo results and also to the non-reciprocal proposal of previous work to use $p_c(s)$ with $s$ initialized to $0$ at the boundary to enter the medium.  The exponential BRDF is also shown as a reference.  In the case of no particle correlation, we see an agreement between the classical model and simulation \new{with an error comparable to the MC noise}.  An exception to this alignment happens at the coherent back-scattering peak, which is expected.  As the correlation in the scattering particles increases, we see more energy leaving the medium than in the uncorrelated case, which \new{we also see in our model}.  As the separation distance $\smin$ becomes large compared to the classical mean free path we see valleys adjacent to the coherent backscattering peak not seen in the uncorrelated case.  The use of the correlated free path distribution to enter the media is less accurate than our model, pushing energy too far into the medium before the first collision due to the free path distribution being $0$ for $s < \smin$.

      For the case of double-scattering we consistently saw good profile shapes with our model over-predicting the total energy leaving the medium in a way that was also seen in the exponential case due to extra energy in the back-scattering peak leaving less for future scattering orders.
    
  \section{Additional Deterministic Analysis}\label{sec:determ}

  In this section we briefly examine several other forms of non-exponential transport, the distributions needed to apply them with our formalism and derive some new deterministic reciprocal results.

  \subsection{Power-law transmittance}

    Davis and Xu~\shortcite{davis14} proposed a family of free-path distributions for generalized transport with power law transmittance defined by shape parameter $a$.  We can apply this form of non-exponential transport by setting
    \begin{equation}
	    \Sigma_{tc}(s) = \frac{1 + a}{s + a \ell},
	  \end{equation}
	  with $a>0$, which, in the limit $a \rightarrow \infty$ becomes the classical $\Sigma_t = 1 / \ell$ and exponential transport follows.
      From here, free paths between scatterers use
	  \begin{equation}
	  	p_c(s) = \frac{ a (a+1) \ell (a \ell)^a } { (a \ell+s)^{a+2} }
	  \end{equation}
	  with correlated mean $\mfp_c = \ell$ and correlated transmittance
	  \begin{equation}\label{eq:davisXc}
	  	X_c(s) = \left(\frac{a \ell}{a \ell+s}\right)^{a+1}.
	  \end{equation}
	  Sampling is via
	  \begin{equation}
	  	s = a \ell \left((1-\xi)^{-\frac{1}{a+1}}-1\right).
	  \end{equation}
	  For entering the medium or birth from uncorrelated sources, we use
	  \begin{equation}
	  	p_u(s) = \frac{X_c(s)}{\ell},
	  \end{equation}
	  \new{where $X_c$ is given by Equation~\ref{eq:davisXc} and we use} an uncorrelated transmittance
	  \begin{equation}
	    X_u(s) = \left(\frac{a \ell}{a \ell+s}\right)^a
	  \end{equation}
	  for attenuating light paths inside the atmosphere, such as sunlight reflecting off of the earth and reaching a sensor.

	  We could not derive single-scattering generally but for integer or half-integer values of $a$ we see interesting reciprocal closed-form results, such as for $a = 1/2$,
	  \begin{equation}
	  	f_1(\mu_i,\mu_o) = \frac{c}{4 \pi} \frac{1}{2 \sqrt{\mu_i \mu_o}+\mu_i+\mu_o}
	  \end{equation}

	  Alternatively, we could begin with
	  \begin{equation}
	    \Sigma_{tc}(s) = \frac{a}{s + a \ell}
	  \end{equation}
	  to get Davis's attenuation for the correlated free paths instead of for the uncorrelated paths.

\subsection{Gamma-2 steps}

      In 3D, choosing the correlated free-path distribution
      \begin{equation}
      	p_c(s) =s e^{-s}
      \end{equation}
      leads to a density of collision rates about a correlated emitter described exactly by a diffusion equation~\cite{deon14,frank15}.  To apply this to bounded volumes we find the uncorrelated free-path distribution must be
      \begin{equation}
      	p_u(s) = \frac{1}{2} e^{-s} (s+1)
      \end{equation}
      with single-scattering from a half space given by
      \begin{equation}
		f_1(\mu_i,\mu_o) = \frac{c}{4 \pi}  \frac{\mu_i^2+3 \mu_i \mu_o+\mu_o^2}{(\mu_i+\mu_o)^3}.
      \end{equation}
      While the collision density about an isotropic point source is described exactly by diffusion if the emission is correlated to medium collisions, in the case of an uncorrelated point source, diffusion is no longer exact.  Thus, for uncorrelated emission or reciprocal imbeddings in finite media, transport methods cannot be used to exactly solve diffusion problems~\cite{frank15}.  In neither case is the scalar flux / fluence about the point source described exactly by diffusion.

  \section{Conclusion}
  
  We have proposed a novel, reciprocal, equilibrium imbedding of the generalized linear Boltzmann equation into bounded homogeneous domains.  In order to attain reciprocal transport over all transport subpaths whose end points are not correlated medium events requires using the equilibrium distribution of free paths when sampling any path that begins on the surface of a medium boundary, inclusion or uncorrelated emitter.  The uncorrelated free-path distribution is computed easily from the correlated free-path distribution provided is has a finite mean.  We have studied multiple scattering in bounded domains with negatively correlated scattering centers and shown consistent improvement over alternative proposals for bounded non-exponential transport.  The analysis has highlighted the need to precisely understand how particle correlation near boundaries influences free paths in those regions.  

  The approximate, analytic forms we derived for blue noise free-path statistics may prove useful for predictive \new{condensed-history~\cite{moon07,meng2015multi,muller2016efficient}} applications of our transport formalism and might also generalize to include power-law transmittance tails.

  In the following parts of this work we present a complete formalism with considerations for many of the advanced complexities and subtleties of building robust Monte Carlo codes with multiple variance reduction techniques and their combinations.  Beyond this, there is also need to define what heterogeneous densities of correlated centers are and how they should be treated.  Finally, we have not included a study of reflective boundary conditions and measured the errors due to angular memory in this case.  \remove{It may indeed be more accurate or appropriate to reflect off of a boundary and continue with a correlated free path, perhaps remembering $s$ post reflection.}

\section{Acknowledgements}
  \new{We would like to thank Johannes Hanika and the anonymous reviewers for their recommendations for improving the paper.}

\bibliographystyle{acmsiggraph}
\bibliography{sigg18talk}

\end{document}